\documentclass[article,12pt]{JHEP3}
 \usepackage{graphicx}
 \usepackage{bm}
 \usepackage{braket}
\usepackage{epsf,
shadow,pifont}
 \usepackage{amsmath,epsfig}
 \usepackage{amssymb,amsfonts,amsthm,amstext,amscd,dsfont}
\usepackage{latexsym}
\usepackage{slashed}
\usepackage{float}
\usepackage{epsfig}
\usepackage{xcolor}

\def\hri#1#2{\href{http://arxiv.org/abs/#1}{[ArXiv:#1]#2}}
\def\hre#1#2{\href{http://arxiv.org/abs/#1/#2}{[ArXiv:#1/#2]}}

\def\hsp#1#2{\href{https://doi.org/#1}{#2}}
\relax
\renewcommand{\theequation}{\arabic{section}.\arabic{equation}}

\def\be{\begin{equation}}
\def\ee{\end{equation}}

\newcommand{\bear}{\begin{eqnarray}}
\newcommand{\bea}{\begin{eqnarray}}
\newcommand{\eear}{\end{eqnarray}}
\newcommand{\eea}{\end{eqnarray}}
\newcommand{\unit}{\mathds{1}}
\newcommand{\Tr}{\mathrm{Tr}\,}
\newcommand{\morder}[1]{\mathcal{O}\left(#1\right)}
\newbox\pippobox

\def\II{\relax{\rm I\kern-.18em I}}

 \def\m{\mu}
\def\n{\nu}

 \def\s{\sigma}
 \def\pa{\partial}

 \def\sp{\;\;\;,\;\;\;}

 \def\a{\alpha}

\def\tr{\ensuremath{\mathrm{Tr}}}

\def\l{\lambda}

\def\II{{\cal I}}

\title{Non-derivative Axionic Couplings to Nucleons at large and small N}

\author{Francesco Bigazzi$^{a}$, Aldo L. Cotrone$^{a,b}$, Matti  J\"arvinen$^{c,d}$ and Elias Kiritsis$^{e,f}$\\
 ~\\
 $^a$ \href{http://theory.fi.infn.it/index/}{INFN, Sezione di Firenze}; Via G. Sansone 1; I-50019 Sesto Fiorentino (Firenze), Italy\\
$^b$ \href{https://www.fisica.unifi.it/mdswitch.html}{Dipartimento di Fisica e Astronomia}, Universit\'a di Firenze; Via G. Sansone 1; I-50019 Sesto Fiorentino (Firenze), Italy\\
$^c$  \href{https://www.uu.nl/en/research/institute-for-theoretical-physics}{Institute for Theoretical Physics}, Utrecht University, Princetonplein 5,\\ 3584 CC Utrecht, The Netherlands\\
$^d$ \href{https://www.helsinki.fi/en/faculty-of-science/faculty/physics}{Department of Physics} and \href{https://www.hip.fi/}{Helsinki Institute of Physics}, P.O. Box 64, FI-00014 University
of Helsinki, Finland\\
$^e$ \href{http://hep.physics.uoc.gr/}
 {Crete Center for Theoretical Physics}, Department of Physics, University of Crete,
 71003 Heraklion, Greece\\
$^f$ \href{http://www.apc.univ-paris7.fr}{APC, Universit\'e Paris 7}, CNRS/IN2P3, CEA/IRFU, Obs. de Paris, Sorbonne Paris Cit\'e, B\^atiment Condorcet, F-75205, Paris Cedex 13, France (UMR du CNRS 7164).\\
}

\preprint{CCTP-2019-6\\ITCP-IPP 2019/6}

\abstract{
Among the possible CP-odd couplings of the axion to ordinary matter, the most relevant ones for phenomenology are the Yukawa couplings to nucleons. We analyze such non-derivative couplings within three different approaches: standard effective field theory, the Skyrme model and holographic QCD.  In all the cases, the couplings can be related to the CP-odd non-derivative couplings to nucleons of the low-lying mesons and the $\eta'$. Using the effective field theory approach we discuss how to derive the expressions for the CP-odd interaction terms as functions of the parameters of the effective Lagrangian
at generic number of colors $N_c$  and flavors $N_f$.
Then, we compute the CP-odd couplings to nucleons of the axion, the $\eta'$ and the pseudo-Goldstone mesons in both the Skyrme and the holographic QCD model with $N_f=2,3$.
We present model-independent expressions for the coefficients of the non-derivative axion-nucleon couplings. This allows us to provide quantitative estimates of these couplings.
}

\makeindex
\begin{document}

\section{Introduction and results}
The Peccei-Quinn proposal \cite{PQ} for a natural solution of the strong CP problem - the unnaturally small value of the QCD $\theta$ angle - implies the existence of a new light neutral pseudoscalar boson, the axion \cite{Wilczek:1977pj,Weinberg}. The original theory, severely constrained by data, was not renormalizable because of the axion coupling to the QCD instanton density \cite{Witten:1979vv, Veneziano:1979ec}. In renormalizable axion models, successively proposed in \cite{Kim,SVZ,DFS}, the axions were also made very weakly interacting with ordinary matter, as required by experimental constraints. These ``invisible axions'', whose couplings to matter have been deduced using anomalies and the chiral Lagrangian \cite{Kaplan}-\cite{DiVecchia:2017xpu}, are nowadays considered among the most promising candidates as dark matter constituents, and also as possible realizations of the inflaton \cite{Kim2}-\cite{IR}.

As of today, axion-like particles (ALPs) are ubiquitous and serve various purposes. The nomenclature has also evolved but still remains sometimes murky. Axions that solve the strong CP problem are typically called QCD axions.
The term ``axion-like particle'' is often used to refer (only) to other types of axions.

Generically, axions display a perturbative shift symmetry and couple to instanton densities. In any reliable quantum field theory (QFT) realization, the symmetry is broken (at best) to a discrete symmetry due to non-perturbative effects. Such effects, related to instantons in weakly-coupled models, induce a mass and, more generally, a potential term, for the QCD axion (see {\it e.g.} \cite{villadoro}).

ALPs arise very commonly in string theory \cite{SW}, the simplest example being the Ramond-Ramond (RR) axion of type IIB string theory. They can also arise, after compactification, from internal components of antisymmetric form gauge fields as well as from off-diagonal components of the metric. In these cases, ALPs are related to generalized gauge fields \cite{pol,k}. The corresponding gauge symmetries provide the perturbative Peccei-Quinn (PQ) symmetries in string theory \cite{s}.

Continuous shift symmetries in string theory can be broken by non-perturbative effects.
The argument is general:\footnote{In fact it also applies to NS-NS axions that couple to world-sheet or NS5-brane instantons.} RR axions couple to the world volume of D-branes \cite{pol2}. The same D-branes, wrapped around some appropriate Euclidean cycle, provide instanton effects in string theory \cite{BBS}-\cite{revi}. The nature of these effects depends on the amount of supersymmetry. In the case of maximal supersymmetry they do not generate a potential, but affect higher derivative terms like the $R^4$ corrections, \cite{gg,kp}.
In all cases, the end result is that the original shift symmetry is broken to a discrete subgroup.\footnote{There can be subtleties that arise when axionic symmetries are coupled to anomalous U(1)'s \cite{Bianchi, Bianchi:2007wy}, but the final result is analogous.}

In the above descriptions (and in other QFT constructions), ALPs are fundamental fields in the theory. However this is not the only possibility. ALPs can also be composite or emergent.  In its most common realizations \cite{Kim1}-\cite{Redi}, \cite{Kim2}, the composite axion is identified with a neutral pseudo-Goldstone boson of an extra, strongly coupled, confining gauge theory, whose dynamical scale $\Lambda_a$ determines the axion coupling $f_a$. In these models, the axion can be rendered ``invisible'' only if  $\Lambda_a\gg \Lambda_{QCD}$, \cite{Kim2}. As a common feature, the Peccei-Quinn symmetry in these models acts on some extra fermion fields. Moreover, as a difference with fundamental axions, in simple composite axion models there is no direct coupling between the axion and the lepton sector.

In a more general setup, recently considered in \cite{axions} and motivated in \cite{1}, ALPs emerge as composites of a hidden sector with a large number of colors ($N_c\gg1$), which couples to the Standard Model (SM) via interactions that are irrelevant in the IR. This fact ensures that one can have an invisible emergent axion even if  the strong coupling scale $\Lambda_h$ of the hidden gauge theory is much smaller than $\Lambda_{QCD}$. These emergent/composite  axions do not rely necessarily on fermionic Peccei-Quinn like symmetries, but rather on the approximate Peccei-Quinn symmetry associated to instanton densities in gauge theories.

The holographic correspondence \cite{Malda} provides a paradigm for such a type of composite/emergent axions. In the holographic context, in fact, composite operators in a QFT are mapped into fundamental fields in the dual string/gravity theory.
Hence, standard string theory axions can be related to composite operators in QFT.
This is the case for the QCD topological $\theta$ term \cite{Witten-t}-\cite{Hamada}.

The holographic QCD axion model recently proposed in \cite{bccdv} provides a further different way of realizing a composite axion. In the model there is no hidden gauge theory sector, hence no extra dynamical scale. The Peccei-Quinn symmetry acts on an extra massless quark flavor with a (non-local) Nambu-Jona-Lasinio (NJL) quartic interaction. The latter has a suitable higher dimensional UV completion in the holographic model and it induces the condensation of the extra quark at a scale $M_a$, of the order of $f_a$, set by the NJL coupling. If $M_a$ is taken to be much higher than the QCD chiral symmetry breaking scale, the extra quarks and the related massive hadrons essentially decouple from the Standard Model. The only exception is the pseudo-Goldstone boson of the extra chiral symmetry breaking. This is identified with the composite QCD axion. Its interactions with the low energy QCD modes are precisely described by the axion-dressed chiral Lagrangian \cite{DiVecchia:2017xpu}.

The couplings of the QCD axion and other ALPs to ordinary matter are crucial in axion searches. They typically receive contributions both from the UV realization of the ALP theory and from the QCD chiral Lagrangian and the structure of the axial anomaly \cite{Sre,Kaplan,GKR,villadoro}. These are combined with some input from QCD matrix elements that can be extracted from experiments or, more recently, from lattice data \cite{lattice}. State of the art derivative axion couplings to nucleons are obtained in this way.

There exist also non-derivative couplings to nucleons that so far have not received much attention (see however \cite{Wilczek}).
There are two reasons for this. The first is that such couplings, when they arise from the CP-preserving effective Lagrangian for the nucleons, are effectively derivative couplings. This is the case when the nucleons are near the mass-shell, as it happens at low enough energies. If the couplings arise in the CP-violating sector of the effective Lagrangian, then they are suppressed by the small CP violation due to a residual $\theta$ angle and/or to SM effects.\footnote{The CP-violation due the $\theta$-term is constrained to be tiny by the neutron electric dipole moment measurements. The CP-violation appearing in the CKM matrix trickles also down to the hadronic sector at high loop order, \cite{Khri,EG,GM}. Similar remarks apply to the leptonic CKM matrix, \cite{DP}.} The second reason is that such non-derivative couplings to nucleons depend on non-trivial and non-perturbative QCD physics, and chiral symmetry considerations (and the chiral Lagrangian) are not helpful enough to pin them down.

In this paper we will undertake an analysis of such couplings. There are two main motivations behind our analysis. The first is that ALPs are generically speaking light particles and the range of allowed masses towards the lower end has expanded substantially in recent years, especially as candidates for dark matter, \cite{Marsh,Graham,IR,Hui}. In such cases, non-derivative couplings, even if small, can become important in coherent collections of ALPs. The second is that we now have a new theoretical tool to address non-perturbative nucleon physics and this is the holographic correspondence.

In the holographic approach, QCD is modeled by a large $N_c$ strongly coupled gauge theory, which can be studied by means of a dual weakly coupled string model on a classical gravity background. The top-down holographic model which better reproduces the main features of low energy QCD is due to Witten, Sakai and Sugimoto (WSS) \cite{witten,SS1}.  Nucleons in this theory have been first studied in \cite{Yi,SS-baryons,hss}. They are realized in a way which resembles the traditional Skyrme picture \cite{Skyrme}, where nucleons arise, at large $N_c$, as solitons of the chiral Lagrangian \cite{anw}. In the WSS model they are instanton solutions of a five dimensional $U(N_f)$ gauge theory ($N_f$ being the number of flavors) on a certain curved background. After reduction to four dimensions this theory encodes not only the chiral Lagrangian (including the Skyrme term) for the QCD-like pseudo-Goldstone bosons and the $\eta'$, but also the effective Lagrangian for the massive axial (vector) mesons. Moreover, couplings and masses for the mesons are determined by the few parameters of the model. The holographic QCD axion in \cite{bccdv} is embedded in the WSS model.

There are moreover bottom up models for holographic meson and baryon physics that have been analyzed in the past. The first and crudest of them named AdS/QCD, \cite{son,DP1}, reproduced the physics of the chiral Lagrangian to leading order and has given rather good predictions of parameters for the chiral Lagrangian beyond the two-derivative level.
There are more sophisticated bottom-up models for QCD, like V-QCD, \cite{VQCD1,theta} that improves importantly on AdS/QCD.

In this paper we embark in an effort to determine the non-derivative couplings of the
axion to nucleons under three different perspectives: standard effective field theory, Skyrme model and holographic WSS model.\footnote{Analysing the same effects in the more sophisticated but also more difficult theory, namely V-QCD is beyond the scope of the present work.} In our analysis of the Skyrme and holographic model, the large $N_c$ limit is taken by construction.
In all the three approaches, the axion couplings can be related to those of the $\eta'$ and the pions to the nucleons, after considering the mixing between the axion and these mesons. One loop corrections in the effective theory of nucleons and mesons can contribute too. The mixing arises diagonalizing the mass matrix in the low energy effective action.
It is worth emphasizing that meson-nucleon couplings have been largely studied in chiral perturbation theory. For instance, the CP-odd pion-nucleon coupling has been computed in {\it e.g.}~\cite{crewther, Mereghetti:2010tp,deVries:2015una,Shindler:2015aqa} for $N_f=2,3$ at various orders in the perturbative expansion. We will show how our results in the Skyrme and holographic model compare with those findings.

A word of caution is in order.
The Skyrme and holographic models, while reproducing many features of low-energy QCD, are certainly not suitable for precision physics at percent level as there is no formal parameter that controls their accuracy.
Nevertheless, their use alongside the standard Chiral Lagrangian approach is justified by the following considerations.
To begin with, the CP-odd sector of the axion physics is difficult to be analyzed directly within any first-principle model (e.g.~one has a sign problem on the lattice).
Moreover, while in the original Skyrme model the nuclear binding energies are notoriously overestimated, its extension including vector mesons improves significantly this issue \cite{Sutcliffe:2010et}.
Finally, bottom-up holographic models for meson physics have been surprisingly successful in reproducing experimental data, and organizing (subleading) parameters that are theoretically unconstrained in the chiral Lagrangian.
The holographic WSS model is a top-down model and  has a proven discrepancy with experimental results of less than 30\% (in average) when dealing with static properties of nucleons, including couplings similar to the ones computed in this paper \cite{Hashimoto:2008zw}.  It is also worth mentioning that if, in principle, Chiral Lagrangian calculations (combined with large $N_c$ arguments) can be systematically improved, in practice they are limited by unknown low-energy constants. Large $N_c$ models like Skyrme and WSS can provide model-dependent estimates of such unknowns.\footnote{Moreover, in principle, the top-down holographic QCD approach can also be systematically improved, up to the point of providing a string dual of real world QCD: however, in practice, this reveals to be a very hard task.}

Our main results are those for the CP-odd non-derivative couplings of the pseudoscalar mesons and the axion to the nucleons.
They can be summarized as follows.
\begin{itemize}
\item Using chiral perturbation theory, we derive the expressions for the CP-odd interaction terms, at any number of colors $N_c$ and flavors $N_f$, between $\eta'$ or the axion and nucleons (transforming under generic spin-half representations of the gauge group) in terms of the parameters of the effective Lagrangian. The main results are the Lagrangian terms in equations~\eqref{thetacouplings1}--\eqref{thetacouplings2} (for contributions arising through a ``residual'' $\theta$-angle, possible even in the presence of the axion\footnote{There are many possible contributions to $\theta$ in our formulae, and include uncanceled contributions from the strong sector or induced by the CKM CP violation.}).
\item We estimate, in turn, the one-loop contributions to the non-derivative couplings in effective field theory using previous results,  \cite{Gutsche:2016jap} and extending them to general $N_c,N_f$.
\item In chiral perturbation theory, we write the CP-odd non-derivative couplings of the axion to the nucleons both in terms of the CP-odd couplings of the pions and $\eta'$, equation (\ref{34}), and, alternatively, in terms of the pion-nucleon sigma term and the strong contribution to the neutron-proton mass splitting, equation (\ref{geneform}), extending the formula in \cite{Wilczek}.
\item We compute the CP-odd couplings (due to the effective $\theta$ angle and other sources of CP-violation in the QCD sector)
 of the axion, the $\eta'$ meson and pions to nucleons  in the Skyrme model. The main results for $N_f=2$ are given in equations~\eqref{cngen} and~\eqref{cpoddskyrme}, and for $N_f=3$ in equations~\eqref{bargaNNS3},~\eqref{bargetap3} and~\eqref{bargeta3}. To the best of our knowledge, these couplings have not been computed in the literature so far.
\item We compute in turn novel results for the same couplings
 in the WSS model. The results for $N_f=2$ are given in equations~\eqref{cngenwss},~\eqref{bargetawss}, and~\eqref{gbarpiNNWSS}. The generalization to $N_f=3$ is analogous to the computation in the Skyrme model.
\end{itemize}

In section \ref{sectionnumbers} we will compare the numerical estimates of the axion couplings to nucleons obtained in model-independent expressions from chiral perturbation theory plus lattice and phenomenological inputs, with the ones obtained in the Skyrme and holographic WSS models.
The first ones are affected by some uncertainties on the value of the sigma term, while the latter depend on the specific models.
Although there is a certain discrepancy in the results, a reasonable range of the value of the couplings appears to be
\be\label{firstbis}
{\bar c}_N = (14-21)\ \mathrm{MeV}\times \frac{\theta}{f_a}\ ,
\ee
having in mind an effective Lagrangian term of the form $\delta{\cal L}=  {\bar c}_N a \bar N N$, where $N$ is the nucleon doublet and $a$ is the axion.
Here we do not distinguish the proton and neutron couplings, whose difference is smaller than the proposed window of values.
Also, we have taken into account the one loop corrections in the effective theory, discussed in section \ref{loop}. They are somewhat smaller compared to the numbers in (\ref{firstbis}) for $N_f=2$, $N_c=3$.

%

The paper is organized as follows. In section \ref{CHPTL} we review the construction of the QCD chiral Lagrangian, containing the $\eta'$ meson and the pions, coupled with the axion. In particular we write down the field redefinitions suitable for diagonalizing the related mass matrix. In section \ref{sec:axionnucleon} we discuss the effective theory describing the coupling of mesons and axions to nucleons and deduce the general structure of the non-derivative axion-nucleon couplings. One loop corrections are discussed here too.
These CP-odd non-derivative axion couplings are expressed in terms of the pion and $\eta'$ couplings to nucleons, and in turn of the pion-nucleon sigma term and proton-neutron mass difference in section \ref{sec:generalIR}.
In section \ref{sectionskyrme} we compute the couplings in the Skyrme model with $N_f=2$ QCD flavors.
In section \ref{sec:WSS}, after a short review of the description of nucleons in the holographic setup,  we perform analogous computations in the WSS model with $N_f=2$ QCD flavors.
A summary and an analysis of the results obtained in sections \ref{sec:generalIR}, \ref{sectionskyrme} and \ref{sec:WSS} is given in section \ref{sectionnumbers}.  Review material, further technical details and the results for $N_f=3$ QCD flavors in the Skyrme and WSS models are provided in the appendices.

\section{The effective Lagrangian couplings of the axion to the $\eta'$}\label{CHPTL}

The general structure of axion models and their UV completions are summarized in appendix \ref{structure}.
Here, we examine the effective Lagrangian (containing the pions and the $\eta'$ meson) for QCD coupled to the axion. Our aim is to write the effective theory for generic axion couplings rather than choosing a specific ``frame'', which allows us to check that the final results, in term of physical states, are frame independent. This will be useful in particular when considering the couplings of the $\eta'$ and the axion to the nucleons as we will do in Sec.~\ref{sec:axionnucleon}.

The QCD Lagrangian with generic renormalizable axion couplings\footnote{There may also be non-renormalizable axion couplings if the theory that generates the axion is an effective theory. This is the case in the context of composite or emergent axions, \cite{axions} but may also be relevant for the case of fundamental axion fields that appear at intermediate scales.}  is given by
\begin{align} \label{LQCDaxion}
  S_\mathrm{a+QCD} = \int d^4 x ~\Big(&-\frac{1}{2}\partial_\m a \partial^\m a- \frac{a}{f_a} \frac{c_s\a_{s}}{8\pi}\mathrm{Tr}[G\wedge G]+\frac{1}{f_a}\pa_{\m} aJ^{\m}_{q5} - \frac{1}{2}\mathrm{Tr}[G^2] &\nonumber\\
  &- i \bar q \gamma^\mu D_\mu q - \bar q_R M(a) q_L - \bar q_L M(a)^\dagger q_R \Big) \,.
\end{align}
Here $q_{L/R} = (1\pm\gamma_5) q/2$ are the quark fields, we suppressed flavor indices, used  mostly plus convention for the Minkowski metric, and
also included the axion dependence in the mass terms $M(a)$ as well as the axion kinetic term. The $\theta$-angle can be absorbed in the constant term of $a$ so we set it to zero -- effects due to finite $\theta$ will be considered later on. We also included the current term $\propto \pa_{\m} aJ^{\m}_{q5}$ here, but as we argue in Appendix~\ref{app:chpt}, it does not affect those mixing terms between the axion and mesons which we are interested in. Therefore we will set the current to zero in the following.

We shall then derive the low energy effective action for pions and the $\eta'$ mesons which contains the axion couplings determined by~\eqref{LQCDaxion}. Axion couplings in effective theories for QCD have been considered in earlier works~\cite{DiVecchia:1980yfw,GKR}, see for example the fairly recent detailed analysis~\cite{villadoro}. In our analysis, it is essential to also include the effect of the $\eta'$ meson for two main reasons. First, we are interested in the couplings of the axion both at small and large $N_c$, and the axial anomaly of QCD is suppressed at large $N_c$ so that the $\eta'$ meson is as light as the pions. Second, as we shall see, the mixing between the $\eta'$ and the axion is particularly important for the non-derivative couplings of the axion to the nucleons which we will analyze below. Earlier work considering the effect of the $\eta'$ in the axion-meson mixing include~\cite{Choi2,Evans:1996eq}.
Since we are mostly interested in the mixing of the axion with the mesons, we will concentrate on the terms giving rise to quadratic terms in the action.

The first interaction term in~\eqref{LQCDaxion} couples, in the effective theory, the axion to the pseudoscalar glueball field~\cite{DiVecchia:1980yfw,schechter}. The glueball can be integrated out, giving rise to the  ``standard'' effective Lagrangian with the $\theta$-angle replaced by the axion field.
The meson fields are included in
\be
 U=\exp\left[\frac{i\sqrt{2}\eta'}{\sqrt{N_f}f_{\eta'}}+\frac{i\pi^aT^a}{f_\pi}\right] = \exp\left[\frac{i\sqrt{2}\eta'}{\sqrt{N_f}f_{\eta'}}+\frac{i\Pi}{f_\pi}\right]\,,
\label{1}\ee
where $\Pi = \pi^a T^a$. The matrix $U$ transforms as
\be
U \to L U R^\dagger ~~~{\rm under}~~~~ q_{L} \to L\, q_{L}\sp q_{R} \to R\, q_{R}\;,
\label{2} \ee
where $L,R$ are $U(N_f)$ matrices. We normalize the group generators as
 \be
 \mathrm{Tr}[T^aT^b] =2\delta^{ab}\;.
\label{3}  \ee
  Notice that $\eta'$ is not a mass eigenstate but will mix with the other fields,  as we will discuss below.

With these conventions, the effective Lagrangian that is the low-energy theory of \eqref{LQCDaxion} (with the axion derivative/current term omitted) can be written as,\footnote{As usual this is based on symmetries only. See Appendix~\ref{app:chpt} for a more detailed discussion.}
\begin{align} \label{chiralL1}
{\cal L}_{\rm chiral} &=-\frac{1}{2}\partial_\m a \partial^\m a-{ f_{\pi}^2\over 4}\left(\mathrm{Tr}[\pa_{\m}U\pa^{\m}U^\dagger]
+\mathrm{Tr}[ M_\pi(a) U^\dagger+ M_\pi(a)^{\dagger}U]\right) &\nonumber\\
&-
\frac{\chi_\mathrm{YM}}{2}\,\left(\frac{ c_s\,a}{f_a}-i\log\det U\right)^2 + \frac{c_{\eta'}}{N_f} \mathrm{Tr}[U^\dagger\pa_{\m}U]\mathrm{Tr}[U\pa^{\m}U^\dagger]\,,
\end{align}
where $M_\pi$ is the pion mass matrix, and $\chi_\mathrm{YM}$ is the Yang-Mills topological susceptibility.

We allowed the $\eta'$ to have a different decay constant than the pion~\cite{Kaiser:1998ds} by including the last term in~\eqref{chiralL1}. Requiring the standard kinetic term normalization fixes
\be
 c_{\eta'} = \frac{f_\pi^2-f_{\eta'}^2}{4}\, .
\label{4}\ee

The meson mass matrix is proportional to the quark mass matrix, $ M_\pi(a) = B_0 M(a)$ up to higher order chiral corrections.

We assume that it is possible to write the mass matrix in the form
\be
M(a) = e^{ia \tilde Q/f_a} M_0 e^{ia \tilde Q/f_a}\,,
 \label{7}\ee
 (possibly after vectorial axion-valued transformations) where\footnote{We do not need an explicit formula for this matrix, but a convenient choice would be $M_0 = \mathrm{diag}(m_1,m_2,\ldots,m_{N_f})$.}  $M_0$ is Hermitean and independent of $a$. It is then straightforward to solve explicitly for the mixing between the mesons and the axion. Effects due to an anti-Hermitean component in the mass matrix are discussed in Appendix~\ref{app:chpt}.
We define the topological susceptibility of full QCD, $\chi$, and the ``gauge invariant'' coupling of the axion, $\tilde c_s$, as
\be
 {1\over \chi} =  {1\over \chi_\mathrm{YM}} + 2 {\mathrm{Tr}\left[ M_0^{-1}\right]\over B_0 f_\pi^{2} }
 \ , \qquad \tilde c_s = c_s+ 2 \mathrm{Tr} \tilde Q \,.
\label{8}\ee
In order to diagonalize the Lagrangian, we first carry out the following field redefinitions (see also~\cite{Choi2}):\footnote{The $_{p}$ stands for ``physical".}
\begin{align} \label{transform1}
 \hat \eta' &= \eta' - \frac{\sqrt{2} f_{\eta'}}{\sqrt{N_f}} \mathrm{Tr}[\tilde Q ]\, \frac{a}{f_a}+ \frac{ \sqrt{2} \tilde c_s\,\chi\,f_{\eta'}\,\mathrm{Tr} \left[ M_0^{-1}\right]}{f_\pi^2\, B_0 \sqrt{N_f}} \frac{a}{f_a} \ , \\
 \label{transform2}
 \hat \pi^b &= \pi^b -  f_\pi \mathrm{Tr}[\tilde Q T^b ]\, \frac{a}{f_a}+ \frac{\tilde c_s\,\chi\,\mathrm{Tr}\left[M_0^{-1} T^b\right] }{f_\pi B_0} \frac{a}{f_a}\,,
 \\
 \label{transform3}
  a_p &= a + \frac{ f_\pi}{f_a} \mathrm{Tr}\left[\tilde Q
  \Pi\right]+\frac{\sqrt{2} f_{\eta'}}{f_a\sqrt{N_f}} \mathrm{Tr}\left[
  \tilde Q\right]\eta'\nonumber\\
  & \quad  \ \ \,
  - \frac{\tilde c_s\,\chi\, }{f_\pi^2 f_a B_0}\left(f_\pi\mathrm{Tr}\left[M_0^{-1} \Pi\right]+\frac{\sqrt{2}f_{\eta'}}{\sqrt{N_f}}\mathrm{Tr}\left[M_0^{-1}\right]\eta'\right)\,,&
\end{align}
which remove the mixing of the pions and $\eta'$ with the axion up to terms suppressed by higher powers of $1/f_a$. More precisely, the mixing terms in Eqs.~\eqref{transform1} and~\eqref{transform2} are determined by the non-diagonal elements of the mass matrix in~\eqref{chiralL1}. Notice that the terms involving $\tilde Q$ are (up to nonlinear terms in the pions) equivalent to a chiral rotation which removes the $a$ dependence of $M(a)$. Then the other terms are sourced by the gauge invariant coupling $\tilde c_s$. After applying \eqref{transform1} and~\eqref{transform2}, the transformation~\eqref{transform3} removes the non-diagonal kinetic terms.

Inserting these relations in the Lagrangian~\eqref{chiralL1}, the quadratic terms (with terms suppressed by $1/f_a$ dropped) are included in
\begin{align}
\hat{\mathcal{L}}_\mathrm{chiral} &= -\frac{1}{2}\partial_\m a_p \partial^\m a_p -\frac{1}{2} m_a^2 a_p^2
-{ f_{\pi}^2\over 4}\mathrm{Tr}[\pa_{\m}\hat U\pa^{\m}\hat U^\dagger]+ c_{\eta'} \mathrm{Tr}[\hat U^\dagger\pa_{\m}\hat U]\mathrm{Tr}[\hat U\pa^{\m}\hat U^\dagger] &\nonumber\\
& \phantom{=}- \frac{N_f\chi_\mathrm{YM}}{f_{\eta'}^2} \hat \eta'^2+{f_{\pi}^2B_0 \over 4}\mathrm{Tr}[M_0 \hat U^\dagger+ M_0^{\dagger}\hat U]\ , &
\label{piLhat}
\end{align}
where
\be
\hat U=\exp\big[i\sqrt{2}\hat\eta'/\sqrt{N_f}/f_{\eta'}+i\hat \pi^aT^a/f_\pi\big]\,,
\label{9}\ee
 and the axion mass squared is given by
\be
 m_a^2 = \chi \frac{\tilde c_s^2}{f_a^2} \ .
\label{10}\ee

The fields $\hat \eta '$ and $\hat \pi^b$ are however not mass eigenstates (for a generic quark mass matrix) because they are subject to the ``usual'' mixing of neutral states  imposed by the quark masses.  The final physical states $\eta'_p$ and $\pi_p$ are found by diagonalizing the mass matrix arising from~\eqref{piLhat} which in general needs to be done numerically.

CP-odd couplings and a finite $\theta$-angle (which may be induced by linear couplings of the axion to other sectors not considered here) introduce small VEVs for the mesons. Such VEVs will be important (apart from the mixing between the axion and the mesons) when considering their couplings to the nucleons in Sec.~\ref{sec:axionnucleon}. We discuss these VEVs in Appendix~\ref{app:vevs}. We also include results for the CP-violating vertices of three pseudoscalar mesons in Appendix~\ref{app:threepi}.

Finally, the scaling of the QCD parameters at large $N_c$ and/or $N_f$ for our conventions is the following. The decay constants scale as
\be
f_\pi^2 \sim f_{\eta'}^2 \sim N_c\;,
\label{15} \ee
while the term $\propto c_\eta$, which induces the splitting of the pion and $\eta'$ decay constants, is suppressed by a power of $1/N_cN_f$ since it is the coefficient of a double trace term. We normalized this coefficient such that it scales as
\be
c_{\eta'}\sim \morder{1}\;.
\label{16} \ee
 In this case,  the difference of the decay constants also obeys
 \be
 f_\pi^2 - f_{\eta'}^2  = \morder{1}\;.
\label{17} \ee
We also note that
\be
\chi_\textrm{YM} \sim  \morder{1} \sp  B_0\sim \morder{1}\sp \widetilde Q \sim \morder{1/N_f}\;.
\label{18}\ee

\section{The effective couplings of $\eta'$ and the axion to nucleons}\label{sec:axionnucleon}

In this section we proceed further and discuss the effective field theory that couples the mesons and the axion to  the nucleons. In particular we analyze the CP-violating non-derivative couplings of the $\eta'$ and the axion to the nucleons.\footnote{The lowest dimension CP-preserving couplings for on-shell nucleons reduce to derivative couplings.}

\subsection{Generic $N_f$ and $N_c$}

We shall discuss here the effective theory for the nucleons and their couplings to the pions, $\eta'$ and the axion for generic values of $N_c$ and $N_f$. Specific results for $N_f=2$
will be given in Sec.~\ref{sec:FiniteNf}.
We start by writing down the couplings in the absence of the axion and the $\theta$-angle, and then discuss how they are introduced in a generic frame.  Working in a generic frame will help us to check that the results behave correctly under chiral transformations. We shall work at any odd value of $N_c$, and for simplicity only consider spin $1/2$ baryons but in arbitrary representations of the flavor group.

In order to describe the couplings of the baryons to the mesons and the axion, the standard method, \cite{Georgi:1985kw},  is to first define a new matrix through $u^2 = U$. Because the square root of a matrix is ambiguous, this is not enough to make the definition precise; we can take
\be
u=\exp\left[i{\eta'\over \sqrt{2N_f}~f_{\eta'}}+i{\pi^aT^a\over 2 f_\pi}\right]\;.
\label{19a} \ee
 This matrix transforms as
\be \label{xitransform}
 u \to L u K^\dagger = K u R^\dagger
\ee
under the chiral transformations  (see, e.g.,~\cite{Scherer:2002tk}). Here $K$ is a unitary matrix depending on the pion fields and defined\footnote{For a precise definition and discussion, see, e.g.,~\cite{Georgi:1985kw,Scherer:2002tk}.} through the equation in~\eqref{xitransform}. The baryons belong to the $N_c$-index representations\footnote{One should keep in mind that in the limit of large $N_c$ there would be more relations than what we impose here: the nucleon states are arranged according to the spin-flavor symmetry $SU(2N_f)$, with the ground state being the fully symmetric $N_c$-index representation in this limit. These representations break into towers of states classified according to their spin and flavor representations separately~\cite{Gervais:1983wq,Bardakci:1983ev,Dashen:1993as,Dashen:1994qi}.} of (the vectorial) $SU(N_f)$ and transform according to~\cite{Jenkins:1995gc}
\be
 B_{i_1\cdots i_{N_c}} = \sum_{\{j_k\}} K_{i_1j_1} \cdots K_{i_{N_c}j_{N_c}}  B_{j_1\cdots j_{N_c}} \ .
\label{19}\ee
The baryon bilinears may therefore couple to meson operators $X$ transforming as $X \to K X K^\dagger$. Such operators include
\begin{align}
 D_{\m}\equiv \pa_\m +\Gamma_\m &\equiv \pa_\m + \frac{1}{2} \left[  u^\dagger \left(\pa_\m u\right) + u \left(\pa_\m u^\dagger\right)\right] \ , \qquad u_\mu \equiv \frac{1}{2} i \left[  u^\dagger \left(\pa_\m u\right) - u \left(\pa_\m u^\dagger\right)\right]\ , & \nonumber\\
  \chi_\pm &\equiv u M_0^\dagger u\pm u^\dagger M_0 u^\dagger  \ ,
\label{20}\end{align}
where $M_0$
is the quark mass matrix and $u$ was defined in (\ref{19a}).

The derivative couplings of the pions and the $\eta '$ arise from the terms
\be\label{derivativecouplings}
\frac{\hat g_a}{N_f} \mathrm{Tr}\left[ \,u_\m \,\right]\ \mathrm{Tr}\left[\, \bar B \gamma^\m \gamma_5 B\,\right] + \sum_{k\sim \mathrm{contractions}} N_c \,g_a^{(k)}\,\mathrm{Tr}\left[ \bar B
\gamma^\m \gamma_5 \left(u_\m\right)B\right]\,,
\ee
where ``$k\sim \mathrm{contractions}$" stands for all possible (invariant) contractions of the flavor indices of the two nucleons and the meson matrix.
 We have also scaled out the leading expected behavior of the coupling constants on $N_c$ and $N_f$ so that the expectation is
\be
\hat g_a =\morder{1}~~~{\rm  and}~~~ g_a^{(k)} =\morder{1}\;.
 \label{c20}\ee
 Notice however that we may still have subleading dependence involving terms $\sim 1/N_c$ and $\sim 1/N_f$. We will follow similar conventions for other couplings below, unless stated otherwise.

In the second term of (\ref{derivativecouplings}) we include a sum over various contractions (with couplings $g_a^{(1)}$, $g_a^{(2)}$,\ldots) of the flavor indices of the baryons and $u_\m$ before taking the trace,  which are consistent with chiral symmetry. Not all of the possible terms are independent, but their relations are nontrivial for baryons transforming under generic representations of the chiral symmetry. The nontrivial piece amounts to combining the baryon, the part of the meson term transforming under the adjoint representation, and the anti-baryon in the conjugate representation into a singlet.
We will explain the structure in more detail in Appendix~\ref{app:groupth},
considering $N_f=2$ and $N_f=3$ as examples.

The precise knowledge of the structure of these terms is not needed for most of our results.
In the limit of large $N_c$ it has been solved in~\cite{Dashen:1994qi}.
Moreover, for the coupling of the flavor singlet $\eta'$ meson, these terms reduce to the standard trace, i.e., they have the same form as the first term, but are leading by a factor of $N_c$. The suppression of the double trace term is consistent with the expected enhancement of the chiral symmetry\footnote{In the $N_c\to\infty$ limit with $N_f$ finite the U(1)$_A$ anomaly vanishes.} to $U(N_f)_L\times U(N_f)_R$ in the limit of large $N_c$. In other words,~\eqref{derivativecouplings} may be replaced by
 \be\label{derivativecouplings2}
\frac{g_a N_c}{N_f} \mathrm{Tr}\left[\, u_\m\,\right] \ \mathrm{Tr}\left[\, \bar B \gamma^\m \gamma_5 B\,\right] + \sum_{k\sim\mathrm{contractions}} N_c \,g_a^{(k)}\,\mathrm{Tr}\left[\,\bar B
\gamma^\m \gamma_5 \left(u_\m- \frac{1}{N_f} \,\mathrm{Tr}\, u_\m\right)
B\,\right] \,,
\ee
where we projected to the adjoint representation in the second term by rearranging the singlet contribution.
Therefore
\be
N_c~g_a=\hat g_a+\sum_{k\sim {\rm contractions}}N_c~ g_a^{(k)}
\label{c21}\ee
and according to (\ref{c20})
\be
g_a\sim {\cal O}(N_c)\,.
\label{c22}\ee
 In this article, however, we are interested in the non-derivative couplings to the nucleons, which we will discuss next.

 \subsubsection{Non-derivative couplings}

We go on to discuss the non-derivative couplings of the nucleons to the $\eta'$, the pions, and the axion.
  Such couplings are absent when CP is conserved, but they can arise through a nonzero $\theta$-angle
which in turn can result from additional UV terms in the action, which drive the expectation value of the axion away from zero.
Other CP-violating contributions than those appearing effectively through the $\theta$-angle are also possible, we analyze them in Appendix~\ref{app:baryoncouplings}.

We first write down the relevant terms in the effective Lagrangian in the absence of the axion and the $\theta$-angle, and then discuss how the axion and $\theta$ are included. A class of non-derivative couplings arises from the mass terms involving $\chi_\pm$ (that were defined in (\ref{20})):
\begin{align}\label{massterms1}
  &\frac{c_{m1}N_c }{N_f}\mathrm{Tr}\left[\, \chi_+\,\right]\,\mathrm{Tr}\left[\, \bar B B\,\right] + \sum_\mathrm{contractions}\,N_c \, c_{m1}^{(k)}\,\mathrm{Tr}\left[\, \bar B \left(\chi_+-\frac{1}{N_f} \mathrm{Tr} \chi_+\right) B \,\right]&\\
  \label{massterms3}
  +\,&\frac{c_{m2}N_c}{N_f}\mathrm{Tr}\left[\, \chi_-\,\right]\,\mathrm{Tr}\left[\, \bar B\gamma_5 B\,\right] + \sum_\mathrm{contractions}\,  N_c\,c_{m2}^{(k)}\,\mathrm{Tr}\left[\, \bar B \gamma_5\left(\chi_--\frac{1}{N_f} \mathrm{Tr} \chi_-\right) B \,\right]\ .&
\end{align}
Several comments are in order.

The terms in~\eqref{massterms1} give rise to corrections to the nucleon masses, and they are $\sim {\cal O}(N_c)$, in agreement  with the scaling of the nucleon mass.  Notice that we separated again a piece from the flavor singlet contributions in order to write the more complicated non-singlet terms in terms of the combinations $\chi_\pm - \mathrm{Tr}\, \chi_\pm/N_f$ which vanishes (up to higher order interaction terms in the pions) for flavor independent quark masses.

The terms containing an extra $\gamma_5$ are suppressed by a power of $1/N_c$. Considering the singlet term as an example, this can be seen from the identity
 \be \label{deridentity}
- i \mathrm{Tr}\left[\, \bar B\gamma_5 B\,\right] = \frac{1}{2m_B} \partial_\mu \mathrm{Tr}\left[\, \bar B\gamma^\mu \gamma_5 B\,\right] \sim \frac{1}{N_c} \partial_\mu \mathrm{Tr}\left[\, \bar B\gamma^\mu \gamma_5 B\,\right]\,,
 \ee
 where the first equality is exact when the nucleons are on-shell and we used the fact that the baryon mass $m_B \sim N_c$ (see also Appendix~\ref{app:spinors}).

 From~\eqref{deridentity} we also observe that the terms involving $\gamma_5$ are,  moreover, derivative terms in disguise and they are suppressed in the non-relativistic limit.
  This is not entirely true for the single trace ``contraction'' terms
  in~\eqref{massterms3}, because of the flavor structure: the single trace terms of~\eqref{massterms1} give rise to flavor dependence of the nucleon mass which leads to a non-derivative contribution from
  ~\eqref{massterms3}. These contributions are of the second order in the differences $\chi_\pm - \mathrm{Tr}\, \chi_\pm/N_f$. The single trace term in~\eqref{massterms3} will not, however, contribute to the linear axion couplings because the axion dependence also cancels at linear order due to the subtraction of the singlet piece  after substituting~\eqref{transform1} and~\eqref{transform2} as we shall show below. This holds true also for the mixing effects between the pions (arising from expanding the $\chi_-$) and the axion. Therefore we shall not discuss the terms in~\eqref{massterms3} further.

Another interesting class of terms are the flavor singlet terms which couple the $\eta' \propto \log \det U$ to the nucleon singlets. Possible terms are
\begin{align}\label{etaterms}
 &d_1  \log \det U\,\mathrm{Tr}\left[\, \bar B\gamma_5 B\,\right]
 + \frac{d_2}{N_c} \left(i \log \det U\right)^2 \,\mathrm{Tr}\left[\,  \bar B B\,\right]\\
 \label{etaterms2}
 &- \frac{ d_3}{N_f} \log \det U\,\mathrm{Tr}\left[\,\chi_-\,\right]\,\mathrm{Tr}\left[\,  \bar B B\,\right] \ .
\end{align}
The scaling with $N_c$ was determined as follows: every factor of $i\det\log U$ brings a power of $1/N_c$ with respect to the nucleon mass term and the terms in~\eqref{massterms1} and in~\eqref{massterms3}~\cite{HerreraSiklody:1996pm}.
Recall that the term $d_1$ is in addition suppressed by $1/N_c$ due to the $\gamma_5$ factor.

In order to extract the non-derivative couplings from these Lagrangian terms, we consider the axion couplings induced by the axion terms in the QCD Lagrangian~\eqref{LQCDaxion} and turn on a finite $\theta$-angle. We replace\footnote{One can also introduce the $\theta$-angle by shifting $a$ (after these transformations) or by adding the $\theta$-angle in the quark mass matrix: all these will lead to the same results due to chiral symmetry.}
\be \label{axrepl1}
 i \log \det U \mapsto  i \log \det U -c_s a/f_a-\theta
  \ee
  and
  \be \label{axrepl2}
  M_0\mapsto  M(a) = e^{ia \tilde Q/f_a} M_0 e^{ia \tilde Q/f_a}\,,
   \ee
   where $M_0$ is assumed to be Hermitean. In terms of the axion, the $\theta$-angle is therefore introduced effectively by replacing
\be \label{thetarepl}
c_s a/f_a \mapsto \theta + c_s a/f_a\;.
 \ee

Notice that also the meson mixing in~\eqref{transform1} and in~\eqref{transform2} will be affected: a finite $\theta$-angle leads to a small displacement of vacuum expectation values (VEVs) of the pions and the $\eta'$ from zero.\footnote{Our convention is that after the shift, the VEV of the axion is zero so that the constant mode of the axion is included in $\theta$. This is equivalent to introducing $\theta$ through an extra term in the Lagrangian which is linear in the axion and independent of the QCD fields.} Other sources of CP-violation than the $\theta$-angle can also lead to such VEVs. We analyze them in Appendix~\ref{app:vevs} and as it turns out, they do not lead to terms with essentially different structures than those terms which we derive in this section. In Appendix \ref{app:chpt} we also analyze the effects of non-hermiticity of the quark mass matrix and show that they can be absorbed in the $\theta$-angle.

In order to extract the contributions involving $\theta$, we therefore need to study the nucleon couplings which are quadratic in the pion fields, the axion, and $\eta'$ (at vanishing $\theta$). First we notice that
\begin{align}
\label{detUafactor}
&\!\!\!\!\!\!\!\!\! i \log \det U -\frac{c_s a}{f_a} = -\frac{\sqrt{2N_f}}{f_{\eta'}}\left(\hat\eta' + \frac{\tilde c_s f_{\eta'}}{\sqrt{2N_f}}\frac{\chi}{\chi_\mathrm{YM}}\frac{a_p}{f_a} +\morder{\frac{1}{f_a^2}}\right) \ ;&\\
\chi_-(a) &= \left\{\frac{i\sqrt{2}\eta'}{\sqrt{N_f}f_{\eta'}}+\frac{i\Pi}{f_\pi}-2i\tilde Q
\frac{a}{f_a},M_0\right\} + \mathrm{nonlinear} &\\
\label{chiminusresult}
 & = \left\{\frac{i\sqrt{2}\hat\eta'}{\sqrt{N_f}f_{\eta'}}+\frac{i\hat \pi^bT^b}{f_\pi},M_0\right\} - \frac{4 \tilde c_s i \chi}{f_\pi^2B_0} \frac{a_p}{f_a} + \mathrm{nonlinear}\,; & \\
\label{trchiplus}
 \mathrm{Tr}\,\left[\chi_+(a)\right] &= 2 \mathrm{Tr}\,\left[M_0\right]  -\mathrm{Tr}\,\left[\left(\frac{\sqrt{2}\hat\eta'}{\sqrt{N_f}f_{\eta'}}+\frac{\hat\pi^b T^b}{f_\pi}- \frac{2\tilde c_s \chi M_0^{-1}}{f_\pi^2 B_0}  \frac{a_p}{f_a}\right)^2 M_0\right] + \cdots &
\end{align}
where $\chi_\pm(a) = u M(a)^\dagger u\pm u^\dagger M(a) u^\dagger$, the anticommutator is normalized as $\{A,B\}=AB+BA$, and
we already inserted the relations~\eqref{transform1}--\eqref{transform3} but so far kept $\theta=0$.
We remark that in~\eqref{chiminusresult} the axion dependence is proportional to the identity matrix in flavor space.\footnote{In Appendix~\ref{app:chpt} we point out that this proportionality is broken by higher order terms in the chiral Lagrangian. This breaking is however only present for terms which are quadratic or higher in the quark masses, which we are neglecting in this article.}
This verifies the fact that it is indeed absent in the second term of~\eqref{massterms3}
(which we have already omitted). Notice that we discuss here only the trace term $\sim \mathrm{Tr}\,\left[\chi_+(a)\right]$, see Appendix~\ref{app:chip} for the analysis of the second term in~\eqref{massterms1}.

The terms which give rise to quadratic couplings of $\eta'$ and the axion to the nucleons are those in~\eqref{massterms1}, the second term in~\eqref{etaterms}, and the term in~\eqref{etaterms2}. Inserting~\eqref{detUafactor},~\eqref{chiminusresult} and~\eqref{trchiplus} in these expressions, and adding $\theta$ through the replacement~\eqref{thetarepl}, we find the following couplings:
\begin{align}
\label{thetacouplings1}
\mathcal{L}_{\eta'aB} ^{(\theta)} & = \frac{4 N_c c_{m1}\chi\theta}{N_f f_\pi^2 B_0}\left(\frac{\sqrt{2N_f}}{f_{\eta'}}\hat \eta' -\frac{2\tilde c_s \chi \mathrm{Tr}\,\left[M_0^{-1}\right]}{f_\pi^2 B_0}\frac{a_p}{f_a}\right) \mathrm{Tr}\,\left[\bar BB\right]&\\
\label{thetacouplings2}
&\ \ \ + \left(\frac{2 d_2}{N_c \chi_\mathrm{YM}}-\frac{4d_3}{f_\pi^2B_0}\right)\frac{\sqrt{2N_f}\chi\theta}{f_{\eta'}}\hat \eta'\ \mathrm{Tr}\,\left[\bar BB\right]\ , &
\end{align}
where we neglected second-order terms in the quark masses. Since the axion coupling is proportional to $c_{m1}$,  which also controls the nucleon mass corrections, there is a relation between the two (in the absence of other sources of CP-violation than the $\theta$-angle). We will discuss this in more detail in Sec.~\ref{sec:generalIR}. Notice also that the $\hat \eta'$ couplings on the second line~\eqref{thetacouplings2} are suppressed by $N_f/N_c$ with respect to the couplings on the first line. Therefore the couplings on the first line should be compared to the Skyrme and WSS model analyses carried out in sections~\ref{sectionskyrme} and~\ref{sec:WSS}, which employ the 't Hooft limit.

Contributions from the single trace term in~\eqref{massterms1} are analyzed in Appendix~\ref{app:chip}: these include the CP-odd nucleon-pion couplings~\eqref{thetapinucleon}, and a correction to the nucleon-axion coupling~\eqref{thetaanucleonad}, which is nonzero if quark masses depend on flavor. The CP-odd nucleon-$\eta'$ coupling is independent of these terms. Recall also that the mixing of $\eta'$ and the pions due to flavor dependent quark masses is not included in these results.

\subsection{ Results for $N_f=2$
}\label{sec:FiniteNf}

We now discuss the structure of the couplings at finite, fixed $N_f$, taking $N_f=2$ (and any odd $N_c$) as an example.
This will be useful in order to compare to the results of sections~\ref{sec:generalIR},~\ref{sectionskyrme} and~\ref{sec:WSS}.
 In addition, results for $N_f=3$ are given in Appendix~\ref{app:baryoncouplings}.
First, we analyze the derivative couplings in~\eqref{derivativecouplings}. The three pions and the $\eta'$\footnote{{In the case $N_f=2$, what we call $\eta'$ is, as usual, a ``mathematical" isosinglet pseudoscalar which does not have any $s\bar s$ component, see e.g. \cite{meissner}. This  $\eta'$ at $N_f=2$ does not correspond to the physical one, but it can be related to the physical {$\eta_P$ and $\eta'_P$ fields. In \cite{meissner} for instance, taking into account the pseudoscalar mixing angle, it is taken as $\eta'\approx \frac45 \eta_P + \frac35 \eta'_P$, so its mass is around 549 MeV.}}} form a four-dimensional representation of $U(N_f=2)$ which breaks into the iso-singlet $\eta'$ and the iso-triplet of pions at finite $N_c$. The ground state, spin-$1/2$ fermions, are in the fundamental isospin-$1/2$ representation:
\be
 N=\left(\begin{array}{c}p \\ n\end{array}\right) \ .
\label{22}\ee In this case the derivative couplings of~\eqref{derivativecouplings} and~\eqref{derivativecouplings2} read
\begin{align}
&\phantom{=} \frac{\hat g_a}{2} \mathrm{Tr}\left[ \,u_\m \,\right]\ \bar N \gamma^\m \gamma_5 N + N_c\, g_a^{(1)} \bar N
\gamma^\m \gamma_5  u_\m N & \\
\label{nf2eq1}
&=  \frac{N_c g_a}{2} \mathrm{Tr}\left[ \,u_\m \,\right] \ \bar N \gamma^\m \gamma_5 N  +  N_c\,g_a^{(1)} \bar N
\gamma^\m \gamma_5  \left(u_\m -\frac{1}{2} \mathrm{Tr}\, u_\m\right) N & \\
\label{nf2eq2}
& =  - \frac{N_c g_a}{2 f_{\eta'}}\partial_\mu \eta' \ \bar N \gamma^\m \gamma_5 N - \frac{N_c}{2f_\pi}  g_a^{(1)} \partial_\mu\pi^a\, \bar N
\gamma^\m \gamma_5 \tau^a N + \mathrm{higher\ order}\,,
\end{align}
where $g_a = g_a^{(1)} +\hat g_a/N_c$, and $\tau^a$ are the Pauli matrices.\footnote{In this action we neglected interactions with other nucleon representations like the $\Delta$.} There is only one independent term in the sums over contractions in \eqref{derivativecouplings} and \eqref{derivativecouplings2}.  The flavor structures of the mass terms~\eqref{massterms1}--\eqref{massterms3} are analogous to the derivative terms.

We note that in the limit of large $N_c$, the derivative couplings of the $\eta'$ and the pions become identical since $f_{\eta'}/f_\pi \to 1$ in this limit, signaling the enhancement of the chiral symmetry from $SU(2)$ to $U(2)$.

When $N_f=2$, it is natural to take the light quark masses to be equal, $m_u=m_d\equiv m_{ud}$, so that isospin is unbroken and $\eta'$ does not mix with the pions.
In this case the couplings  of the previous subsection apply directly for the physical mass eigenstates $a_p$, $\eta'_p = \hat \eta'$, and $\pi_p^b = \hat \pi^b$.
The couplings of the axion and the mesons to the nucleons therefore read
\begin{align}
\label{thetacouplings1Nf2}
&  \frac{4 N_c c_{m1}\chi\theta}{ f_\pi^2 B_0}\left(\frac{\eta'_p}{f_{\eta'}} -\frac{2\tilde c_s \chi }{f_\pi^2 B_0 m_{ud}}\frac{a_p}{f_a}\right) \,\bar NN + \left(\frac{4 d_2}{N_c \chi_\mathrm{YM}}+\frac{8d_3}{f_\pi^2B_0}\right)\frac{\chi\theta}{f_{\eta'}}\eta'_p\ \bar NN &\\
&+\frac{4 N_c\chi\theta}{f_\pi^3B_0}
c_{m1}^{(1)}
\pi_p^b\, \bar N \tau^b N \ .&
\label{24}
\end{align}
The terms on the first line arise from~\eqref{thetacouplings1} and~\eqref{thetacouplings2}. We also included the pion coupling on the second line, which depends on the single trace coupling $c_{m1}^{(1)}$ in~\eqref{massterms1}; see eq.~\eqref{thetapinucleon} in Appendix~\ref{app:chip}. Recall that in this case the pion mass is given by $m_\pi^2 \approx B_0 m_{ud}$ in the chiral limit.

We note that the same coupling  also controls the QCD contributions to the mass splitting between the proton and the neutron, so that there is a relation between the mass splitting and the CP-odd pion-nucleon-nucleon coupling~(see, e.g., \cite{deVries:2015una}). We will discuss this relation in more detail below.

Similarly to the derivative couplings above, $c_{m1}$ equals $c_{m1}^{(1)}$
at large $N_c$ up to corrections suppressed by $1/N_c$. Since also $f_{\eta'} \approx f_\pi$ in this limit, the couplings of the pions and that of the $\eta'$ in~\eqref{thetacouplings1Nf2} to the nucleons become equal, signaling the enhanced chiral symmetry.

\subsection{The one-loop corrections\label{loop}}

The meson-nucleon couplings of the effective Lagrangian described above give rise to further contributions to the couplings of interest via quantum effects.
These quantum effects are determined by the effective meson and nucleon fields. In the 't Hooft limit, they are suppressed by powers of $N_c$, while in the Veneziano limit, they may be unsuppressed.

We will discuss such effects below. In the case of $N_c=3$, $N_f=2$,  they were calculated in \cite{Gutsche:2016jap} but we will extend here this calculation to arbitrary $N_c$ and $N_f$.

The starting Lagrangian for the pion-nucleon interactions was given already in the previous subsections, but we will abstract here the couplings that are of interest.
The $\eta$ and $\eta'$ couplings to pions from the chiral Lagrangian  are
\be
{\cal L}_{H\pi\pi}=f_{H\pi\pi}m_H~H~\vec \pi\cdot \vec \pi\sp H=\eta,\eta'\,,
\label{80}\ee
where the couplings behave as
\be
f_{H\pi\pi}m_H \sim\, {\theta m_{ud}\,\over  N_f^{3\over 2}N_c^{1\over 2}}\,,
\label{c24} \ee
 in the chiral and large $N_c,N_f$ limits  (see Appendix~\ref{app:threepi}). Similarly the coupling to the axion can be written as
 \be
 {\cal L}_{a\pi\pi} = f_{a\pi\pi}\frac{a}{f_a} ~\vec \pi\cdot \vec \pi \,,
 \ee
 where $f_{a\pi\pi}$ scales as $\sim \theta m_{ud}/N_f^2$.
The couplings to the nucleons are\footnote{The $\theta$ contribution from~\eqref{thetacouplings1} scales as $\sim \theta m_{ud} N_c^{1/2} N_f^{-3/2}$.}
\be
{\cal L}_{HNN}=\bar g_{HNN}~H\bar N N\,.
\label{81}\ee
We also need the kinetic terms of the spin 1/2 nucleons and the spin-3/2 $\Delta$, which in the conventional $N_f=2$ case are
\be
{\cal L}_N=\bar N\left(-i~\slash \hskip -8pt D-m_N-{g_0\over 2}\slash \hskip -6.5pt u \gamma^5  \right)N \sp {\cal L}_{\Delta N \pi}=-{i h_A\over 2f_{\pi}m_{\Delta}}\bar N T^a\gamma^{\m\n\l}(\pa_{\m}\Delta_{\nu})\pa_{\l}\pi^a+h.c.
\label{82}\ee
$N$ above stands for the spin-1/2 nucleon iso-doublet while $\Delta_{\m}$ for the $\Delta$ iso-quartet.
The covariant derivatives $D_{\m}$ and $u_{\m}$ were defined in (\ref{20}) and the matrix $u$ in (\ref{19a}).
$T^a$ are the Clebsch-Gordan matrices that couple the doublet, the quartet and the triplet of $SU(2)$.
From (\ref{nf2eq1}) we have
\be
g_0={2N_c g_a^{(1)}}\simeq g_A\,,
\label{c23}\ee
where the last equality above with the coupling of the nucleons to the axial current being valid in the chiral limit.

At arbitrary $N_c$ and $N_f$, the field $N$ represents the lowest-lying flavor representation of spin {1/2}, that is the $\left({N_c+1\over 2},{N_c-1\over 2},0,\cdots,0\right)$ in Dynkin notation.
The spin 3/2 representation is the $\left({N_c+3\over 2},{N_c-3\over 2},0,\cdots,0\right)$ flavor representation and so on.
In the general case, there may be also higher-spin baryons with spin larger than 3/2 and up to $N_c/2$. The parameters $m_N$,  $g_0$ and $h_A$ scale as ${\cal O}(N_c)$.
The matrix $u$ is a unitary $U(N_f)$ matrix while the pion current $u_{\m}$ is a Hermitean $N_f\times N_f$ matrix.

The flavor structure of the pion-nucleon interaction involves $T^a_{ij}$
where $a$ is an adjoint index of $U(N_f)$ and $i,j$ are indices in the flavor representations $R, R'$ of the nucleons. At general $N_f$, these can be any representation in the tensor product $(\otimes \square)^{N_c}$.
 We will be assuming here exact flavor symmetry so that all quark masses are the same.

 \begin{figure}[t]
 \begin{center}

  \includegraphics[width=.9\textwidth]{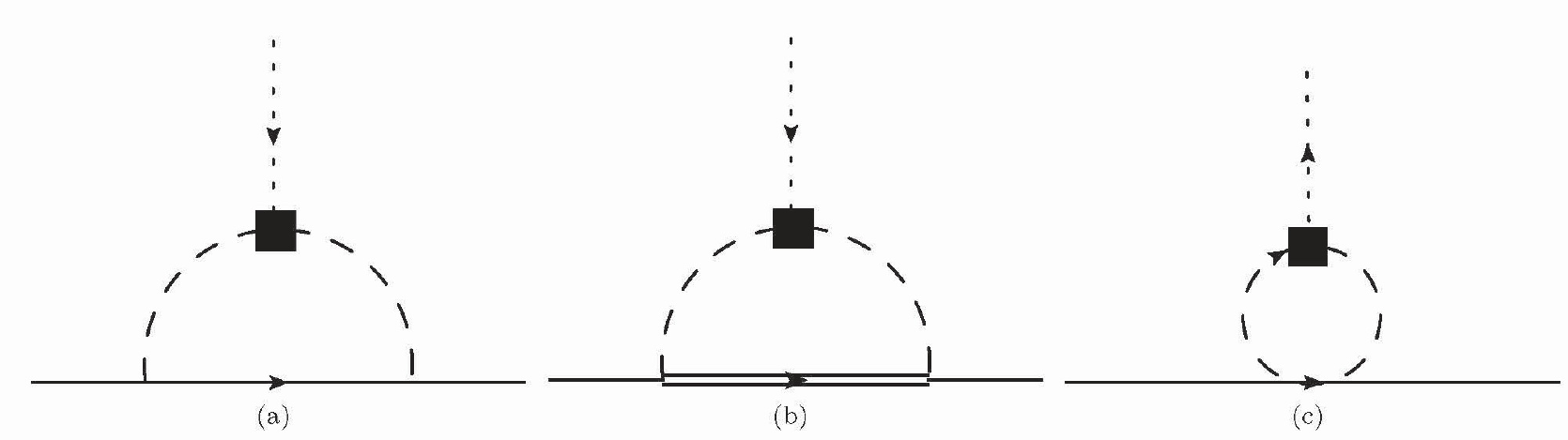}
 \end{center}
   \caption{Diagrams contributing at one-loop to the coupling between $\eta$, $\eta'$ and ground-state nucleons. (a) With a ground-state nucleon intermediate state. (b) With a higher-spin nucleon intermediate state. (c) From pions running in the loop. This last diagram does not contribute due to isospin symmetry. Figure from \cite{Gutsche:2016jap}.}
  \label{fig1}
 \end{figure}

 We discuss first the one-loop couplings of the $\eta'$ and $\eta$ mesons to the nucleons and consider the axion coupling below.
 The diagrams that contribute at one loop are shown in figure \ref{fig1}.
 Diagram (a) has ground-state nucleons as intermediate states while diagram (b) has higher spin nucleons. Diagram (c) gives a vanishing contribution due to isospin symmetry as charge conjugate pions give opposite sign contributions. {Notice that the CP-violation must arise from the three-meson vertex because this vertex vanishes if CP is conserved.}

For the rest, we will call $R$ the flavor representation of the ground-state nucleons and $R_I$ all the others that give rise to the analogues of  $\Delta$.
The flavor couplings that enter in ${\cal L}_{\Delta N \pi}$ in (\ref{82}), $F^a_{i,\alpha}$, are  the Clebsch-Gordan coefficients
linking the adjoint, $R$ and $\tilde R_I$ of the flavor group.
The index $i$ is that of the ground-state nucleon while $\alpha$ is that of the $\Delta$. Of course for higher spin nucleons, $\Delta_{\m_1\m_2\cdots}$ is a tensorial spinor, and the space-time factors in the couplings in (\ref{82}) will change appropriately. These are ${\cal O}(1)$ changes though and we will not worry about them at the moment.

{ For the computation of the loop diagrams we follow Ref.~\cite{Gutsche:2016jap}.} The covariant extended on-mass shell scheme is used
for renormalization~\cite{Gegelia:1999gf,Fuchs:2003qc}. The diagram (a) gives  at $N_f=2$, \cite{Gutsche:2016jap}
\be
\bar g_{HNN}=-{2f_{H\pi\pi}g_A^2 m_{H}\over 16\pi^2 f_{\pi}^2 m_N}\left[m_\pi^2\log{m_N\over m_{\pi}}+m_N^2+{m_{\pi}(m_{\pi}^2-3m_N^2)\over \sqrt{4m_N^2-m_{\pi}^2}}\times \right.
\label{83}\ee
$$
\left.\times \left(\arctan{m_{\pi}\over \sqrt{4m_N^2-m_{\pi}^2}}+\arctan{2m_N^2-m_{\pi}^2\over m_{\pi}\sqrt{4m_N^2-m_{\pi}^2}}\right)\right]\,,
$$
where { the renormalization scale was set to the nucleon mass}
and the external nucleons were put on-shell.
In the general $N_f$, $N_c$ case, where the coupling is between the nucleon states
$N^i$ and $N^j$ we must multiply (\ref{83}) by $(T^aT^a)_{ij}=I(R)\delta_{ij}$ of the flavor group where $I(R)$ is the standard quadratic Casimir of the relevant representations.
The axial coupling in~\eqref{83} scales as $g_A \sim N_c (N_f)^0$ at large $N_c$ while the scaling of $f_{H\pi\pi}$ is in (\ref{c24}).
From (\ref{83}) we deduce that\footnote{This estimate is valid if we take first the chiral limit and then the large $N_c$ limit. It is known that the chiral limit and the large $N_c$ do not commute for generic observables.}
\be
\bar g_{HNN}\sim N_c^{3\over 2}\,,
\label{c25}\ee
as $m_N\sim {\cal O}(N_c)$ and $I(N)\sim {\cal O}(1)$.

The $\Delta$-mediated diagram has the contribution
\be \label{Deltaloopresult}
\bar g^{\Delta}_{HNN}={f_{H\pi\pi} h_A^2 m_N^3 m_H\over 1152\pi^2 f_{\pi}^2 m_{\Delta}^2}Z\left({m_{\pi}\over m_N},{m_{\Delta}\over m_N}\right)\,,
\ee
where the dimensionless function $Z$ can be found in \cite{Gutsche:2016jap}.
In the large $N_c$ limit it scales as $Z\sim {\cal O}(1)$ while $h_a\sim {\cal O}(N_c)$.
We obtain
\be
\bar g^{\Delta}_{HNN}\sim N_c^{3\over 2}\,.
\ee

At general $N_c$ and $N_f$, the diagrams of type (b) corresponding to the coupling of external nucleon states $N^i$ and $N^j$ are multiplied by
\be
\sum_{a,\alpha}(F^a_I)_{i,\alpha}(F^a_I)_{\alpha,j}\,,
\ee
where $(F^a_I)_{i,\alpha}$ are the Clebsch-Gordan coefficients linking the adjoint ($a$ index), the ground-state flavor representation ($i$ index) and the higher spin intermediate flavor representation $I$ (index $\alpha$).
There are also ${\cal O}(1)$ differences due to the different spin structure of the intermediate $\Delta$ states.
The conclusion is that all such individual contributions scale at large $N_c$ as in (\ref{c25}).\footnote{Notice that at large $N_c$ both the nucleons and the $\Delta$ (as well as the higher spin states) are actually packed into the completely symmetric representation of the combined spin-flavor group $SU(2N_f)$~\cite{Dashen:1993as,Dashen:1994qi}.}

For $N_f=2$, $N_c=3$ from \cite{Gutsche:2016jap} we obtain
\be
{\bar g_{\eta NN}\over f_{\eta \pi\pi}}\simeq 1.33\sp {\bar g_{\eta' NN}\over f_{\eta' \pi\pi}}\simeq 2.32\,,
\ee

\be
{\bar g^{\Delta}_{\eta NN}\over f_{\eta\pi\pi}}\simeq 0.38\sp {\bar g^{\Delta}_{\eta' NN}\over f_{\eta' \pi\pi}}\simeq 0.64\,.
\ee
{The results in the extended on-mass shell scheme are suppressed by about 30\% with respect to the heavy-baryon limit~\cite{Gutsche:2016jap}.}

The one-loop couplings to the axion can also be extracted from the above results: this amounts to replacing $f_{H\pi\pi}m_H$ by $f_{a\pi\pi}/f_a$ in~\eqref{83} and in~\eqref{Deltaloopresult}. We find that
\be
 \frac{\bar g_{aNN}\,f_a}{f_{a\pi\pi}} \simeq 2.4 \times 10^{-3}\, \mathrm{MeV}^{-1}\,,\qquad \frac{\bar g_{aNN}^{\Delta}\,f_a}{f_{a\pi\pi}} \simeq 6.7 \times 10^{-4}\, \mathrm{MeV}^{-1} \,.
\ee

Finally, we can estimate these loop contributions as a function of the $\theta$ angle as follows.
By using the expressions for the three meson couplings in chiral perturbation theory,~\eqref{3pioncoupling} and~\eqref{etaprimepipicoupling}, we find that
\be
 f_{\eta' \pi\pi} \simeq - \frac{\theta m_\pi^2}{m_{\eta'}f_\pi}\left(\frac{1}{2\sqrt{6}}\cos\theta_{\eta\eta'} + \frac{1}{4\sqrt{3}}\sin\theta_{\eta\eta'}\right) \simeq - 0.029\ \theta\,,
\ee
where we took the limit $m_u+m_d \ll m_s$, and used $m_\pi \simeq 135$~MeV, $m_{\eta'} \simeq 958$~MeV, $f_\pi\simeq 92$~MeV, and $\theta_{\eta\eta'} \simeq -\pi/9$ for the mixing angle defined in~\eqref{12}.
We obtain for the CP-violating one-loop couplings
\be \label{loopbargestimates}
 \bar g_{\eta' NN} \simeq -0.07 \ \theta \ , \qquad \bar g_{\eta' NN}^{\Delta} \simeq -0.02 \ \theta \ .
\ee

Similarly, from~\eqref{apipicoupling}
(setting $\tilde c_s =1$ and taking the chiral limit with two light quarks)
we find that
\be
 f_{a\pi\pi} \simeq \theta \times 2300\, \mathrm{MeV}^2
\ee
so that the one-loop axion couplings can be estimated as
\be
 \bar g_{a NN} \simeq  \theta \left( \frac{5.5\,\mathrm{MeV}}{f_a} \right)  \ , \qquad \bar g_{a NN}^{\Delta} \simeq   \theta \left( \frac{1.5\,\mathrm{MeV}}{f_a}\right) \ .
\label{barganneft}
\ee
\section{General IR relations for the couplings}\label{sec:generalIR}

Before considering large $N_c$ models in detail in the following sections, let us consider model-independent low energy relations between the couplings valid also at small $N_c$ (partially recollecting what we have learned in the previous sections).

We consider the chiral Lagrangian (\ref{chiralL1})
choosing the scheme where the axion is entirely contained in the topological term \cite{DiVecchia:2017xpu}
\bea\label{skyrme2flav}
{\cal L} &= & \frac{f_{\pi}^2}{4}{\rm Tr} \left(U^{\dagger} \partial_{\mu} U \cdot U^{\dagger} \partial^{\mu} U  \right) - \frac12 \partial_{\mu}a \partial^{\mu}a + \frac{f_{\pi}^2}{4}B_0{\rm Tr}  \left(M U + M^{\dagger} U^{\dagger}  \right)
\\
&&  - \frac{\chi_{YM}}{2}\left[-\frac{i}{2}{\rm Tr} \left(\log{U} - \log{U^{\dagger}}\right)  + \frac{a}{f_a}\right]^2 + \frac{f_{\pi}^2-f_{\eta'}^2}{4 N_f} \mathrm{Tr}[U^\dagger\pa_{\m}U]\mathrm{Tr}[U\pa^{\m}U^\dagger]\,,\nonumber
\eea
where $U$ is defined in (\ref{1}).
We concentrate on $N_f=2$ flavors, where the quark mass matrix reads
$M={\rm diag}(m_u, m_d)$ and $T^a=\tau^a$ are the Pauli matrices. The GMOR relation sets
\be
2m_{\pi}^2 = \tr[M]B_0\;.
\label{28} \ee
As we have seen in section \ref{CHPTL}, (\ref{skyrme2flav}) generates a mixing of the $\eta'$ with the axion,  and in turn the latter with the pions. Neglecting the mixing terms between $\eta'$ and pions, the mass eigenstates at leading order in $1/f_a$ are now (see equations (\ref{transform1})-(\ref{transform3}))
\bea\label{mixingswpions}
&&\hat \eta' = \eta' + \frac{\chi f_{\eta'}}{B_0 f_{\pi}^2}\tr [M^{-1}] \frac{a}{f_a}\,,\nonumber \\
&&\hat \pi^a = \pi^a + \frac{\chi}{B_0 f_{\pi}}\tr [\tau^a M^{-1}] \frac{a}{f_a}\,,\nonumber \\
&& a_p = a - \frac{\chi f_{\eta'}}{B_0 f_{\pi}^2}\tr [M^{-1}] \frac{\eta'}{f_a} - \frac{\chi}{B_0 f_{\pi}}\tr [\tau^a M^{-1}] \frac{\pi^a}{f_a}\,,
\eea
where
\be
\chi= \frac{\chi_{YM} B_0 f_{\pi}^2}{B_0 f_{\pi}^2 + 2\chi_{YM} \tr[M^{-1}]}
\label{29} \ee
is the full topological susceptibility of the theory. The previous expressions show that the mixing of the axion with pions only occurs if the flavors are non-degenerate in mass.

The couplings of the $\eta'$ and the pions to nucleons are defined via the following effective Lagrangian terms
\be
\delta {\cal L} = ig_{\eta' NN} \eta' \bar N \gamma_5 N + ig_{\pi NN} \pi^a \bar N \gamma_5 \tau^a N + {\bar g}_{\eta' NN} \eta' \bar N N + {\bar g}_{\pi NN} \pi^a \bar N\tau^a N\,,
\label{effel}
\ee
where $N$ is the two component vector for the spin $1/2$ nucleons $N=(p,n)$, the first two couplings are CP-even, while the last two are CP-odd and arise in the presence of any effective non zero $\theta$-angle or any other CP-violating source.

The axion couplings to nucleons, to leading order in $1/f_a$, can be obtained from the ones above, simply replacing $\eta'$ and $\pi^a$ with the respective physical fields defined in (\ref{transform1}), (\ref{transform2}). Notice that in the non-relativistic limit, which is justified at energies much smaller than the nucleon mass $m_N$, the non-derivative CP-even couplings can be traded for derivative ones since, for any scalar field $\phi$
\be
i \partial_{\mu}\phi \bar N \gamma^{\mu} \gamma^5 N \approx 2 m_N \phi \bar N\gamma^5 N\,.
\label{nonre}
\ee
These meson-nucleon couplings were already analyzed using effective field theory in Sec.~\ref{sec:axionnucleon}; the results for derivative couplings are given in equation~\eqref{nf2eq2} and for the non-derivative CP-violating couplings in equations~\eqref{thetacouplings1Nf2} and~\eqref{24}. 

If we define the derivative axion-nucleon couplings as
\be
\delta{\cal L}_{aNN,{\rm der}} = -\frac{\partial_{\mu} a}{2f_a} c_N \bar N \gamma^{\mu}\gamma^5 N\,,
\label{30} \ee
we find, using (\ref{nonre}), (\ref{effel}) and (\ref{mixingswpions})
\be
c_p = -{\frac{g_{\eta'NN}}{m_N} \frac{\chi f_{\eta'}}{f_{\pi}^2B_0} \tr[M^{-1}] - \frac{g_{\pi NN}}{m_N} \frac{\chi}{f_{\pi}B_0} \tr[M^{-1}\tau^3]}\,,\label{c26}\ee
\be
c_n = -\frac{g_{\eta'NN}}{m_N} \frac{\chi f_{\eta'}}{f_{\pi}^2B_0} \tr[M^{-1}] +\frac{g_{\pi NN}}{m_N} \frac{\chi}{f_{\pi}B_0} \tr[M^{-1}\tau^3]\,.
\label{31} \ee
To leading order in the chiral limit, where the mass of the pion is much smaller than that of the $\eta'$ (i.e. when $f_{\pi}^2 B_0 \ll 2\chi_{YM} \tr [M^{-1}]$), the expressions above can be rewritten as
\be
\label{dercup1}
c_p  \approx - \frac{1}{2}\hat g_A -\frac{1}{2} g_A \frac{m_d - m_u}{m_d+m_u}\,,
\ee
\be
c_n  \approx - \frac{1}{2}\hat g_A +\frac{1}{2} g_A \frac{m_d - m_u}{m_d+m_u}\,,
\label{c27}\ee
where the isoscalar and isovector axial couplings, obtained from the matrix elements of the axial current between baryon states (the axial form factors), are related to those of the $\eta'$ and the pion by means of the generalized Goldberger-Treiman relations
\be
g_{\eta' NN} = \frac{m_N}{f_{\eta'}}\hat g_A\,,\quad g_{\pi NN} = \frac{m_N}{f_{\pi}}g_A\,.
\label{32} \ee
The couplings (\ref{dercup1}), (\ref{c27}) are, for the case $N_f=2$, precisely the ``universal" ones obtained in any axion model in the KSVZ class \cite{Kim,SVZ}.

The non-derivative axion-nucleon couplings
\be
\delta{\cal L}_{aNN, {\rm non-der}} =  {\bar c}_N a \bar N N\,,
\label{33} \ee
are present whenever the Peccei-Quinn mechanism is not perfect (or if weak interaction CP-breaking effects are taken into account).
As already recalled, explicit breaking of the Peccei-Quinn symmetry by higher-dimensional operators shifts the VEV of the axion.
These operators are generically present in most UV completions of axion models. Even if suppressed by powers of the Planck mass, up to dimension ten their coefficient must be very small, in order to cope with the experimental bounds (see e.g.~\cite{NEDMexp,pendlebury}) setting $\theta  \lesssim 10^{-10}$, \cite{Holman:1992us,Kamionkowski:1992mf}.
The extent of fine-tuning needed for these coefficients determines the so-called ``quality'' of the axion.
Therefore, on general grounds one can expect that the axion does not cancel completely the $\theta$-angle.
In such a case, we can study linear CP-breaking couplings of the axion and $\eta'$ to nucleons, with a (non-vanishing) $\theta$-dependent coefficient.

Taking the axion-meson mixing (\ref{mixingswpions}) into account, we can write
\bea
&& \bar c_p = -\bar g_{\eta' NN} \frac{\chi f_{\eta'}}{f_a f_{\pi}^2B_0} \tr[M^{-1}] - \bar g_{\pi NN} \frac{\chi}{f_a f_{\pi}B_0} \tr[M^{-1}\tau^3]\,,\nonumber \\
&& \bar c_n =  -\bar g_{\eta' NN} \frac{\chi f_{\eta'}}{f_a f_{\pi}^2B_0} \tr[M^{-1}] + \bar g_{\pi NN} \frac{\chi}{f_a f_{\pi}B_0} \tr[M^{-1}\tau^3]\,.
\label{34} \eea
Therefore, to leading order in the above mentioned chiral limit
\bea\label{hatcpcnchi}
&& \bar c_p \approx  -\frac12\bar g_{\eta' NN}\frac{f_{\eta'}}{f_a} - \frac12\bar g_{\pi NN}\frac{f_{\pi}}{f_a}\frac{m_d - m_u}{m_d+m_u}\,,\nonumber \\
&& \bar c_n \approx -\frac12\bar g_{\eta' NN}\frac{f_{\eta'}}{f_a} + \frac12\bar g_{\pi NN}\frac{f_{\pi}}{f_a}\frac{m_d - m_u}{m_d+m_u} \,.
\eea


Let us introduce the shorthand notation for the first and second term in each line of (\ref{hatcpcnchi})
\be
\bar c_{p} \equiv \bar c_1 - \bar c_2\,, \qquad \bar c_{n} \equiv \bar c_1 + \bar c_2\,.
\ee
In \cite{Wilczek} a relation between the common term $\bar c_1$ in (\ref{hatcpcnchi}) and the pion-nucleon sigma term $\delta M_N$ in the chiral limit has been derived.
We will re-derive it in the degenerate case in the Skyrme and WSS models in sections \ref{sectionskyrme}, \ref{sec:WSS}.
We denote by $M_N$ the nucleon mass in the isospin-conserving limit, where the two light quarks have a common mass equal to ${m_u+m_d\over 2}$.
By virtue of the Feynman-Hellmann theorem, the sigma term is then
\be
\delta M_N = 
m_u \frac{\partial M_N}{\partial {m_u}} + m_d \frac{\partial M_N}{\partial {m_d}}\,.
\ee
The relation of the sigma term with the $\bar c_1$ part of the axion coupling\footnote{Note that this implies a relation between the coupling $\bar g_{\eta' NN}$ and the sigma term.} reads \cite{Wilczek}\footnote{In \cite{Wilczek} the formula is obtained by considering the nucleon matrix element of the interaction operator $\frac{\theta}{f_a} \frac{m_u m_d}{m_u+m_d}(\bar u u +\bar d d +\bar s s)$. The latter is extracted, in the $m_{u,d} \ll m_s$ limit, from the expansion of the quark mass term in the Hamiltonian, once the $\theta$-term has been rotated in the mass sector and the energy has been minimized.}
\be\label{relMW}
\bar c_1 = \frac14 (1-\epsilon^2) \delta M_N \frac{\theta}{f_a}\,,
\ee
where
\be
\epsilon\equiv \frac{m_d-m_u}{m_d+m_u}\,.
\label{defepsilon}
\ee
Moreover, working again in the chiral limit, the pion coupling (and so the term $\bar c_2$ defined above) is known in chiral perturbation theory to be related at tree level to the strong contribution to the neutron-proton mass splitting, which we denote as $(M_n-M_p)_{({\cal L}_M)}$ (see e.~g.~\cite{Mereghetti:2010tp} and sections \ref{sectionskyrme}, \ref{sec:WSS}), as\footnote{In the $N_f=3$ case an analogous relation holds involving the mass difference of $\Xi$ and $\Sigma$ baryons \cite{crewther}.}
\be \label{gpiDeltaM}
\bar g_{\pi NN} = -\theta \frac{(M_n-M_p)_{({\cal L}_M)}}{4 f_{\pi} \epsilon}(1-\epsilon^2)\,.
\ee
Note that the term $(M_n-M_p)_{({\cal L}_M)}$ is of ${\cal O}(\epsilon)$.

Thus, using (\ref{hatcpcnchi}), (\ref{relMW}) and (\ref{gpiDeltaM}), we can write universal low energy relations, valid for $N_f=2$ flavors in the chiral limit ($m_\pi \ll m_{\eta'}$) at tree level in chiral perturbation theory and up to third order in the quark mass difference $\epsilon$, generalizing the expression in \cite{Wilczek},
\bea
\bar c_p &=& \frac14\delta M_N(1-\epsilon^2) \frac{\theta}{f_a} + \frac18 (M_n - M_p)_{{\cal L}_M}(1-\epsilon^2) \frac{\theta}{f_a} \,,\nonumber\\
\bar c_n &=& \frac14 \delta M_N (1-\epsilon^2) \frac{\theta}{f_a}- \frac18 (M_n - M_p)_{{\cal L}_M}(1-\epsilon^2) \frac{\theta}{f_a}\,.
\label{geneform}
\eea
In summary, in these relations the CP-odd couplings of the axion to neutrons and protons are expressed in terms of the pion-nucleon sigma term and the strong contribution to the neutron-proton mass splitting.
The latter can be estimated with phenomenological extractions or lattice calculations.
We are going to comment on the numerical evaluation of these couplings in section \ref{sectionnumbers}.

Notice that what we have previously denoted by ${\bar g}_{a NN}$ is the value of the $\bar c_N$ couplings in the quark mass degenerate case, which corresponds to $\epsilon=0$ and $(M_n - M_p)_{{\cal L}_M}=0$. We will keep this definition in the rest of the paper.

\section{The Skyrme model picture}
\label{sectionskyrme}

In this section we consider the couplings discussed above in the context of the Skyrme model, where the nucleons can be seen explicitly as solutions in the low-energy Lagrangian.
In the Skyrme model, one treats baryons as solitons of the chiral Lagrangian, stabilized with the aid of a quartic term in $U$, the so-called Skyrme term \cite{Skyrme}. The axion-dressed Lagrangian (\ref{skyrme2flav}) thus contains now a further term
\be
\delta{\cal L}_{\rm{Skyrme}}=\frac{1}{32 e^2}{\rm Tr}\left([U^{\dagger} \partial_{\mu} U, U^{\dagger} \partial_{\nu} U][U^{\dagger} \partial^{\mu} U, U^{\dagger} \partial^{\nu} U]\right) \,.
\ee
The coupling in the Skyrme term scales with $N_c$ as $e\sim 1/f_{\pi} \sim 1/\sqrt{N_c}$.
Moreover, we work at large $N_c$ and finite $N_f$, so that $f_{\eta'}=f_{\pi}$ and double trace terms are discarded in (\ref{skyrme2flav}). Hence, for instance, in the $N_f=2$ mass-degenerate case $m_u=m_d=m_{ud}$, the relations (\ref{mixingswpions}) reduce to
\be\label{mixingS}
\eta'_{p} = \eta' + \frac{2 \chi }{f_{\pi} m_{\pi}^2} \frac{a}{f_a}\,, \quad a_{p} = a -\frac{2 \chi }{f_{\pi} m_{\pi}^2} \frac{\eta'}{f_a}\,,
\ee
where
\be
\chi = \frac{m_{\pi}^2 f_{\pi}^2 \chi_{YM}}{f_{\pi}^2 m_{\pi}^2 +4 \chi_{YM}}\,.
\label{35} \ee
Up to corrections of order ${\cal O}(1/f_a^2)$, the masses of the physical states are
\bea\label{masseshere}
&& m^2_{\eta_p'} = m^2_{\pi} +m^2_{WV}\,, \quad {\rm with}\,\, \,m^2_{WV}\equiv \frac{2N_f \chi_{YM}}{f_{\pi}^2}=\frac{4\chi_{YM}}{f_{\pi}^2} \,, \\
&& m^2_{a_p} = \frac{\chi}{f_a^2}\,,
\eea
where in the first line we have introduced the Witten-Veneziano mass $m_{WV}$.

The Skyrme model without the axion allows to compute the meson-nucleon couplings defined in eq. (\ref{effel}). From these, using the general formulae collected in the previous section, the axion-nucleon couplings can be deduced.

Notice that in the Skyrme model, both $g_{\pi NN}$ \cite{anw} and $g_{\eta' NN}$ \cite{meissner} have been computed (the latter one in a suitable extension of the Skyrme model including some vector mesons). To the best of our knowledge, the CP-odd couplings $\bar g_{\eta' NN}$ and $\bar g_{\pi NN}$ have not been computed in the Skyrme model so far. The coupling $\bar g_{\pi NN}$ in the Skyrme model with a finite $\theta$-angle has been argued to be suppressed in the large $N_c$ limit \cite{riggs}. In the chiral limit (which in general does not commute with the large $N_c$ one) it has been computed in {\it e.g.} \cite{crewther, Mereghetti:2010tp,deVries:2015una,Shindler:2015aqa}.

In the following subsection, considering the $N_f = 2$ non-degenerate case, we will directly extract the non-derivative axion-nucleon couplings and, in turn, $\bar g_{\eta' NN}$ and  $\bar g_{\pi NN}$, showing in particular that the relations (\ref{relMW}), (\ref{gpiDeltaM}) and (\ref{geneform}) are indeed reproduced. 
\subsection{Nucleon mass terms and CP-odd couplings for $N_f=2$}\label{sectionthecouplings}
The starting point of our analysis is the computation of the quark mass contribution to the nucleon mass in the $N_f=2$ non-degenerate case. It can be extracted from the mass term of (\ref{skyrme2flav}), set on-shell on the Skyrme solution, once we subtract the vacuum configuration $U=\unit_2$ and using the GMOR relation $B_0 (m_u+m_d) = 2m_{\pi}^2$. This gives
\be
{\cal L}_M = \frac{f_{\pi}^2 m_{\pi}^2}{4}\Tr[(\unit_2-\epsilon\,\tau^3)(e^{i\varphi} \mathbf a(t) U_s \mathbf a^{-1}(t)-\unit_2) + h.c. ]\,,
\label{sklag}
\ee
where
\be
\frac{\hat\eta'}{f_{\pi}}\equiv \varphi \equiv \frac{\eta(r)}{f_{\pi}} \frac{\vec J \cdot \hat x}{2\, \Theta}\,,\quad U_s = \unit_2 \cos{F(r)} + i \vec\tau \cdot \hat x \sin{F(r)}\,,
\label{exts}
\ee
describe the ansatz for the extended Skyrmion solution of \cite{Jain0,Jain:1989kn} which we truncate to the $\eta'$-pion sector. The generalized pseudoscalar-vector meson Lagrangian considered in \cite{Jain:1989kn} contains in fact various mixing and isospin-breaking terms involving the pion, the $\eta'$ and other vector mesons. These terms contribute to the Skyrmion equation of motion and to the nucleon mass terms. However they are expected to provide subleading corrections (see e.g. comments in \cite{reviewSk}) and we will neglect their effect here.

In eq. (\ref{exts}) $\vec J$ is the angular momentum and $\Theta$ the moment of inertia of the nucleon. Moreover  $r^2 = \sum_{n=1}^3 x_i^2$, $\hat x = \frac{\vec x}{|\vec x|}$, $\tau^i$ ($i=1,2,3$) are the Pauli matrices and $\unit_2$ is the two-dimensional identity matrix. To leading order in $1/N_c$ and in the small quark mass limit, the function $F$ is the solution of the equation \cite{anw}
\be\label{skyrsol}
\left(\frac{y^2}{4} +2 \sin^2{F} \right) F'' +\frac{y}{2} F' + \left( F'^2 -\frac14 \right) \sin{2F} -\frac{\sin^2{F} \sin{2F}}{y^2}=0\,,
\ee
where $F=F(y)$ with $y= 2 e f_{\pi} r$, and the conditions for having baryonic number equal to one are $F(0)=\pi$, $F(\infty)=0$.
The equation can be solved numerically. The solution asymptotes  as $y^{-2}$ at large $y$.

The static Skyrmion solution is described by the matrix $U_s$, and the quantization is achieved in a non-relativistic limit passing to a slowly rotating time-dependent solution ${\mathbf a}(t) U_s {\mathbf a}(t)^{-1}$ where the matrix ${\bf a}$ accounts for the time-dependent $SU(2)$ (for $N_f=2$) moduli. Some detail on the instanton moduli quantization, valid {\it mutatis mutandis} for both the Skyrme and the holographic WSS model, can be found in appendix \ref{instantonmoduli}.

In the extended Skyrme model of \cite{Jain0,Jain:1989kn}, where the Skyrmion emerges as a soliton solution of an effective Lagrangian including $\rho$ and $\omega$ vector mesons, the quantization includes a time-dependent solution for the $\eta'$ soliton component \cite{Jain0}, whose form is given in eq. (\ref{exts}) above. The function $\eta(r)$ is extracted from the $\eta'$ equations of motion. The latter include (at least) two parameters, related to the vector-vector-pseudoscalar coupling $g_{VV\phi}$ and the vector-pseudoscalar$^3$ coupling $h$. These have been extracted in the model at hand in \cite{meissner}.

The $\eta'$ soliton component is proportional to $\vec J$ and thus to the soliton angular velocity $\vec \chi \sim$Tr$[\dot {\mathbf a} \vec\tau {\mathbf a}^{-1}]$ which clearly vanishes in the static limit. Time derivatives of the moduli are usually $1/N_c$ suppressed in the model and therefore the contribution of this term will be generically suppressed too. However, this term provides the leading order contribution to the CP-even $g_{\eta' NN}$ coupling. The explicit result can be found in section E of \cite{meissner} for the $N_f=2$ case. By standard parameter fitting the authors of \cite{meissner} obtain $g_{\eta' NN}=1.61$.

Since $\varphi\sim 1/N_c$, we can expand the Lagrangian term (\ref{sklag}) around $\varphi\approx0$. To first order in $\varphi$ we get
\be
{\cal L_M}\approx -f_{\pi}^2 m_{\pi}^2 (1-\cos F) + \epsilon\, \frac{f_{\pi}^2 m_{\pi}^2}{2} \sin F \frac{x^a}{r}\Tr\left[\tau^3 \mathbf a \tau^a \mathbf a^{-1}\right] \varphi\,.
\ee
The corresponding Hamiltonian term is given by
\be
H_M = - \int d^3 x\, {\cal L_M} = f_{\pi}^2 m_{\pi}^2 \int d^3 x (1-\cos F) + \frac{2\pi}{3}\epsilon\, \frac{f_{\pi} m_{\pi}^2}{\Theta}\int dr r^2 \sin F\,\eta\, I^3\,,
\label{hms}
\ee
where $I^3$ is the third component of the isospin operator and we have used the relation
\be
\Tr\left[\tau^3 \mathbf a \tau^a \mathbf a^{-1}\right]J^a = -2 I^3\,.
\ee
From (\ref{hms}), recalling that $I^3=1/2$ (resp.~$-1/2$) on the proton (resp. neutron) state, we get the pion-nucleon sigma term and the quark mass contribution to the neutron proton mass splitting \cite{an,Jain:1989kn}
\be
\delta M_N = f_{\pi}^2 m_{\pi}^2 \int d^3 x (1-\cos F)\,,\quad (M_n-M_p)_{{\cal L}_M} =  -\frac{2\pi}{3}\epsilon\, \frac{f_{\pi} m_{\pi}^2}{\Theta}\int dr\, r^2\, \sin F\, \eta\,.
\label{massesky}
\ee
Notice that the sigma term can be given explicitly as
\be
\delta M_N = \frac{\alpha \pi m_{\pi}^2 }{2 e^3 f_{\pi}}\,,
\label{37} \ee
where the constant $\alpha$ is determined by the numerical solution for $F(y)$ through
\be\label{alpha}
\alpha \equiv \int_0^{\infty} dy y^2 \left( 1- \cos{F(y)} \right) \approx 18.25 \,.
\ee
Moreover, the neutron-proton mass difference can be obtained by numerically solving the equation for $\eta(r)$ \cite{Jain:1989kn}.
Notice that in the more general case of $N_f$ degenerate quark masses, one would get the same expression as in (\ref{37}),
as a result of the standard embedding of the $SU(2)$ Skyrmion solution into $SU(N_f)$.

Taking the previous results into account, there is a quick way to get the CP-odd axion-nucleon couplings.

Following the procedure outlined in section \ref{CHPTL},
we can remove the axion-meson mixing terms in the vacuum solution in such a way that the net effect is just a redefinition of the quark mass matrix. In the ``chiral limit" $m_\pi \ll m_{WV}$ (with $m_{WV}$ defined in (\ref{masseshere})),  treating the axion as an external field, a convenient choice is\footnote{Following the notations in section \ref{CHPTL}, a natural choice of frame 
which indeed reproduces the nucleon Hamiltonian mass term in eq. (\ref{amae}), consists in taking $c_s=0$ and $\tilde Q = M^{-1}/(2\Tr[M^{-1}]) = (1/4) (\unit_2+\epsilon\, \tau^3)$, so that $\tilde c_s=1$.}
\be
M(a,\varphi) = m\, {\rm diag} \left[(1-\epsilon) e^{i \frac{a}{2 f_a}(1+\epsilon)}\,, (1+\epsilon) e^{i \frac{a}{2 f_a}(1-\epsilon)}\right] e^{i\varphi}\equiv m\,{\tilde M}(a,\varphi) \,,
\ee
where $2m=m_u+m_d$. Hence, in the Skyrme plus axion case, after shifting the axion by $f_a \theta$ to account for a residual $\theta$ angle, the nucleon mass term arises from the Lagrangian term
\be
{\cal L}_M = \frac{f_{\pi}^2 m_{\pi}^2}{4}\Tr[{\tilde M}(a+f_a\theta, \varphi) (\mathbf a U_s \mathbf a^{-1}-\unit_2) + h.c. ]\,,
\ee
where $\varphi$ and $U_s$ are given in (\ref{exts}). Working to first order in $\varphi$, expanding for small $a,\theta$ and keeping only the terms in $\theta a$, we get
\be
{\cal L}_{M}|_{a\theta} = -\frac{f_{\pi}^2 m_{\pi}^2}{4}\left[ (\cos F -1) - \frac{\epsilon}{2}\sin F \frac{x^a}{r}\Tr\left[\tau^3 \mathbf a \tau^a \mathbf a^{-1}\right] \varphi\right]\theta \frac{a}{f_a}(1-\epsilon^2)\,.
\label{skyaxi}
\ee
Treating the axion as an external field, the corresponding Hamiltonian term reads
\be
H_M|_{a\theta} = -\int d^3x {\cal L}_{M}|_{a\theta}= - \left[ \frac{\delta M_N}{4}+\frac{M_n-M_p}{4}I^3\right] (1-\epsilon^2)\frac{a}{f_a}\theta\,,
\label{amae}
\ee
which can be expressed in terms of the Skyrme model parameters using (\ref{massesky}) and (\ref{37}). This expression gives the Lagrangian CP-odd axion couplings $\bar c_N$ precisely as in eq. (\ref{geneform}). In terms of the model parameters
\bea
\bar c_{p} =\left[ \frac{\alpha \pi m_{\pi}^2 }{8 e^3 f_{\pi}} -\frac{\pi}{12}\epsilon\, \frac{f_{\pi} m_{\pi}^2}{\Theta}\int dr\, r^2\, \sin F\, \eta \right]\,(1-\epsilon^2)\frac{\theta}{f_a}\,,\nonumber \\
\bar c_{n} =\left[ \frac{\alpha \pi m_{\pi}^2 }{8 e^3 f_{\pi}} +\frac{\pi}{12}\epsilon\, \frac{f_{\pi} m_{\pi}^2}{\Theta}\int dr\, r^2\, \sin F\, \eta \right]\,(1-\epsilon^2)\frac{\theta}{f_a}\,.
\label{cngen}
\eea
For the more general case of $N_f$ degenerate quark flavors, the common factor $4$ in the denominator is replaced by $N_f^2$.

Taking into account the axion-meson mixing, and the general IR relations (\ref{hatcpcnchi}), the previous expressions allow us to immediately find $\bar g_{\eta' NN}$ and $\bar g_{\pi NN}$. In the chiral limit they read
\bea
\bar g_{\eta' NN} &=& - \frac{\delta M_N}{2 f_{\pi}}(1-\epsilon^2) \theta= -\frac{\alpha \pi m_{\pi}^2 }{4 e^3 f_{\pi}^2}(1-\epsilon^2) \theta\,,\nonumber \\
\bar g_{\pi NN} &=& -\frac{(M_n - M_p)_{({\cal L}_M)}}{4 f_{\pi} \epsilon}(1-\epsilon^2)\theta= \frac{\pi}{6}\, \frac{m_{\pi}^2}{\Theta}\int dr\, r^2\, \sin F\, \eta\,(1-\epsilon^2)\theta\,,
\label{cpoddskyrme}
\eea
where we have used eqns. (\ref{massesky}) and (\ref{37}). As a countercheck, we have controlled that the expressions above can be also obtained directly starting from the $\theta$-dependent Skyrme Lagrangian\footnote{See e.g. \cite{dixon,salomonson} and improve their results with the extensions in \cite{Jain0,Jain:1989kn}.} and considering the equations of motion for the $\eta'$ and pion fluctuations around the Skyrmion solution.

If the Peccei-Quinn mechanism works perfectly, so that no residual $\theta$ term is left over, the results in (\ref{cngen}) can be interpreted as providing the quartic axion-axion-nucleon couplings $\bar g_{a^2 NN}$ and those in (\ref{cpoddskyrme}) the meson-axion-nucleon couplings: it just suffices to trade back $\theta$ for $a/f_a$. In practice in the chiral limit this gives
\be
\bar g_{a^2 NN} =\frac{\bar c_N}{2\theta f_a}\,,\quad \bar g_{H a N N}= \frac{\bar g_{H N N}}{\theta f_a}\,, \quad (H=\eta'\,,\pi)\,.
\label{44}
\ee
In appendix \ref{nf3skyrme} we will present related results in the $N_f=3$ non-degenerate case.

\section{Large $N_c$ estimates in the WSS holographic model}
\label{sec:WSS}
In this section we consider a second model which allows to estimate the non-derivative couplings to nucleons of axion, $\eta'$ and pions. It is the most successful top-down holographic cousin of planar QCD, know as the  Witten-Sakai-Sugimoto model (WSS)  \cite{witten, SS1}.
Although it has a higher-dimensional UV completion, it gives reasonable quantitative results for many QCD observables.
Similarly to the Skyrmions, baryons are solitons of this theory.
Moreover, the model allows to include systematically the contribution of the (axial) vector mesons.

In the next subsection, we will briefly review the model for the reader who is  unfamiliar with it.
In the following subsections we will compute the above mentioned CP-odd couplings in analogy to what we have done for the Skyrme model.

\subsection{The WSS model}

The WSS model is based on a $D_4-D_8$ brane setup in type IIA string theory. There are $N_c$ $D_4$-branes wrapped on a circle of radius $R_4 = 1/M_{KK}$ where antiperiodic boundary conditions for the fermions are imposed. Then $N_f$ $D_8$-anti-$D_8$-branes are placed at antipodal points on that circle.\footnote{This is not the only possibility. More general non-antipodal configurations can be also considered. However, it is only in the antipodal case that the glueball and meson mass scales coincide.} At energies much smaller than $M_{KK}$, the dynamics on such branes is given by a $3+1$ dimensional large $N_c$ $SU(N_c)$ gauge theory with $N_f$ massless quarks. The massive sector contains scalars and fermions (with mass scaling as $M_{KK}$) in the adjoint representation of the gauge group. In the limit where a simple holographic gravitational dual description of the model can be given, this massive sector cannot be decoupled from the QCD-like one. The classical gravity regime, in fact, amounts on taking $N_c\gg1$ and $\lambda\gg1$, where the 't Hooft coupling $\lambda$ sets the ratio between the confining $SU(N_c)$ string tension and $M_{KK}^2$.

The flavor sector in the model is described by the low energy modes of the $D_8$-branes.
When $\epsilon_f \sim\lambda^2(N_f/N_c)\ll1$ the latter can be treated as probes of the background sourced by the $D_4$'s: this corresponds to the quenched approximation for the quarks.\footnote{See \cite{wsssmeared} for an account of the flavor backreaction to first order in $\epsilon_f$.} In this limit, the $D_8$-anti-$D_8$-branes actually join in the background, realizing geometrically chiral symmetry breaking.
Their effective action reduces to a $U(N_f)$ Yang-Mills theory with Chern-Simons terms on a curved space-time in five dimensions
\bea\label{actions}
&& S_{WSS}= S_{YM} + S_{CS}\,,\\
&& S_{YM} = -\kappa\int d^4x d z\,\Tr\left(\frac{h(z)}{2} \mathcal{F}_{\mu\nu}\mathcal{F}^{\mu\nu} + k(z)\mathcal{F}_{\mu z}\mathcal{F}^\mu_{\;\; z}\right)\,,\nonumber\\
&& S_{CS} = \frac{N_c}{24\pi^2}\int  \Tr\left(\mathcal{A}\mathcal{F}^{ 2}- \frac{i}{2}\mathcal{A}^{ 3}\mathcal{F} - \frac{1}{10}\mathcal{A}^{ 5}\right)\,,
\nonumber
\eea
where (in units\footnote{The correct dimensions in the formulae of this section are recovered by inserting powers of $M_{KK}$ where needed.} with $M_{KK}=1$) \be
\kappa = \frac{\lambda N_c}{216\pi^3}\sp  h(z) = (1+z^2)^{-1/3}\sp  k(z) = (1+z^2)\sp z\in(-\infty,\infty)\;.
\label{51}\ee
Here $z$ is the radial holographic direction and we have omitted the wedge product symbol ``$\wedge$" in the CS terms.
We shall mostly focus on such an action in the $N_f=2$ case.

The matrix for the Goldstone modes is given by the path ordered holonomy matrix
\be
U= {\mathcal P} {\rm exp} \left[i \int dz {\mathcal A}_z \right]\,.\label{52}
\ee

The effective action for $U$ precisely reduces to the chiral Lagrangian with the Skyrme term \cite{SS1}, with the pion decay constant $f_{\pi}$ and the coupling $e$ given by
\begin{equation}
f_{\pi} = 2 \sqrt{\frac{\kappa}{\pi}}\,, \qquad e^2 \sim \frac{1}{2.5\kappa}\,.
\label{fpaie}
\end{equation}
The pseudoscalar meson $\eta'$ acquires a mass according to the Witten-Veneziano relation
\be
m_{\eta'}^2 = m_{WV}^2 = \frac{2N_f}{f_{\pi}^2}\chi_{YM}\,,
\label{53}\ee
where $\chi_{YM}$ is the topological susceptibility of the pure $SU(N_c)$ theory, given by
\be
\chi_{YM} = \frac{\lambda^3 M_{KK}^4}{4(3\pi)^6}\,.
\label{tsusc}
\ee
This implies that in the WSS model the Witten-Veneziano mass,
\be
m_{WV}^2 = \frac{1}{27\pi^2}\frac{N_f}{N_c} \lambda^2 M_{KK}^2 \sim \epsilon_f M_{KK}^2\,,
\label{mWV}
\ee
is much smaller than $M_{KK}$.

Baryons in the model are instanton solutions \cite{SS-baryons} with baryon number $n_B$ given by
\be
n_B=\frac{1}{8\pi^2}\int_B \Tr \mathcal{F}\wedge\mathcal{F}\,,
\label{54}\ee
where $B$ is the space spanned by $(x^{1,2,3},\,z)$ and $\mathcal{F}$ is the $U(N_f)$ field strength. Nucleons correspond to $n_B=1$ solutions.

In the $N_f=2$ case with massless flavors, using an $SU(2)\times U(1)$ notation for the gauge field
\be
{\cal A} = A + \widehat A \frac{\unit_2}{2} = A^a \frac{\tau^a}{2} + \widehat A \frac{\unit_2}{2}\,,
\label{55}\ee
a simple static instanton solution can be given around $z=0$ where the curvature of the background can be neglected. The related solution corresponds to a charged BPST instanton
\begin{equation}
A_M^{\mathrm{cl}} = -i f(\xi) g \partial_M g^{-1}\,,\quad \widehat{A}^{\mathrm{cl}}_0 = \frac{N_c}{8\pi^2\kappa}\frac{1}{\xi^2}\left[1-\frac{\rho^4}{(\rho^2+\xi^2)^2}\right]\;,\quad A_0^{\mathrm{cl}}=\widehat{A}_M^{\rm cl}=0\,,
\label{Ainst}
\end{equation}
where $M=1,2,3,z$ and
\begin{equation}
f(\xi) = \frac{\xi^2}{\xi^2+\rho^2}\;,\quad g(x) = \frac{(z-Z)\unit_2 - i (\vec{x}-\vec{X})\cdot \vec{\tau}}{\xi}\label{eqg}\;, \quad \xi^2 \equiv (\vec{x}-\vec{X})^2+(z-Z)^2\,.
\end{equation}
This solution depends on eight parameters: the instanton center of mass position $X^{M}=(\vec X, Z)$, the instanton size $\rho$ and implicitly, three $SU(2)$  ``angles" related to the fact that the solution can be rotated by means of a global gauge transformation. Precisely as in the Skyrme case of section \ref{sectionthecouplings}, they are encoded in the matrix
\begin{equation}
\mathbf{a}=a_4 \unit +i\sum_{i=1}^3a_i\tau^i, \quad\quad \sum_{I=1}^4 a^2_I = 1\,.
\label{56}\end{equation}
Substituting the solution (\ref{Ainst}) into the action (\ref{actions}) one finds
\be
S_{\rm{on\,shell}} = -\int d t ~M_B\;,
\label{57} \ee
 where, up to $\mathcal{O}(\lambda^{-2})$ corrections
\begin{equation}\label{MB}
M_B(\rho,Z) =  M_{B0}\left[1 +  \left(\frac{\rho^2}{6}+ \frac{N_c^2}{320 \pi^4 \kappa^2}\frac{1}{\rho^2} + \frac{Z^2}{3}\right)\right]\,,\quad M_{B0}\equiv8\pi^2 \kappa\,,
\end{equation}
where $M_{B0}$ gives the baryon mass in the $\lambda\rightarrow\infty$, $N_c\rightarrow\infty$ limit.

This implies that $\rho$ and $Z$ are not genuine moduli; in fact they are classically fixed by minimizing $M_B$ as
\begin{equation}\label{rhocl}
\rho_\mathrm{cl}^2 = \frac{N_c}{8\pi^2 \kappa}\sqrt{\frac{6}{5}}= \frac{27\pi}{\lambda}\sqrt{\frac{6}{5}}\;,\qquad
Z^\mathrm{cl} = 0\,.
\end{equation}
These relations imply that the instanton size $\rho \sim 1/\sqrt{\lambda}$ is very small (but not zero) in the $\lambda\gg1$ regime, and that the center of the instanton is classically localized at $Z=0$.  The full expression for the static instanton solution valid for any value of $z$ in the $\lambda\gg1$ limit can be found in \cite{hss}.

In \cite{wsstheta1,wsstheta2}, the solution above has been modified by including a (small) mass term $m_q$ for the flavor fields (taken to be degenerate in mass) as well as a (small) non-zero topological $\theta$ term. The deformation is driven (in units $M_{KK}=1$) by the (small) parameter
\be\label{ci}
\frac{c m_q}{\kappa} \theta \equiv \frac{m_{\pi}^2}{\pi} \theta\,.
\ee
In the mass-and-$\theta$-deformed instanton solution, two new gauge field components are turned on, to first order in the above parameter. One is the Abelian component $\widehat A^{\rm{mass}}_z$, which
gives a non trivial vacuum expectation value to the $\eta'$ meson, while the other is the non-Abelian component $A_0^{\rm mass}$ which can be interpreted as a dipole electric potential in the bulk. It is the latter component which is responsible for the nucleon electric dipole moment to be different from zero \cite{wsstheta1,wsstheta2}, but it is the former which will be useful for our purposes below.

In appendix \ref{instantonmoduli} we review the instanton moduli quantization, which will be necessary for the computation of the CP-odd couplings in section \ref{WSSbg}.


\subsection{The derivative axion-nucleon couplings}

As we have pointed out in section \ref{sectionskyrme} the derivative axion coupling to nucleons, for the $N_f=2$ case, can be simply deduced once the CP-even couplings $g_{\eta'NN}$ and $g_{\pi NN}$ are known. In the WSS model in the chiral limit these couplings have been already computed in \cite{hss}. They read
\bea\label{dercouplingsWSS}
&&g_{\eta' NN} = \frac{m_N}{f_{\pi}}\frac{N_c}{16\pi^3\kappa} \langle k(Z)^{-1}\rangle = \frac{m_N}{f_{\pi}} \hat g_A\,,\nonumber \\
&&g_{\pi NN} = \frac{m_N}{f_{\pi}}\frac{16\pi\kappa}{3}\langle \rho^2 k(Z)^{-1}\rangle = \frac{m_N}{f_{\pi}} g_A\,,
\eea
clearly satisfying the generalized Goldberger-Treiman relations. Using (\ref{dercup1}), (\ref{c27}) the derivative axion-nucleon couplings follow.

It is worth noticing that the leading contribution to the isoscalar coupling $g_{\eta' NN}$ comes from the time-dependent term, proportional to the angular velocity $\vec\chi$, in the solution for $\widehat A_z$ (\ref{hataz}).

\subsection{The non-derivative axion-nucleon couplings for $N_f=2$}\label{WSSbg}
In the holographic WSS model, the nucleons correspond to leading order to the static instanton solutions ${\cal A} = {\cal A}^{cl}$  (\ref{Ainst}) for the five-dimensional gauge field  on the flavor branes. Moreover, the quantization of the solution proceeds in analogy with the Skyrme case and, as we review in appendix \ref{instantonmoduli}, it generates (among others) a non-trivial time-dependent profile for the Abelian component $\hat A_z$ which is related to the $\eta'$ field. Furthermore, the quark mass term in the model, in the small quark mass limit, turns out to have precisely the same form as in the Skyrme case \cite{AK,Hashimotomass,wsstheta1,wsstheta2}. Finally, and crucially, the WSS model allows for the explicit introduction of an axion \cite{bccdv}, giving the standard modification of the low energy chiral Lagrangian as written in (\ref{skyrme2flav}).
The holographic set-up provides a UV origin of the axion model, but $f_a$ is an essentially free parameter.

All in all, focusing on the pseudoscalar sector, one can extract the quark mass contribution to the baryon Hamiltonian and the CP-odd couplings precisely as in the Skyrme model in section \ref{sectionskyrme}.

Let us consider the ``chiral limit" $m_\pi \ll m_{WV}$ with $N_f=2$ non-degenerate flavors. The starting point is thus the same Lagrangian mass term as in (\ref{sklag}) where now
\begin{equation}\label{instsol}
U_s\equiv U_{cl}= U({\cal A}^{cl}) = \exp\left[i\pi\frac{\vec{\tau}\cdot\vec{x}}{|\vec{x}|}\left(1-\frac{1}{\sqrt{1+\rho^2/|\vec{x}|^2}}\right)\right]\,,
\end{equation}
and, from the time-dependent solution (\ref{hataz}),
\be
\varphi(r) = \frac12\int dz \widehat A_z = F(r) P_Z - \frac{N_c}{64\pi\kappa} \frac{\rho^2}{(r^2+\rho^2)^{3/2}}\vec\chi\cdot\vec x\,,
\label{63a}\ee
where $P_Z$ is the momentum conjugate to $Z$ and $\vec\chi$ is the instanton angular velocity defined in eq. (\ref{chij}). We shall not need the explicit expression for $F(r)$ since, when computing the VEV of the above expression on the nucleon ground state, this will not contribute. We thus set $P_Z=0$ in the following. Notice that $\varphi$ scales like $1/N_c$, just as in the Skyrme case.

The above ingredients are sufficient to extract the pion-nucleon sigma term and the neutron-proton mass splitting following precisely the same steps as in the Skyrme model case. To leading order in the holographic limit they are given by \cite{hashimoto:2009hj,bini}
\begin{equation}
\delta M_{N}= \frac{2^5 \gamma  \rho_{cl}^3 m_{\pi}^2 \kappa^{3/2}}{\pi^{1/2} f_{\pi}} \,,\quad (M_n-M_p)_{({\cal L}_M)}=\frac{\hat \gamma \rho_{cl} N_c}{24\,\pi^{7/2}} \frac{m_{\pi}^2}{\kappa^{1/2}f_{\pi}}\epsilon
\label{her}
\end{equation}
where
\be
\gamma\equiv\int_0^\infty d y\, y^2\left(1+\cos \frac{\pi}{\sqrt{1+1/y^2}}\right)\approx 1.10 \,,
\label{60}\ee
and
\be
\hat \gamma \equiv \int_0^{\infty} dy (1+y^{-2})^{-3/2} \sin\left(\frac{\pi}{\sqrt{1+y^{-2}}}\right) \approx 1.05\,.
\label{66}\ee
Since at low energy the effective axion-pseudoscalar Lagrangian in the WSS model \cite{bccdv} has precisely the same form as in the Skyrme case, the CP-odd axion-nucleon couplings are expressed in terms of the nucleon mass properties (\ref{her}) precisely as in (\ref{geneform}). In terms of the model parameters we thus get
\bea
\bar c_{p} = \left[\frac{8 \gamma  \rho_{cl}^3 m_{\pi}^2 \kappa^{\frac32}}{\pi^{1/2} f_{\pi}}+\, \frac{\hat \gamma \rho_{cl} N_c}{3\cdot2^6\,\pi^{7/2}} \frac{m_{\pi}^2}{\kappa^{1/2}f_{\pi}}\epsilon \right] (1-\epsilon^2)\frac{\theta}{f_a}\,,\nonumber \\
\bar c_{n} = \left[\frac{8 \gamma  \rho_{cl}^3 m_{\pi}^2 \kappa^{\frac32}}{\pi^{1/2} f_{\pi}}-\, \frac{\hat \gamma \rho_{cl} N_c}{3\cdot2^6\,\pi^{7/2}} \frac{m_{\pi}^2}{\kappa^{1/2}f_{\pi}}\epsilon \right] (1-\epsilon^2)\frac{\theta}{f_a}\,.
\label{cngenwss}
\eea
In the ``chiral limit", the CP-odd meson-nucleon couplings are given in turn by precisely the same relations as in (\ref{cpoddskyrme}). In terms of the WSS model parameters, they are given by
\be\label{bargetawss}
{\bar g}_{\eta' NN}=- \frac{\delta M_N}{2 f_{\pi}}(1-\epsilon^2) \theta = - \frac{2^4 \gamma \rho_{cl}^3  m_{\pi}^2 \kappa^{3/2}}{\pi^{1/2} f_{\pi}^2}(1-\epsilon^2)\theta\,,
\ee
and by
\be\label{gbarpiNNWSS}
\bar g_{\pi NN} = -\frac{(M_n - M_p)_{({\cal L}_M)}}{4 f_{\pi} \epsilon}(1-\epsilon^2)\theta= - \frac{\hat \gamma \rho_{cl} N_c}{3\cdot 2^5 \pi^{7/2}} \frac{m_{\pi}^2}{\kappa^{1/2}f_{\pi}^2}(1-\epsilon^2)\theta\,.
\ee
Note that
\be
\bar g_{\pi NN}\sim N_c^{-1/2}\,,
\label{67} \ee
which is precisely the minimal scaling argued in \cite{riggs} to occur in the Skyrme model.

It is worth noticing that the above expressions can be equivalently obtained by considering the instanton solution for the gauge field components corresponding to the pseudoscalar mesons in the WSS model at finite $\theta$ angle \cite{wsstheta1,wsstheta2}.

Let us notice that, as in the Skyrme model case, the above results have been obtained by neglecting the effects of vector mesons which are expected to be subleading.\footnote{This expectation appears to be confirmed by explicit computations in the complete model. We thank Lorenzo Bartolini for discussions on this point and for communication of his preliminary results.}

In appendix \ref{wssnf3} we will collect analogous results for the $N_f=3$ case.

\section{Discussion and numerical estimates of the couplings}
\label{sectionnumbers}

Let us summarize  what we have learned in the previous sections.
As a first result, we have observed that the non-derivative CP-odd axion couplings $\bar c_{p,n}$ with protons and neutrons can be expressed, purely from low energy considerations valid for any number of colors $N_c$, in two equivalent ways.
The first one is in terms of the CP-odd couplings $\bar g_{\eta' NN}$ and $\bar g_{\pi NN}$ of the $\eta'$ and pions to the nucleons, as in formula (\ref{34}) or its chiral limit (\ref{hatcpcnchi}).
Then, by exploiting known results, the axion couplings can also be expressed in terms of the pion-nucleon sigma term and the strong contribution to the neutron-proton mass splitting, as in formula (\ref{geneform}).

These have to be thought of as tree level relations in the language of chiral perturbation theory. Their relevance is that the axion-nucleon couplings are expressed in a model-independent way in terms of nucleon properties which, if not directly measurable from experiments, can be computed on the lattice or can be deduced from phenomenology. There is a vast literature on lattice QCD results for the pion-nucleon sigma term. The most recent papers, from {\it e.g.} the Wuppertal-Marseille \cite{wuppertal} and the ETMC collaborations \cite{etcm} report values of $\delta M_N$ of $38(3)(3)$ and $41.6(3.8)$ MeV, respectively. These lattice results exhibit a certain tension with phenomenological derivations from pionic atom data (see e.g.~\cite{pionat}) which essentially point towards a value of the sigma term which is around $60$ MeV.\footnote{Further discussions on the sigma term issue can be found in {\it e.g.} \cite{Bali:2011ks, Ellis:2018dmb}.} Lattice QCD results on the strong force contribution to the neutron-proton mass splitting can be found in {\it e.g.} \cite{borsanynp}, giving $(M_n - M_p)_{{\cal L}_M}\approx 2.52(17)(24)\, {\rm MeV}$. Using this result, the value $\epsilon  \equiv (m_d-m_u)/(m_d+m_u)\approx 0.35(3)$ (see e.g.~\cite{villadoro}) and the most recent ETMC computation of the sigma term we get, from our formulae (\ref{geneform}), the estimates in the first column of table \ref{tablenumbers}.
The values are expressed in MeV times $\theta/f_a$.
Instead, by employing the phenomenological extraction of the sigma term in the second paper in \cite{pionat}, i.e.~$\delta M_N \sim 58(5)$ MeV, we obtain the values in the second  column of table \ref{tablenumbers}.

\begin{table}[h!]
\centering
\begin{tabular}{ |c||c|c|c|c|c|c| }
 \hline
     & CL + lattice & CL + pheno & Skyrme & Skyrme alt. & WSS & WSS alt.\\
 \hline
 $\bar c_p$   & 9.4(9) & 13.0(1.1) & 18.7(4) & 27(5) & 16.3(4) &  21(4)\\
 $\bar c_n$    & 8.8(9) & 12.4(1.1) & 18.7(4) & 27(5) & 16.3(4) & 20(4)\\
 \hline
\end{tabular}
\caption{Values of the couplings of the axion with protons and neutrons, expressed in MeV times $\theta/f_a$. In the first and second columns the tree level Chiral Lagrangian (CL) result (\ref{geneform}) is employed with the lattice results for the neutron-proton mass splitting and respectively lattice and phenomenological extractions for the pion-nucleon sigma term. The one-loop corrections increase these values by 5.5 MeV times $\theta/f_a$. In the third and fourth columns are reported the results (\ref{cngen}) and (\ref{79}) in the Skyrme model.
In the fifth and sixth columns there are the values from (\ref{cngenwss}) and (\ref{63}) in the holographic WSS model. The latter results receive corrections from the instanton pseudo-moduli quantization, see appendix \ref{appendixnumbers}.}
\label{tablenumbers}
\end{table}

Moreover, we estimated the contributions for $N_f=2$ mass degenerate flavors to $\bar c_{N}$  ($\bar g_{a NN}$) arising through CP-violating $a-\pi-\pi$ coupling at one-loop order in chiral perturbation theory in section~\ref{loop}, equation~\eqref{barganneft}.\footnote{A similar estimate of ${\bar g}_{\eta' NN}$ for $N_f=2$ mass degenerate flavors is given in eq.~(\ref{loopbargestimates}).}
These contributions are smaller than the estimates for the ``direct'' contributions which we discussed so far, but they affect the results in the first two columns of table \ref{tablenumbers} by increasing their values by $5.5$ MeV times $\theta/f_a$.

\vspace{0.5cm}
In view of the tension between lattice and phenomenological data for the sigma term, we find it interesting to write down other possible expressions for the CP-odd axion couplings. This is one of the reasons why it is interesting to have found explicit expressions for those couplings, in terms of the model parameters, in Skyrme and WSS models.
Moreover, trying to improve the leading order relations we have found in chiral perturbation theory is in practice problematic, due to unknown low energy constants.
In this situation, one could view the Skyrme and WSS results as providing model-dependent estimates of the latter.

Note that our computations of the CP-odd meson-nucleon couplings in both the Skyrme and WSS model have shown agreement,
to leading order in the large $N_c$ limit, with formulae (\ref{relMW}) and (\ref{gpiDeltaM}) already known from chiral effective field theory.
But below we want to avoid the explicit use of the value of the sigma term and the neutron-proton mass difference.
Instead, we use the Skyrme and holographic models for direct estimates of the couplings. Although, as mentioned in the Introduction, we cannot expect these models to have an accuracy at percent level (e.g.~in WSS it is expected to be around 30\% on average in our type of observables), at the moment we cannot claim to have a better precision with other means. A-posteriori, the numerical values obtained in these models appear to be comparable with the Chiral Lagrangian ones, once we include the one-loop corrections in the latter.

Let us begin with the Skyrme model. In the $N_f=2$ case the coefficient of the Skyrme term is usually fixed to $e \approx 4.84$ by fitting the masses of baryons and the pion; the fit also sets $f_{\pi} \approx 54$ MeV \cite{an}.
With these values of parameters we obtain the estimate for the magnitude of the axion couplings (\ref{cngen}) in the third column of table \ref{tablenumbers}. Up to the first decimal digit, isospin breaking corrections are not visible.

One can also proceed in an alternative way as follows. Let us first notice that the numerical value of $\bar g_{\pi NN}$ in (\ref{bargpiNNSnum}) is so small, as compared to $\bar g_{\eta' NN}$ (see equation (\ref{bargetapNNSnum})), that we can safely discard its contribution.\footnote{We do not expect an order of magnitude correction to these couplings by using slightly different values of the parameters.}
Therefore, since
\be
g_{\pi NN}= -{D \pi m_{N}\over 3 e^2 f_{\pi}}~~~{\rm  with}~~~ D \simeq -17.2
\label{77}\ee
\cite{anw}, we can rewrite the CP-odd $\eta'$-nucleon coupling in (\ref{cpoddskyrme}), with $\delta M_N$ given in (\ref{37}), to leading order in the $\epsilon\ll1$ limit, as
\be
\bar g_{\eta' NN} = - \theta \frac{3 \sqrt{3} \alpha}{4 \sqrt{\pi} (-D)^{3/2}} \frac{m_{\pi}^2}{f_{\pi}^{1/2} m_N^{3/2}}\ g_{\pi NN}^{3/2} \,,
\label{78}\ee
and
\be
\bar c_p = \bar c_n = \frac{\theta}{f_a} \frac{3 \sqrt{3}\alpha}{8 \sqrt{\pi} \epsilon^{3/2} (-D)^{3/2}}\frac{m_{\pi}^2}{f_{\pi}} (c_n-c_p)^{3/2} \,. 
\label{79}\ee
The advantage of expressions (\ref{79}) with respect to the form presented in the previous section is that they do not explicitly depend on the parameters of the Skyrme model anymore, so they are suitable to provide a straightforward estimate for the couplings.

We conform the values of the physical quantities to the ones in \cite{villadoro}, where $c_p, c_n$ are calculated\footnote{The error on $m_{\pi}$ is irrelevant.}
\bea
&& m_{\pi}= 134.98\ {\rm MeV}\,, \qquad f_{\pi}= 92.21(14)\ {\rm MeV}\,, \qquad \epsilon = 0.35(3)\,, \nonumber \\
&& c_p = -0.47(3)\,, \qquad \qquad c_n = -0.02(3)\,.
\label{74}\eea
Remembering that $\alpha\approx18.25$ (see eq. (\ref{alpha})), using relations (\ref{74}) and (\ref{79}), we get the numerical values in the fourth column of table \ref{tablenumbers}. Working in the ``chiral limit'', with the leading order mass term and at first order in $\epsilon$ increases the error in table \ref{tablenumbers}, from $5$ MeV to $6$ MeV.

Coming to the Holographic WSS model, fitting the model parameters to the values for $f_{\pi}$ and $m_{\pi}$ reported in (\ref{74}) and to the $\rho$-meson mass $m_{\rho}\approx776\ {\rm MeV}$
implies setting \cite{SS1}
\be\label{parametersh}
\lambda\approx 16.63\,,\quad M_{KK} \approx 949\ \rm{MeV}\,.
\ee
The axion non-derivative coupling (\ref{cngenwss}), including the known quantum corrections (see appendix \ref{appendixnumbers}), is reported in the fifth column of table \ref{tablenumbers}.

Moreover we can obtain an alternative estimate of the couplings in analogy with what we have done for the Skyrme model above.
As we have seen, the WSS model allows for the computation of the four couplings $g_{\eta'NN}, g_{\pi NN},$ $\bar g_{\eta' NN}, \bar g_{\pi NN}$ - see formulae (\ref{dercouplingsWSS}), (\ref{bargetawss}), (\ref{gbarpiNNWSS}) for the $N_f=2$ case.
At leading order in $N_c$, such that the modulus $\rho$ is treated as classical, and to leading order in the $\epsilon\ll1$ limit, these couplings satisfy the relations
\bea
&& \bar g_{\eta' NN} = - \theta \frac{3 \sqrt{3} \gamma}{4 \pi^2} \frac{m_{\pi}^2}{f_{\pi}^{1/2} m_N^{3/2}}\ g_{\pi NN}^{3/2} \,,    \\
&& \bar g_{\pi NN} = - \theta \frac{\hat \gamma}{8 \sqrt{3}\pi} \frac{m_{\pi}^2}{f_{\pi}^{1/2} m_N^{3/2}}\ g_{\pi NN}^{1/2}\ g_{\eta'NN} \,.
\label{72}\eea
where $\gamma\approx 1.10$ was defined in (\ref{60}) and $\hat \gamma\approx 1.05$ in (\ref{66}).

Therefore, from relations (\ref{dercup1}), (\ref{c27}), (\ref{hatcpcnchi}) we can express the non-derivative axion-nucleon couplings $\bar c_{p,n}$ in (\ref{cngenwss}) in terms of the derivative ones $c_{p,n}$
\bea
&& \bar c_p = \frac{\theta}{f_a} \frac{1}{16 \sqrt{3}\pi^2 \epsilon^{3/2}}\frac{m_{\pi}^2}{f_{\pi}} \left[18 \gamma (c_n-c_p)^{3/2} - \hat \gamma \pi \epsilon^2 (c_n-c_p)^{1/2} (c_n+c_p)    \right]\,,     \label{63} \\
&& \bar c_n = \frac{\theta}{f_a} \frac{1}{16 \sqrt{3}\pi^2 \epsilon^{3/2}}\frac{m_{\pi}^2}{f_{\pi}} \left[18 \gamma (c_n-c_p)^{3/2} + \hat \gamma \pi \epsilon^2 (c_n-c_p)^{1/2} (c_n+c_p)    \right]\,.
\label{73}\eea
Taking into account the values reported in (\ref{74})  we obtain the results in the sixth column of table \ref{tablenumbers}.

On top of the errors reported in table \ref{tablenumbers}, we can give an estimate of the errors implied by our approximations.
The estimate is going to be rough, since lacking an explicit calculation of the discarded contributions, we do not control their coefficients.
Nevertheless we can get an idea of the magnitude of these errors for order one coefficients.
Working in the ``chiral limit'' implied discarding corrections of order $(m_{\pi}/m_{\eta'})^2 \sim 0.02$.
Moreover, the mass term in the Lagrangian is a valid approximation at leading order in $m_{ud}/\Lambda_{QCD} \lesssim 0.04$.
Finally, we have discarded corrections in $\epsilon^2 \approx 0.12$.\footnote{The other higher order quantities, e.g.~in $\epsilon_f$ or in $1/f_a$, are small and can be discarded.}
The three effects mentioned above increase the error of $\bar c_p$
from $4$ MeV to $5$ MeV.\footnote{The three effects alone cause errors of $3$ MeV and $2$ MeV to $\bar c_p$ and $\bar c_n$ respectively. These errors are dominated by the $\epsilon^2$ effects, e.g. for $\bar c_p$ they are of order $0.12\cdot 22 \sim 3$ MeV.}

\newpage
\vskip .2cm
\section*{\bf Acknowledgments}
\addcontentsline{toc}{section}{Acknowledgments}

\noindent
We thank Asimina Arvanitaki, Lorenzo Bartolini, Panayiotis Betzios, Yuta Hamada, Maria Paola Lombardo, Francesco Nitti, Maxim Pospelov, Gabriele Veneziano, Giovanni Villadoro and Lukas Witkowski for useful discussions.
We also thank the anonymous referee whose suggestions and criticism helped us improve this presentation.

This work was supported in part by  the Advanced ERC grant SM-grav, No 669288,  by the Netherlands Organisation for Scientific Research (NWO)
under the VIDI grant 680-47-518, and the Delta-Institute for Theoretical Physics ($\Delta$-ITP), both funded by the Dutch Ministry of Education, Culture and Science (OCW).

We thank the Galileo Galileo Institute for Theoretical Physics (Florence) and the organizers of the workshop ``String Theory from a Worldsheet Perspective'' during which part of this work has been completed.

\newpage
\appendix

\renewcommand{\theequation}{\thesection.\arabic{equation}}
\addcontentsline{toc}{section}{Appendices}
\section*{APPENDIX}

\section{The structure of axion models\label{structure}}
In this appendix we review the general structure of axion models.

An axion field may have several meanings in the literature, but for our purposes it has two properties: it has a perturbative shift symmetry and couples linearly to an instanton density (here the QCD density) of the Standard Model (SM).

Such a field was first introduced by Peccei and Quinn to solve the strong CP-problem and render the $\theta$-angle of the strong interactions a dynamical variable.

There are several realizations of axions in QFT, with two extreme classes: the fundamental axions, \cite{Kim}-\cite{GKR}, and the emergent or composite axions, \cite{axions}. We will produce here a typical simple model of a fundamental axions. A general prescription of theories of composite axions and further references can be found in \cite{axions}.

\subsection{A typical UV model for elementary axions}

A simple class of UV complete examples of axions can be constructed as follows.
A new sector that realizes two global $U(1)$ symmetries is used and has two scalar field transforming under them. An anomaly free combination of the two $U(1)$s is gauged.
The scalars obtain a vev so that they break the $U(1)$ gauge symmetry but leave the orthogonal global $U(1)$ symmetry  unbroken. This will play the role of the PQ symmetry when coupled to the SM.

We therefore  consider a set of new chiral fermions $\psi_i$ (we take them to be all left-handed) that are charged under two $U(1)$ symmetries, $U(1)_1$ and $U(1)_2$ with charges $(Q^1_i,Q^2_i)$
and two Higgs-like scalars, $H_1$, $H_2$ with charges ($q^1_1,q^2_1$) and $(q_2^1,q_2^2)$.

We gauge $U(1)_1$ with a corresponding gauge boson $A_{\m}$ and we assume that the full spectrum is anomaly free.

This implies that
\be
\sum_i Q^1_i=\sum_i (Q^1_i)^3=0\,,
\label{a1}\ee
that guarantees the absence of the gauge anomaly and mixed anomaly in four dimensions.

The bosonic part of the action is\footnote{Strictly speaking the simple model we describe is not UV complete as the $U(1)$ couplings are marginally irrelevant in the UV. However, the gauge theory  can be embedded in a non-Abelian group, \cite{Kim}-\cite{GKR}.}
 \be
 S_{UV}=-\int d^4x\left[{1\over 4g^2}F_A^2+{1\over 2}|DH_1|^2+{1\over 2}|DH_2|^2+V(H_1,H_2)\right]\,.
\label{a2}\ee
We can obtain symmetry breaking where both $H_1$ and $H_2$ condense with the same vacuum expectation value for simplicity
\be
H_1=v~e^{i\phi_1}\sp H_2=v~e^{i\phi_2}\,.
\label{a3}\ee
We will assume that the remaining scalar degrees of freedom, beyond the phases, have high masses and we will neglect them in the low energy dynamics.

The bosonic part of the action now becomes
\be
 S_{UV}\simeq \int -{1\over 4g^2}F_A^2-{v^2\over 2}(\pa\phi_1+q^1_1A)^2 -{v^2\over 2}(\pa\phi_2+q^1_2A)^2+\cdots
\label{a4}\ee
Changing basis to
\be
\phi={q^1_1\phi_1+q^1_2\phi_2\over (q^1_1)^2+(q^1_2)^2}\sp \tilde\sigma=v{q^1_2\phi_1-q^1_1\phi_2\over \sqrt{(q^1_1)^2+(q^1_2)^2}}\,,
\label{a5}\ee
\be
\phi_1=q_1^1\phi+q^1_2{\tilde\sigma\over v\sqrt{(q^1_1)^2+(q^1_2)^2}}
\sp
\phi_2=q_1^2\phi-q^1_1{\tilde\sigma\over v\sqrt{(q^1_1)^2+(q^1_2)^2}}\,,
\label{a6}\ee
we may rewrite the action as
\be
 S_{UV}\simeq \int -{1\over 4g^2}F_A^2-{m_A^2\over 2}(A+\pa \phi)^2 -{1\over 2}(\pa\tilde \sigma)^2+\cdots
\label{a7}\ee
where
\be
m_A^2=g^2f_a^2\sp f_a=v\sqrt{(q^1_1)^2+(q^1_2)^2}\,.
\label{a8}\ee
The field $\tilde\s$ will become the axion in the low energy theory and as we defined it, it has mass-dimension 1. We may normalize it as
\be
\s={\tilde \s\over f_a}={q^1_2\phi_1-q^1_1\phi_2\over (q^1_1)^2+(q^1_2)^2}\,.
\label{a9}\ee
Consider now the Yukawa couplings of the fermions
\be
S_{Y_1}=\int H_1\sum_{i,j} g^{i,j}_1\bar \psi_i\psi_j+H_2\sum_{i,j}  g^{i,j}_2\bar \psi_i\psi_j+cc
\label{a10}\ee
with the selection rules
\be
g_I^{i,j}=0~~~{\rm unless}~~~~ q_I^1+Q^1_i-Q^1_j=0\,.
\label{a11}\ee
 We would like to have $v\gg M_{EW}$ where $M_{EW}$ is a typical EW scale.
 For finite and small couplings $g, g_I^{ij}$, it is clear that we can take $m_A\gg M_{EW}$, and that we can arrange the charges so that  all the fermions $\psi_i$ acquire  masses that are also well above the EW scale.

The field $\phi$ is the longitudinal component of the very massive $U(1)$ gauge boson, and can be set to zero by a gauge transformation. The only particle from this sector that remains in the IR theory (the SM) is the massless PQ scalar $\s$.

The scalar $\s$ will have direct couplings to the SM fermions (quarks and leptons) if they are charged under the two UV $U(1)$ symmetries and if the high energy Higgs $H_{1,2}$ couple to them. The remaining global symmetry shifts $\s$ by constants.

We have also assumed that the gauged $U(1)$ symmetry is anomaly free.\footnote{It may however have mixed anomalies with the SM hypercharge. This is generic in string theory realizations, \cite{anom}.}
However the remaining global $U(1)$ symmetry associated with $\s$  may have mixed anomalies with the standard model.
We  denote the relevant charge traces as
\be
T_3=Tr[Q_2 T^a_{SU(3)} T^a_{SU(3)}]\sp T_2=Tr[Q_2 T^a_{SU(2)} T^a_{SU(2)}]
\sp T_1=Tr[Q_2 T^a_{U(1)_Y} T^a_{U(1)_Y}]\,,
\label{a16}\ee
where $Q_2$ is the charge generator of $U(1)_2$, $T^a$ are the gauge generators of the SM and the trace is on the SM fermions.
If such traces are non-zero, then in the effective field theory we have the following axionic couplings
\be
S_{\rm anomaly}=\int d^4 x\left[-{f_a^2\over 2}\pa_{\m}\s\pa^{\m}\s-q^1_1\sigma\left(T_3\a_{s}Tr[G\wedge G]+T_1\a_{em}F\wedge F\right)\right]\,,
\label{a17}\ee
where $G$ is the $SU(3)$ Yang-Mills field strength and $F$ is the $U(1)$ electromagnetic one. We have ignored the coupling to the W-instanton density, as its effects are heavily suppressed.

Therefore we can parametrize the effective couplings of $\s$  to the SM matter as in \cite{villadoro}
\be
S_{\rm axion}=\int d^4 x\left[-{f_a^2\over 2}\pa_{\m}\s\pa^{\m}\s-\frac{\s}{8\pi}\left(c_s\a_{s}Tr[G\wedge G]+c_{em}\a_{em}F\wedge F\right)\right.
\label{a18}\ee
$$
\left.\phantom{f^2\over 2}+\pa_{\m} \sigma~J^{\m}+V_{\rm mass-terms}(\s)+\cdots\right]\,.
$$
Here $\s$ plays the role of the PQ axion for the SM.
$V_{\rm mass-terms}$ is a non-derivative interaction coupling the axions to mass terms of fermions and bosons.
 The ellipsis in (\ref{a18}) stands for higher order and higher derivative interactions that have been omitted.

 We have defined the axion to be dimensionless, and its effective interaction strength is set by the axion scale $f_a$ that is assumed to be much larger than the SM scales. Therefore all axion interactions are weak. This is the reason why higher non-linear interactions are not so important in the effective action.

$S_{\rm axion}$ does not contain quantum corrections associated with the SM.
Due to the protected nature of the axion such corrections are very special. At weak coupling they arise because of YM instantons, that will generate a potential for the axion, \cite{villadoro}.
This is the analogue of the $U(1)_A$ anomaly for the PQ symmetry.

The couplings of the axion to leptons and or photons are straight-forward as they can be treated in perturbation theory.
However, the couplings to the strong interaction operators are trickier.

With a chiral rotation of quarks,  the coupling to the instanton density can be
transferred to the mass matrix of the quarks.
Therefore, an invariant collection of couplings involve the quark masses and the derivative coupling to $J^{\m}$. The quark component of $J^{\m}$ generates out of the QCD vacuum various vector mesons. Their content is axion-model dependent.
The axion dependence of the quark couplings generates axion couplings to scalar mesons. Their content is also axion-model dependent.

There is however a subset of model independent couplings, if we want the axion to solve the strong CP problem. Namely $c_{s}\not=0$ and then there is a universal coupling of the axion to all the quark states, that amounts to a coupling to the so-called singlet $\eta'$ meson.
It is this class of couplings that we address in this work.

\subsection{Other axion models}

Generically, the origin of the axion coupled to the SM may vary. In the fundamental axion case, like the simple model described in the previous subsection, the axion involves a high energy degree of freedom protected by the PQ symmetry.

The effective Lagrangian of the axion coupling it at low energies to the SM degrees of freedom is the same as in (\ref{a18}) as it is dictated by (anomalous) symmetry alone. Therefore different fundamental theories yielding axions will have a similar effective action, albeit with different charge assignments and coefficients, \cite{Sre,Kaplan}.

The situation is somewhat different when the axion is composite/emergent.
In such cases the axion is a composite generated by the instanton density.\footnote{Due to anomalies an analogue of the $\eta'$ will be as good. In general,  at finite $N_c$ and $N_f$ these two mix strongly, \cite{theta}. At weak mixing it is usually the (mostly) $\eta'$ that is lighter in QCD-like theories but in more general theories it could be otherwise. Which of the two is lighter depends    of a hidden QFT coupled at high energies to the SM.}
The general quadratic effective action of the emergent axion was derived in \cite{axions}. In that case there is also a mass contribution to the axion effective action originating in the UV theory. In a QCD-like UV theory this is controlled by the topological susceptibility of the hidden sector. The axion in that case has a mass with two contributions that are a priori independent: one from the hidden sector and the standard one from QCD.

\section{Current terms and $\eta'$-axion mixing in chiral Lagrangians}\label{app:chpt}

In this Appendix we address the mixing between the axion and mesons in a somewhat more general case than that described in  section ~\ref{CHPTL}.

In general, the derivative of the axion couples to a current $J^{\m}$  involving contributions from leptons, vectorial quark current and the axial quark current as $J^{\m} = J^{\m}_\ell + J^\m_{q} +  J^\m_{q5}$. We write the axial current already appearing in~\eqref{LQCDaxion} as
\be \label{Jq5}
J^{\m}_{q5} = \bar q \hat Q \gamma^\mu \gamma_5  q\,,
\ee
with $\hat Q$ a Hermitean matrix in flavor space. The vectorial current is conserved and therefore couplings to such current play no role in our analysis:  we can neglect them without loss of generality.

It is however simpler to start with the action where the derivative coupling to the current $J^{\m}_{q5}$ is absent (i.e., the case already considered in Sec.~\ref{CHPTL}). Therefore we first perform a (flavor dependent) axion valued chiral rotation
\be \label{Qhatrot}
 q \mapsto \exp(i\gamma_5 a \hat Q/f) q\,,
\ee
which removes the derivative coupling. It will be reinstated using a chiral rotation in the effective theory below.

After this transformation, the
Lagrangian obtained from \eqref{LQCDaxion} is without the current term
\begin{align} \label{couplings}
  S_\mathrm{a+QCD} = \int d^4 x ~\bigg(&-\frac{1}{2}\partial_\m a \partial^\m a-\frac{a}{f_a}\frac{\hat c_s\a_{s}}{8\pi}\mathrm{Tr}[G\wedge G]- \frac{1}{2}\mathrm{Tr}[G^2]  &\nonumber\\
  &- i \bar q \gamma^\mu D_\mu- \bar q_R \hat M(a) q_L - \bar q_L \hat M(a)^\dagger q_R \bigg)\,,
  \end{align}
but with the effect of the current included via $ \hat c_s = c_s -2\mathrm{Tr}[\hat Q]$ and $\hat M(a) = e^{-ia \hat Q/f_a} M(a) e^{-ia \hat Q/f_a}$.

The effective field theory Lagrangian  corresponding to~\eqref{couplings} is the one in ~\eqref{chiralL1} of section ~\ref{CHPTL}, but with the understanding that  $c_s$ and $M(a)$ are replaced by $\hat c_s$ and $\hat M(a)$, respectively. It can  be generalized to include explicitly the effect of the derivative coupling of the axion to the current $J^{\m}_{q5}$ as follows.
We perform the inverse of~\eqref{Qhatrot}, which in the chiral effective theory means
\be
 U \mapsto e^{ia \hat Q/f_a} U e^{ia \hat Q/f_a}\, , \qquad U^\dagger \mapsto e^{-ia \hat Q/f_a} U^\dagger e^{-ia \hat Q/f_a}\, .
\ee
After this transformation, the final form of the chiral Lagrangian, corresponding to the QCD Lagrangian of~\eqref{LQCDaxion} with the current term included, is 
\begin{align} \label{chiralkinetic}
{\cal L}_{\rm chiral} &=-\frac{1}{2}\partial_\m a \partial^\m a-
{f_{\pi}^2\over 4}\mathrm{Tr}\left[\left(\pa_{\m}U+i \frac{\pa_\m a}{f_a}\{U, \hat Q\}\right)\left(\pa^{\m}U^\dagger-i \frac{\pa^\m a}{f_a}\{U^\dagger, \hat Q\}\right)\right]
&\nonumber\\
&+{f_{\pi}^2B_0 \over 4}\mathrm{Tr}[M(a) U^\dagger+ M(a)^{\dagger}U]-\frac{\chi_\mathrm{YM}}{2}\,\left(\frac{ c_s\,a}{f_a}-i\log\det U\right)^2  & \\
& + c_{\eta'} \mathrm{Tr}\left[U^\dagger \left(\pa_{\m}U+i \frac{\pa_\m a}{f_a}\{U, \hat Q\}\right)\right]\mathrm{Tr}\left[U\left(\pa^{\m}U^\dagger-i \frac{\pa^\m a}{f_a}\{U^\dagger, \hat Q\}\right)\right] & \nonumber\\
& + \frac{ic_{a\eta'}}{N_f}\left(\frac{c_s\,a}{f_a}-i\log\det U\right) \mathrm{Tr}[  M(a)^{\dagger}U - M(a) U^\dagger]\,, & \nonumber
\end{align}
where the last term is suppressed by both the flavor masses $m_q$ and $1/N_c$~\cite{HerreraSiklody:1996pm,Kaiser:2000gs}.
We show below that its effect on the axion mixing can be absorbed into redefinitions of other parameters up to quadratic terms in $c_{a\eta'}$.
We also observe that the chiral rotation gives rise to kinetic mixing between the axion and the pions.
Comparing the derivatives of the QCD Lagrangian and the effective Lagrangian with respect to $M_{ij}$ at $a=0$ we obtain
\be
 \left.\langle \bar q^i_R q^j_L \rangle\right|_{a=0} = -\frac{f_\pi^2 B_0}{4} \delta^{ij}
 \,.
\label{6}\ee

We then discuss how the mass eigenstates  are computed to leading nontrivial order in $1/f_a$.
First we make an observation about the kinetic mixing. We can apply a shift
\be
 a \mapsto a - \sqrt{2} f_{\eta'} \mathrm{Tr}[\hat Q] \eta'/\sqrt{N_f}-  f_\pi \mathrm{Tr}[\hat Q \Pi]/f_a\qquad \text{with} \qquad \Pi =
\pi^aT^a \,,
\ee
so that the change of the axion kinetic term exactly cancels the kinetic mixing and the remaining terms in \eqref{chiralkinetic} are unchanged up to corrections suppressed by $1/f_a$. The outcome is that at leading nontrivial order in $1/f_a$ the only effect due to the $\mathcal{O}(f_a^{-1})$ current terms, is the mixing of the physical axion with the pions implied by the above shift. Because the non-derivative couplings involving the axion in~\eqref{chiralkinetic} are already suppressed by $1/f_a$, the non-derivative mixing terms introduced by this shift are $\morder{1/f_a^2}$, i.e., negligibly small. Therefore the effect of the kinetic mixing will be irrelevant for the analysis of the axion-nucleon couplings in Sec.~\ref{sec:axionnucleon}. Higher order terms in the effective action also include a derivative coupling between the axion and $\eta'$ which is present even in the absence of the current~\eqref{Jq5} but suppressed by $1/N_c$~\cite{Kaiser:2000gs}. The same applies to such a coupling as the kinetic mixing in~\eqref{chiralkinetic}: its effect is irrelevant for our studies.

We then discuss the diagonalization of the quadratic terms in~\eqref{chiralkinetic}. We assume that $M(a) = e^{ia \tilde Q/f_a} M_0\, e^{ia \tilde Q/f_a}$ where $M_0$ is independent from the axion and Hermitean. We consider the effects due to a non-Hermitean matrix below. In order to eliminate the $\morder{1/f_a}$ non-derivative mixing terms with the axion (up to higher order corrections in $1/f_a$), we first carry out a field redefinition
\be
  \frac{\sqrt{2}\hat \eta'}{\sqrt{N_f}f_{\eta'}} +\frac{\hat \pi^b T^b}{f_\pi} = \frac{\sqrt{2}\eta'}{\sqrt{N_f}f_{\eta'}} +\frac{\pi^b T^b}{f_\pi} -2 \tilde Q \frac{a}{f_a} + k_1 M_0^{-1} \frac{a}{f_a} + k_2 \frac{a}{f_a} \ .
\ee
Inserting this in the action~\eqref{chiralkinetic} and requiring that the non-diagonal mass terms involving the axion vanish, fixes
\begin{align}
 k_1 &= \frac{2 \tilde c_s \left(B_0 f_\pi^2 \chi_\mathrm{YM}-8 c_{a\eta'}^2N_f^{-2}\mathrm{Tr}\left[M_0\right]\right)}{B_0 f_\pi^2 \left(B_0 f_\pi^2+2 \mathrm{Tr}\left[M_0^{-1}\right]\! \chi_\mathrm{YM}\right)+8 c_{a\eta'} B_0 f_\pi^2+16 c_{a\eta'}^2 \left(1-N_f^{-2}\mathrm{Tr}\left[M_0\right] \mathrm{Tr}\left[M_0^{-1}\right]\right)}\,, &\\
 k_2& = \frac{4 c_{a\eta'} \tilde c_s \left(B_0 f_\pi^2+4 c_{a\eta'}\right)/N_f}{B_0 f_\pi^2 \left(B_0 f_\pi^2+2 \mathrm{Tr}\left[M_0^{-1}\right]\!  \chi_\mathrm{YM}\right)+8 c_{a\eta'}B_0  f_\pi^2+16 c_{a\eta'}^2 \left(1-N_f^{-2}\mathrm{Tr}\left[M_0\right] \mathrm{Tr}\left[M_0^{-1}\right]\right)}\,. &
\end{align}
The nondiagonal kinetic terms are then removed by a subsequent redefinition
\begin{align}
  a_p &= a + \frac{ f_\pi}{f_a} \mathrm{Tr}\left[ \left(\hat Q+\tilde Q\right)
  \Pi\right]+\frac{\sqrt{2} f_{\eta'}}{f_a\sqrt{N_f}} \mathrm{Tr}\left[\hat Q+
  \tilde Q\right]\eta'\nonumber\\
  & \quad  \ \ \,
  - \frac{k_1f_\pi}{2f_a}\mathrm{Tr}\left[M_0^{-1} \Pi\right]-\frac{k_1f_{\eta'}}{\sqrt{2N_f}f_a}\mathrm{Tr}\left[M_0^{-1}\right]\eta' -\frac{k_2f_{\eta'}}{f_a} \sqrt{\frac{N_f}{2}}\eta' \ .&
\end{align}
If $c_{a\eta'}$ is treated as a small correction and quadratic terms in this parameter are neglected, we may write the result for the coefficients $k_i$ in a compact way in terms of the topological susceptibility (which now includes  a $1/N_c$ correction):
\be \label{kiapprox}
 k_1 \approx \frac{2 \tilde c_s \chi}{B_0f_\pi^2} \ , \qquad k_2 \approx \frac{4 c_{a\eta'}\tilde c_s\chi}{B_0f_\pi^2\chi_\mathrm{YM}}\,,
\ee
with
\be
\frac{1}{\chi} = \frac{1}{\chi_\mathrm{YM}}+\frac{2}{B_0f_\pi^2}\mathrm{Tr}\,\left[M_0^{-1}\right]+\frac{8c_{a\eta'}}{B_0f_\pi^2\chi_\mathrm{YM}} \ .
\ee
Within this approximation, the rules~\eqref{transform1}--\eqref{transform2} therefore generalize to
\begin{align}
 \label{transformgen1}
 \hat \eta' &= \eta' - \frac{\sqrt{2} f_{\eta'}}{\sqrt{N_f}} \mathrm{Tr}[\tilde Q ]\, \frac{a}{f_a}+ \frac{ \sqrt{2} \tilde c_s\,\chi\,f_{\eta'}\,\mathrm{Tr} \left[ M_0^{-1}\right]}{B_0f_\pi^2\ \sqrt{N_f}} \frac{a}{f_a} +\frac{2 \sqrt{2N_f} c_{a\eta'}\tilde c_sf_{\eta'}\chi}{B_0f_\pi^2 \chi_\mathrm{YM}} \frac{a}{f_a} \ , \\
 \label{transformgen2}
 \hat \pi^b &= \pi^b -  f_\pi \mathrm{Tr}[\tilde Q T^b ]\, \frac{a}{f_a}+ \frac{\tilde c_s\,\chi\,\mathrm{Tr}\left[M_0^{-1} T^b\right] }{f_\pi B_0} \frac{a}{f_a}\,.
 \end{align}
Notice that the last (fourth) term in~\eqref{transformgen1} is suppressed with respect to the third term by the quark mass. The mass eigenstates are then found by diagonalizing the pion mass matrix as explained in Sec.~\ref{CHPTL}.
The axion mass squared is given by
\be
 m_a^2 = \frac{\tilde c_sB_0f_\pi^2 k_1}{2 f_a^2} \approx \chi \frac{\tilde c_s^2}{f_a^2}\ .
\ee

We then discuss the effect of $c_{a\eta'}$ for the axion-nucleon couplings studied in Sec.~\ref{sec:axionnucleon}. It is sufficient to include it in the expressions~\eqref{detUafactor} and~\eqref{chiminusresult} using the approximation in~\eqref{kiapprox}. For the former expression we find that
\be
  i \log \det U -\frac{c_s a}{f_a} = -\frac{\sqrt{2N_f}}{f_{\eta'}}\left(\hat\eta' + \frac{\tilde c_s f_{\eta'}}{\sqrt{2N_f}}\frac{\chi}{\chi_\mathrm{YM}}\left(1+\frac{12 c_{a\eta'}}{B_0f_\pi^2}\right)\frac{a_p}{f_a} +\morder{\frac{1}{f_a^2}}\right) \ .
\ee
Since $f_\pi^2 \sim N_c$ the correction due to $c_{a\eta'}$ is subleading in $1/N_c$.
For the latter expression we find
\be
\chi_-(a) = \left\{\frac{i\sqrt{2}\hat\eta'}{\sqrt{N_f}f_{\eta'}}+\frac{i\hat \pi^bT^b}{f_\pi},M_0\right\} - \frac{4 \tilde c_s i \chi}{f_\pi^2B_0}\left(1 + \frac{2 c_{a\eta'}}{\chi_\mathrm{YM}}M_0\right) \frac{a_p}{f_a} + \mathrm{nonlinear} \ .
\ee
In this case, the correction due to $c_{a\eta'}$ is suppressed by the quark masses. Since $\chi$ also contains a factor proportional to the quark masses, this correction is actually of quadratic order in the quark masses and can be safely ignored.

Notice that the last term in~\eqref{chiralkinetic}
also affects the pion mixing and masses even in the absence of the axion. For example, for a flavor independent quark mass $m_q$, the masses of the pions and the $\eta'$ are given by
\be
 m_\pi^2  = B_0 m_q \ , \qquad m_{\eta'}^2 = \frac{f_\pi^2B_0 m_q}{f_{\eta'}^2} +\frac{2 N_f \chi_\mathrm{YM}}{f_{\eta'}^2}+ \frac{8 c_{a\eta'}m_q}{f_{\eta'}^2} \ .
\ee

Finally, we recall that for $N_f=3$, the diagonalization of the pion mass matrix
boils down to the ``standard'' analysis in the literature (see, e.g.~\cite{Witten:1980sp,DiVecchia:1980yfw}). The $\eta'$ meson, the $\eta$ meson, the pions and the kaons form a nonet of the vectorial $U(3)$, which is broken to various subgroups both by $1/N_c$ effects and by finite quark masses. Taking the light quark masses to be the same, $m_u=m_d\equiv m_{ud}$, only the $\eta$ and $\eta'$ mesons mix. The relevant mass matrix is given as
\be
  \mathcal{L}_\mathrm{mass} = - \frac{1}{2}\left(\hat \pi^8 \ \hat \eta'\right)
  \left(\begin{array}{cc}
         \frac{m_{ud}+2m_s}{3} B_0 & \frac{\sqrt{2}f_\pi(m_{ud}-m_s)}{3f_{\eta'}}B_0 \\
\frac{\sqrt{2}f_\pi(m_{ud}-m_s)}{3f_{\eta'}}B_0 & \frac{f_\pi^2 (2m_{ud}+m_s)B_0+18 \chi_\mathrm{YM}}{3 f_{\eta'}^2}
         \end{array}\right)\left(\begin{array}{c}
                                  \hat \pi^8 \\ \hat \eta'
                                 \end{array}
\right) \ ,
\label{11}\ee
where we dropped the terms $\propto c_{a\eta'}$ for simplicity.
We adopt a convention where the physical states are given by
\begin{align}
 \hat \eta' &=  \eta'_p\,\cos \theta_{\eta\eta'} - \eta_p\,\sin \theta_{\eta\eta'}\ , & \nonumber\\
 \hat \pi^8 &= \eta_p\,\cos \theta_{\eta\eta'}  +\eta'_p\,\sin \theta_{\eta\eta'} \ . &
\label{12}\end{align}

\subsection{VEVs induced by CP-odd couplings} \label{app:vevs}

CP-odd couplings in the meson Lagrangian also give rise to VEVs of the pions and the $\eta'$ which need to be taken into account. First, the quark mass matrix may include CP-odd contributions: we write $M(a) = e^{ia \tilde Q/f_a} \left(M_0+i\overline M_0\right)\, e^{ia \tilde Q/f_a}$ where both $M_0$ and the CP-violating matrix  $\overline M_0$ are Hermitean. We treat all CP-odd couplings as linear perturbations, and also consider the possibility of finite $\theta$-angle (arising from the VEV of the axion) which is obtained by shifting $c_s a/f_a \mapsto c_s a/f_a + \theta$ in the Lagrangian.  The terms driving the VEVs are then
\begin{align}
 \mathcal{L}_\mathrm{CP-odd} = \ & i \frac{f_\pi^2 B_0}{4}\mathrm{Tr}\,\left[\overline M_0 U^\dagger - \overline M_0^\dagger U \right] + \bar c_{\eta'}  \chi_\mathrm{YM} \left(\frac{c_s a}{f_a} - i \log \det U \right) &\nonumber\\
 &-\frac{\chi_\mathrm{YM}}{2}\left(\theta + \frac{c_s a}{f_a} - i \log \det U\right)^2 \ . &
\end{align}
Notice that we chose the normalization of $\bar c_{\eta'}$ similarly as that of the $\theta$-angle: $\bar c_{\eta'}=\morder{N_c}$. This term could arise from the CP-violation in the CKM matrix, \cite{Khri,EG,GM} or BSM sources. In fact, one immediately sees that $\bar c_{\eta'}$ can be absorbed by a shift in the $\theta$-angle (up to highly suppressed contributions) so we can set it to zero without loss of generality.

We then find for the VEVs of the pions after a straightforward computation
\be\label{pionVEVs}
  \left\langle \frac{\sqrt{2}\eta'}{\sqrt{N_f}f_{\eta'}} +\frac{\pi^b T^b}{f_\pi} \right\rangle = \Delta - \frac{2\left(
  \theta+\mathrm{Tr}\,\left[\overline M_0 M_0^{-1}\right]\right)\chi}{f_\pi^2B_0} M_0^{-1} \ ,
\ee
where $\Delta$ is the solution to $\Delta M_0 + M_0 \Delta = 2 \overline M_0$ and satisfies therefore $\mathrm{Tr}[\Delta] = \mathrm{Tr}\,\left[\overline M_0 M_0^{-1}\right]$. If $M_0$ and $\overline{M}_0$ commute, $\Delta = \overline M_0 M_0^{-1}$. Notice that the VEV of the $\eta'$ is given by
\be
 \left\langle\eta'\right\rangle = \frac{f_{\eta'}}{\sqrt{2N_f}}\left(\mathrm{Tr}\,\left[\overline M_0 M_0^{-1}\right] - \frac{2\left(
 \theta+\mathrm{Tr}\,\left[\overline M_0 M_0^{-1}\right]\right)\chi}{f_\pi^2B_0}\, \mathrm{Tr} \,\left[M_0^{-1}\right]\right) \ .
\ee
Moreover, the VEV in the combination appearing in the flavor singlet terms after the inclusion of the $\theta$-angle
\be \label{detUVEV}
 \theta -i \left\langle \log\det U\right\rangle = \frac{\chi}{\chi_\mathrm{YM}}\bar \theta
\ee
only depends on the effective angle $\bar\theta = \theta +  \mathrm{Tr}\,\left[\overline M_0 M_0^{-1}\right]$. In order to analyze the couplings to nucleons, we also need the VEVs in $\mathrm{Tr}\,\left[ \chi_\pm\right]$. For consistency with chiral symmetry, in these couplings we also need to include the contributions from $\overline M_0$. We find that
\be \label{chiminusVEV}
 \left\langle\chi_-\right\rangle = \left\langle u \left(M_0 -i\overline M_0\right) u - u^\dagger \left(M_0 +i\overline M_0\right) u^\dagger \right\rangle  = - \frac{4i\bar\theta\chi}{f_\pi^2B_0} + \cdots \ .
\ee
We also consider the terms arising from $\mathrm{Tr}\,[\chi_+]$ which are linear in the mesons and the axion and are present due to the VEVs. First notice that
\begin{align}
 \mathrm{Tr}\,\left[\chi_+(a)\right] &= \mathrm{Tr}\,\left[u \left(M_0 -i\overline M_0\right) u + u^\dagger \left(M_0 +i\overline M_0\right) u^\dagger \right] & \nonumber\\
 &= 2\, \mathrm{Tr}\,\left[M_0\right] + 2\,\mathrm{Tr}\left[\left(\frac{\sqrt{2}\eta'}{\sqrt{N_f}f_{\eta'}}+\frac{\Pi}{f_\pi}-2\tilde Q
\frac{a}{f_a}\right)\overline M_0\right] \nonumber \\
&\phantom{=}\ - \mathrm{Tr}\left[\left(\frac{\sqrt{2}\eta'}{\sqrt{N_f}f_{\eta'}}+\frac{\Pi}{f_\pi}-2\tilde Q
\frac{a}{f_a}\right)^2M_0\right] + \cdots \ .
\end{align}
The linear contributions are given by the second term on the second line and the terms arising from the third line after inserting the VEVs. After some cancellations, we obtain
\be
 \mathrm{Tr}\,\left[\chi_+(a)\right] = 2\, \mathrm{Tr}\,\left[M_0\right] +\frac{4 \bar\theta\chi}{f_\pi^2B_0}\left(\frac{\sqrt{2N_f}}{f_{\eta'}}\hat \eta'- \frac{2 \tilde c_s\chi\,\mathrm{Tr}\,\left[M_0^{-1}\right]}{f_\pi^2B_0}\frac{a_p}{f_a}\right) + \mathrm{quadratic} \label{chiplusVEVterms}
\ee
Therefore, as expected, all contributions due to $\overline M_0$ can be absorbed in the $\bar \theta$-angle, so it is enough to consider the couplings of the mesons and the axion arising (only) from the VEVs due to the $\theta$-angle as we do in Sec.~\ref{sec:axionnucleon}.

\subsection{Three-meson couplings} \label{app:threepi}

Having extracted the VEVs of the mesons due to a finite $\theta$-angle, we can also extract the (CP-violating) couplings of (non-derivative) vertices with three mesons. Up to corrections suppressed at large $N_c$, these arise from the mass term
\be
 \frac{f_\pi^2B_0}{4} \, \mathrm{Tr}\,\left[M(a) U^\dagger + M(a)^\dagger U\right] \ .
\ee

When computing these couplings it is important to note that the mixing~\eqref{transform1} and in~\eqref{transform2} is (for generic quark mass matrices) only consistent with chiral symmetry if cubic and higher interactions are included. That is, we write the field redefinitions  of the pions and the $\eta'$ in general as
\be \label{nonlineartransf}
 \hat U  =  \exp\left(i\frac{\tilde c_s\chi}{B_0 f_\pi^2}\frac{a}{f_a}M_0^{-1}\right)\exp\left(-i \frac{a}{f_a}\widetilde Q\right)U\exp\left(-i \frac{a}{f_a}\widetilde Q\right) \exp\left(i\frac{\tilde c_s\chi}{B_0 f_\pi^2}\frac{a}{f_a}M_0^{-1}\right)\,,
\ee
where $\hat U$ was defined in~\eqref{9}. This definition differs from~\eqref{transform1} and~\eqref{transform2} if $M_0$, $\widetilde Q$ and $\Pi$ do not commute. Notice that the transformation~\eqref{nonlineartransf} corresponds to an axion-valued axial transformation of the pion fields, which ensures that the field redefinition does not generate undesired higher order interactions that would be inconsistent with chiral symmetry.

After inserting the mixing with the axion  from~\eqref{nonlineartransf} and introducing the $\theta$-angle through~\eqref{thetarepl}
we obtain the three pion couplings~\cite{crewther}:
\be \label{3pioncoupling}
 \mathcal{L}_{\pi\pi\pi} = -\frac{\theta\chi}{6 f_\pi^3}\,\mathrm{Tr}\left[\hat \Pi^3\right]\ ,
\ee
where $\hat \Pi = \hat \pi^a T^a$. The coupling of the $\hat\eta'$ to the pions is
\be \label{etaprimepipicoupling}
 \mathcal{L}_{\eta'\pi\pi} = -\frac{\theta\chi}{\sqrt{2N_f}f_\pi^2f_{\eta'}}\,\hat\eta'\,\mathrm{Tr}\left[\hat\Pi^2\right] \ .
\ee
In the limit of large $N_c$ and in the chiral limit with flavor independent quark masses
the $\eta'$-pion-pion coupling  simplifies to
\be \label{etaprimepipi}
{\cal L}
_{\eta'\pi\pi} = - \theta \frac{m_{\pi}^2}{\sqrt{2}N_f^{3/2}f_{\pi}} \hat \eta' \hat \pi^a\hat \pi^a \equiv \bar g_{\eta'\pi\pi}\, \hat\eta' \hat\pi^a\hat\pi^a\,,
\ee
where we inserted $\chi \approx m_\pi^2 f_\pi^2/2N_f$.
Notice that the expressions of these couplings are identical in the WSS model and in chiral perturbation theory at large $N_c$, so~\eqref{etaprimepipi} also holds in the WSS model.
The coupling between the axion and the pions is found to be
\be \label{apipicoupling}
 \mathcal{L}_{a\pi\pi} = \frac{\tilde c_s \theta \chi^2}{f_\pi^4 B_0} \frac{a_p}{f_a}\,\mathrm{Tr}\left[\hat\Pi^2M_0^{-1}\right]  \ .
\ee

\section{Details on the nucleon couplings in chiral perturbation theory}\label{app:baryoncouplings}

\subsection{Non-derivative couplings from external CP-violation}

In this Appendix we discuss the non-derivative  couplings of nucleons to the mesons and the axion which arise due to such sources of CP-violation that are not included in the $\theta$-angle,  such as the CKM matrix in the quark or lepton sector. For example, the mass terms~\eqref{massterms1} and~\eqref{massterms3} can be generalized to read
\begin{align}\label{appmassterms1}
 &\frac{c_{m1}N_c }{N_f}\mathrm{Tr}\left[\, \chi_+\,\right]\,\mathrm{Tr}\left[\, \bar B B\,\right] + \sum_\mathrm{contractions}\,N_c \, c_{m1}^{(k)}\,\mathrm{Tr}\left[\, \bar B \left(\chi_+-\frac{1}{N_f} \mathrm{Tr} \chi_+\right) B \,\right]&\\
 \label{appmassterms2}
  +\,&\frac{i\bar c_{m1}N_c}{N_f}\mathrm{Tr}\left[\, \chi_+\,\right]\,\mathrm{Tr}\left[\, \bar B\gamma_5 B\,\right] + \sum_\mathrm{contractions}\, iN_c\, \bar c_{m1}^{(k)}\,\mathrm{Tr}\left[\, \bar B \gamma_5\left(\chi_+-\frac{1}{N_f} \mathrm{Tr} \chi_+\right) B \,\right]& \\
  \label{appmassterms3}
  +\,&\frac{c_{m2}N_c}{N_f}\mathrm{Tr}\left[\, \chi_-\,\right]\,\mathrm{Tr}\left[\, \bar B\gamma_5 B\,\right] + \sum_\mathrm{contractions}\,  N_c\,c_{m2}^{(k)}\,\mathrm{Tr}\left[\, \bar B \gamma_5\left(\chi_--\frac{1}{N_f} \mathrm{Tr} \chi_-\right) B \,\right]& \\
  +\,&\frac{i\bar c_{m2}N_c}{N_f}\mathrm{Tr}\left[\, \chi_-\,\right]\,\mathrm{Tr}\left[\,\bar B B\,\right] + \sum_\mathrm{contractions}\, iN_c \, \bar c_{m2}^{(k)}\,\mathrm{Tr}\left[\, \bar B \left(\chi_--\frac{1}{N_f} \mathrm{Tr} \chi_-\right) B\,\right] \ .& \label{appmassterms4}
\end{align}
The coefficients with a bar are CP violating and in the absence of the $\theta$-angle should arise from non-QCD CP-violating sources. We only consider here CP-odd couplings which are singlets under the chiral transformations.\footnote{In Appendix~\ref{app:vevs} we discuss CP-odd contributions arising through the quark mass matrix.}

Similarly, an extra term should be included in the Lagrangian~\eqref{etaterms} containing flavor singlet couplings:
\begin{align}\label{etatermsapp}
 &d_1  \log \det U\,\mathrm{Tr}\left[\, \bar B\gamma_5 B\,\right] +  \bar d_1 \,  i \log \det U\,\mathrm{Tr}\left[\,  \bar B B\,\right]   & \nonumber \\
 + &\frac{d_2}{N_c} \left(i \log \det U\right)^2 \,\mathrm{Tr}\left[\,  \bar B B\,\right] - \frac{ d_3}{N_f} \log \det U\,\mathrm{Tr}\left[\,\chi_-\,\right]\,\mathrm{Tr}\left[\,  \bar B B\,\right] \ .
\end{align}
For the CP-violating term we assumed a scaling which is analogous to the CP conserving terms, as above.

We then consider the axion couplings induced by the axion terms in the QCD Lagrangian~\eqref{LQCDaxion}. 
The  important terms are the first term in \eqref{appmassterms4} and the second term in \eqref{etatermsapp}.We use notation defined in Sec.~\ref{CHPTL} and include the axion through the replacements~\eqref{axrepl1} and~\eqref{axrepl2}.

After this, the terms of interest become
\begin{align}
 & i \bar c_{m2} \frac{N_c}{N_f}\,  \mathrm{Tr}\left[\,\chi_-(a)\,\right]\, \mathrm{Tr}\left[\, \bar B B\,\right]
+ \bar d_1 \left(i \log \det U-\frac{c_s a}{f_a}\right)\, \mathrm{Tr}\left[\, \bar B B\,\right]  \,,&
\label{21}\end{align}
where $\chi_-(a) = uM(a)^\dagger u - u^\dagger M(a) u^\dagger$.
Inserting here the definition of $U$, and using the relations~\eqref{transform1}--\eqref{transform3}, we may extract the non-derivative couplings (i.e., the terms involving $\bar NN$) of the fields $\hat \eta '$ and $a_p$ to the nucleons.

The terms linear in $a$ and $\eta '$ are\footnote{Notice that when the quark masses break the vectorial $SU(N_f)$, the non-singlet terms in~\eqref{massterms3} also contribute to the coupling of $\hat \eta'$ (but not to the coupling of the axion). Therefore, for generic quark masses the result for $\hat \eta'$  here is only an estimate. When all quark masses are equal, it is the precise result for the physical field $\eta'_p = \hat \eta'$.  } then
\begin{align} \label{nonQCDnonderivative}
\mathcal{L}_{\eta'aB} &= - 2\bar c_{m2} \,\left(\frac{\sqrt{2}N_c\hat \eta'}{N_f^{3/2}f_{\eta'}}\, \mathrm{Tr}\left[M_0\right] -  \frac{2 \tilde c_sN_c \chi}{f_\pi^2B_0}\frac{a_p}{f_a}\right)\, \mathrm{Tr}\left[\, \bar B B \,\right]& \\
 &\ \ \ -\bar d_1\left(\frac{\sqrt{2 N_f}}{f_{\eta'}}\hat\eta' + \tilde c_s\frac{\chi}{\chi_\mathrm{YM}}\frac{a_p}{f_a}\right)\, \mathrm{Tr}\left[\, \bar B B  \,\right]\ . & \label{nonQCDnonderivative2}
\end{align}
Notice that since $f_\pi^2 \sim N_c$, both axion terms are of the same order in $1/N_c$. The $\eta'$ term in~\eqref{nonQCDnonderivative2} is suppressed by a factor of $N_f/N_c$ with respect to the term in~\eqref{nonQCDnonderivative} so it is negligible in the 't Hooft limit but should be included in the Veneziano limit.

\subsection{CP-odd nucleon-meson couplings} \label{app:chip}

Here we extend the discussion of the nucleon couplings to mesons in Sec.~\ref{sec:axionnucleon} to include in particular the contributions from the single trace terms in~\eqref{appmassterms1}. As it turns out, for this it is convenient to also redefine the nucleon fields. First we define the transition matrix $K_B$ through (see~\eqref{xitransform})
\be
 \exp\left(i\frac{\tilde c_s\chi}{B_0 f_\pi^2}\frac{a}{f_a}M_0^{-1}\right)\exp\left(-i \frac{a}{f_a}\widetilde Q\right) uK_B^\dagger = K_B u  \exp\left(-i \frac{a}{f_a}\widetilde Q\right) \exp\left(i\frac{\tilde c_s\chi}{B_0 f_\pi^2}\frac{a}{f_a}M_0^{-1}\right)
\ee
with the solution
\be
 K_B = 1 -\frac{1}{2 f_\pi}\frac{a}{f_a} \left[\Pi, \widetilde Q - \frac{\tilde c_s \chi}{B_0f_\pi^2}M_0^{-1}\right] + \cdots
\ee
Then the redefined baryon fields $\hat B$ are obtained by acting on $B$ with $K_B$ as in~\eqref{19}.

Only the couplings involving $\chi_+$ in~\eqref{appmassterms1} are affected by the baryon redefinition at the order we are interested in. The combination coupling to the redefined baryons is
\begin{align}
K_B\chi_+(a)K_B^\dagger &= \hat u M_0 \exp\left(-i\frac{2\tilde c_s \chi}{B_0f_\pi^2}\frac{a_p}{f_a}M_0^{-1}\right) \hat u +\hat u^\dagger M_0 \exp\left(i\frac{2\tilde c_s \chi}{B_0f_\pi^2}\frac{a_p}{f_a}M_0^{-1}\right) \hat u^\dagger & \nonumber\\
& = 2 M_0- \frac{1}{4} \!\left\{\!\frac{\sqrt{2}\hat\eta'}{\sqrt{N_f}f_{\eta'}}+\frac{\hat\pi^b T^b}{f_\pi},  \left\{\frac{\sqrt{2}\hat\eta'}{\sqrt{N_f}f_{\eta'}}+\frac{\hat\pi^b T^b}{f_\pi},M_0\right\} \!\! \right\}&\nonumber\\
&+\frac{4 \tilde c_s \chi}{B_0f_\pi^2}\frac{a_p}{f_a}\left(\frac{\sqrt{2}\hat\eta'}{\sqrt{N_f}f_{\eta'}}+\frac{\hat\pi^b T^b}{f_\pi}\right) -\frac{4 \tilde c_s^2 \chi^2}{B_0^2f_\pi^4}\frac{a_p^2}{f_a^2}M_0^{-1} + \cdots
\label{chiplusresult}
\end{align}
where we also noted that $a=a_p$ up to subleading corrections in $1/f_a$ which are irrelevant here. After introducing the $\theta$-angle through~\eqref{thetarepl}, we obtain the CP-odd couplings
\begin{align} \label{thetapinucleon}
 \mathcal{L}_{a\eta'B}^{(\theta,\mathrm{ad})} &= \sum_\mathrm{contractions}\!\!\frac{4 \chi \theta N_c  c_{m1}^{(k)}}{B_0f_\pi^3}\,\mathrm{Tr}\left[\, \hat {\bar B} \,\hat \Pi\, \hat B \,\right] & \\
 &\quad - \!\!\sum_\mathrm{contractions}\!\!\frac{8 \theta \tilde c_s \chi^2 N_c  c_{m1}^{(k)}}{B_0^2f_\pi^4}
 \frac{a_p}{f_a}\,\mathrm{Tr}\left[\, \hat {\bar B}\left(M_0^{-1}-\frac{1}{N_f}\mathrm{Tr}[M_0^{-1}]\right)\hat B \,\right]\ . & \label{thetaanucleonad}
\end{align}

\subsection{Group theoretic structure of the nucleon couplings to the mesons}\label{app:groupth}

We discuss here the structure of the single trace terms including sum over contractions, which appear in Eqs.~\eqref{derivativecouplings2},~\eqref{massterms1}, and~\eqref{massterms3}. We start by considering the structure at $N_f=2$ as an example.
The number of terms in the sums can be understood in the following way, which can be generalized to higher $N_f$. First, we classify the pions and the $\eta'$ according to the underlying quark representations\footnote{For $SU(2)$ conjugation acts trivially on the representations but we keep the conjugates since this will not be the case for the generalizations to higher $N_f$.} $1 \oplus 3 = 2 \otimes \bar 2$. All possible flavor structures of the $\eta'-\bar N-N$ and $\pi-\bar N-N$ couplings are then obtained by adding the nucleons, and consequently by identifying the singlets in the product state.  We write
\be
2 \otimes 2 \otimes \bar 2 \otimes \bar 2 = \left(2 \otimes 2\right) \otimes \left(\bar 2 \otimes \bar 2\right) = \left(1 \oplus 3\right) \otimes \left(\bar 1 \oplus \bar 3\right) \;.
\label{25} \ee
 When expanding the last expression, only products of a representation with its conjugate contain singlets (and exactly one each),
\be
1 \otimes \bar 1 = 1~~~~{\rm  and}~~~~ 3 \otimes \bar 3 = 1 \oplus 3 \oplus 5\;.
\label{26} \ee
 These two singlets map to the two terms in~\eqref{nf2eq1} and in~\eqref{nf2eq2}.

 We proceed by commenting on
the couplings at $N_f=3$. We also first
fix $N_c=3$ for simplicity. Then, in addition to the nonet of light mesons, the ground state of spin-$1/2$ baryons transforms under the adjoint representation of $SU(3)$. The usual way~\cite{Scherer:2002tk} to write the couplings corresponding to~\eqref{derivativecouplings} and to~\eqref{derivativecouplings2} is
\begin{align}
\label{dercouplingsnf31}
&\phantom{=} \frac{\hat g_a}{3} \mathrm{Tr}\left[ \,u_\m \,\right]\, \mathrm{Tr}\,\left[\bar B \gamma^\m \gamma_5 B\,\right]-\frac{D}{2}\mathrm{Tr}\,\left[\bar B \gamma^\m \gamma_5\left\{u_\m, B\right\}\,\right]-\frac{F}{2}\mathrm{Tr}\,\left[\bar B \gamma^\m \gamma_5\left[u_\m, B\right]\,\right] & \\
&= \frac{ g_a}{3} \mathrm{Tr}\left[ \,u_\m \,\right]\, \mathrm{Tr}\,\left[\bar B \gamma^\m \gamma_5 B\,\right]-\frac{D}{2}\mathrm{Tr}\,\left[\bar B \gamma^\m \gamma_5\left\{u_\m-\frac{1}{3}\mathrm{Tr}\, u_\m, B\right\}\,\right] & \nonumber \\
\label{dercouplingsnf32}
&\phantom{=} \, -\frac{F}{2}\mathrm{Tr}\,\left[\bar B \gamma^\m \gamma_5\left[u_\m, B\right]\,\right] \ , &
\end{align}
where $g_a = - D/2 + \hat g_a$ and with the baryon octet included in the traceless $3 \times 3$ matrix transforming as $B \to K B K^\dagger$. We renamed the couplings $g_a^{(1)}$ and  $g_a^{(2)}$ as $D$ and $F$ following standard conventions.

In order to analyze the number of terms, we may again pack the mesons into $3 \otimes \bar 3$ and consider the product of the nucleon and quark representations: $8 \otimes 3 = 3 \oplus 6 \oplus 15$. When multiplied with the similar antiquark representations each of the representations $3$, $6$, and $15$ again gives rise to a singlet. Therefore we obtain three singlets which map to the three terms of~\eqref{dercouplingsnf31} and~\eqref{dercouplingsnf32}. The same argument actually works for the ground state spin-$1/2$ baryons for any $N_f$ and any odd $N_c$. In this case the baryons belong to a representation defined through a Young diagram having exactly two rows, one with $N_c/2+1/2$ boxes and the other with $N_c/2-1/2$ boxes~\cite{Dashen:1993as,Dashen:1994qi}. For excited states of baryons in generic representations one obtains $\morder{N_c}$ independent terms.

\section{On-shell nucleon vertices in the limit of large mass} \label{app:spinors}

We first consider the bilinears of Dirac spinors
in the limit of large nucleon mass $m_N$.
For leading contributions in this limit we expect that the exchanged three-momenta are $\sim m_N^0$ and the variations in the nucleon energies are $\sim 1/m_N$.
In general the Dirac propagator can be decomposed as
\be
 \frac{\slash\!\!\! p +m_N}{p^2-m_N^2+i\epsilon} = \frac{\sum_s u(\mathbf p,s)\bar u(\mathbf p,s)}{2 E_\mathbf p(p_0-E_\mathbf p +i\epsilon)} +\frac{\sum_s v(-\mathbf p,s)\bar v(-\mathbf p,s)}{2 E_\mathbf p(p_0 +E_\mathbf p -i\epsilon)}\,.
\ee
The second term corresponds to a nucleon propagating backward in time and is suppressed in the limit of large $m_N$ (assuming that we are discussing a nucleon state rather than an anti-nucleon).

This suggests that up to corrections suppressed by $1/m_N$ the nucleon-meson interactions are determined by the (on-shell) amplitudes $\bar u(\mathbf p_2,s_2)\Gamma u(\mathbf p_1,s_1)$ where $\Gamma = \mathbb{I}$, $\gamma_5$, $\gamma^\mu$, and $\gamma^\mu\gamma_5$.
We take the incoming nucleon to be in the rest frame with momentum $p_1=(m_N,\mathbf{0})$ and write the outgoing nucleon momentum as $p_2 = (m_N,\mathbf{k})$ where we dropped terms $\sim 1/m_N$ in the energies. We use a representation where the spinors are given explicitly by
\be
 u(\mathbf p,s) = \sqrt{m_N+E_\mathbf{p}} \left(\begin{array}{c}
                                               \chi(s) \\
                                                \frac{\sigma \cdot \mathbf{p}}{m_N+E_\mathbf{p}} \chi(s)
                                              \end{array}
\right) \ , \qquad v(\mathbf p,s) = \sqrt{m_N+E_\mathbf{p}} \left(\begin{array}{c}
                                               \frac{\sigma \cdot \mathbf{p}}{m_N+E_\mathbf{p}} \chi(s) \\
                                                \chi(s)
                                              \end{array}\right)\,,
\ee
where a possible basis for the spin wave functions $\chi(s)$ is given by the spinors $(1,0)$ and $(0,1)$, but we leave the basis unspecified.
By direct computation we find that
\bea
&& \bar u(\mathbf p_2,s_2) u(\mathbf p_1,s_1) = \sqrt{2m_N(m_N+E_\mathbf{k})} \chi_2^\dagger \chi_1 = 2 m_N \chi_2^\dagger \chi_1 \left( 1 + \frac{\mathbf{k}^2}{8 m_N} + \cdots \right) \,,\\
&& \bar u(\mathbf p_2,s_2) \gamma_5 u(\mathbf p_1,s_1) = \sqrt{\frac{2m_N}{m_N+E_\mathbf{k}}} \mathbf{k} \cdot \chi_2^\dagger \mathbf{\sigma} \chi_1  = \mathbf{k} \cdot \chi_2^\dagger \mathbf{\sigma} \chi_1 \left( 1 - \frac{\mathbf{k}^2}{8 m_N} + \cdots \right)\,,  \\
&& \bar u(\mathbf p_2,s_2) \gamma^\mu u(\mathbf p_1,s_1) = \left(\sqrt{2m_N(m_N+E_\mathbf{k})} \chi_2^\dagger \chi_1,\sqrt{\frac{2m_N}{m_N+E_\mathbf{k}}} ( \chi_2^\dagger \chi_1 \mathbf{k} - i \mathbf{k} \times   \chi_2^\dagger \mathbf{\sigma} \chi_1 )\right) \nonumber \\
 && \quad = \left( 2m_N \chi_2^\dagger \chi_1\left( 1 + \frac{\mathbf{k}^2}{8 m_N} + \cdots \right), ( \chi_2^\dagger \chi_1 \mathbf{k} - i \mathbf{k} \times   \chi_2^\dagger \mathbf{\sigma} \chi_1 )\left( 1 - \frac{\mathbf{k}^2}{8 m_N} + \cdots \right)\right)\,,  \\
&& \bar u(\mathbf p_2,s_2) \gamma^\mu \gamma_5 u(\mathbf p_1,s_1) = \left(\sqrt{\frac{2m_N}{m_N+E_\mathbf{k}}}\mathbf{k} \cdot \chi_2^\dagger \mathbf{\sigma} \chi_1 , \sqrt{2m_N(m_N+E_\mathbf{k})} \chi_2^\dagger \mathbf{\sigma} \chi_1\right) \nonumber \\
 && \quad =  \left( \mathbf{k} \cdot \chi_2^\dagger \mathbf{\sigma} \chi_1 \left( 1 - \frac{\mathbf{k}^2}{8 m_N} + \cdots \right) ,2 m_N \chi_2^\dagger \mathbf{\sigma} \chi_1 \left( 1 + \frac{\mathbf{k}^2}{8 m_N} + \cdots \right)\right)\,,
 \label{avamp}
\eea
where $\chi_i = \chi(s_i)$.
We notice the following identities
\bea
 (p_2-p_1)^\mu \bar u(\mathbf p_2,s_2) \gamma_\mu u(\mathbf p_1,s_1) &=& 0 \\
 (p_2-p_1)^\mu \bar u(\mathbf p_2,s_2) \gamma_\mu \gamma_5 u(\mathbf p_1,s_1) &=& - 2m_N \bar u(\mathbf p_2,s_2) \gamma_5 u(\mathbf p_1,s_1) \\
 (p_2+p_1)^\mu \bar u(\mathbf p_2,s_2) \gamma_\mu u(\mathbf p_1,s_1) &=& -2 m_N \bar u(\mathbf p_2,s_2)  u(\mathbf p_1,s_1) \\
 (p_2+p_1)^\mu \bar u(\mathbf p_2,s_2)\gamma_\mu \gamma_5  u(\mathbf p_1,s_1) &=& 0
 \label{lastid}
\eea
which hold for any value of $m_N$. Notice also that because $v(\mathbf{p},s) = \gamma_5 u(\mathbf{p},s)$ the nucleon creation and annihilation amplitudes immediately follow from the above results.

These identities suggest that the operators $\partial_\mu\eta' \bar N \gamma^\mu\gamma_5 N \propto \partial_\mu\eta' \bar N S^\mu  N$, and $2 m_N\eta' \bar N \gamma_5 N$, where $S^\mu$ is the nucleon spin operator, are identical at least to leading order in the limit of large $m_N$. The spin $S^\mu$ is fixed at least to leading order  as the amplitudes do not flip spin in a basis aligned with $\mathbf k$: the transverse components of the amplitude~\eqref{avamp} contain also potentially spin-flipping elements but these vanish when contracted with the only available momentum $\mathbf k$.

We also notice that at $\morder{1/m_N}$ the couplings involving $\gamma_5$ above are $\propto \mathbf k$. This seems to be due to parity: when $\bar \theta = 0$, the only available parity-odd scalar is $\mathbf k \cdot \mathbf{S}$.\footnote{There are two independent momenta, but invariance of the amplitudes under $\morder{1/m_N}$ boosts requires that at linear order only momentum differences (i.e. $\mathbf k$) appear. Notice also that $\mathbf S$ is only affected by $\morder{1/m_N}$ corrections under such boosts.} Notice however that e.g. the nucleon pair creation amplitude $\propto \bar u(\mathbf p_2,s_2) \gamma_5 v(\mathbf p_1,s_1)$ is not proportional to $\mathbf k \cdot \mathbf{S}$.
\section{CP-odd couplings in the Skyrme model with $N_f=3$}
\label{nf3skyrme}
In this section we will provide a derivation of the CP-odd couplings of the axion and of the $\eta, \eta'$ mesons to the nucleons, in the Skyrme model with $N_f=3$ non-degenerate flavors.
We use the notation in \cite{dixon}, where the Lagrangian, once the substitution $\theta \rightarrow a/f_a$ is performed, is already in a diagonal basis for the axion.
Hence, for the rest of the discussion we drop the subscript ``p" of the fields for notational convenience.
Moreover, in this section we work in the ``chiral limit'' $m_{\pi} \ll m_{WV}, m_{K}$ ($m_{WV}$ is defined in (\ref{masseshere}) and $m_K$ is the kaon mass).

Actually, the Lagrangian in \cite{dixon} is written at linear order in $\theta$.
Generalizing that expression to the non-linear case, the mass term reads
\be\label{lagrS3}
{\cal L}_M = \frac{f_{\pi}^2}{4}{\rm Tr}\left[\tilde M (U- \unit) + h.c.\right]\,,
\ee
where
\be
\tilde M = {\rm diag}\left(M_1 e^{i\frac{\lambda_{\theta}}{M_1}},M_1 e^{i\frac{\lambda_{\theta}}{M_1}}, M_3 e^{i \tilde\gamma \frac{\lambda_{\theta}}{M_3}} \right)\,.
\label{47}\ee
In this expression $M_1 = m_{\pi}^2, M_3 = 2m_{K}^2-m_{\pi}^2$ and $\tilde\gamma$ (called $\alpha$ in \cite{dixon}) is a certain combination of the masses and the topological susceptibility which will be irrelevant in the ``chiral limit''.
Finally $\lambda_{\theta}$, proportional to $\theta$, will be given in a few lines.
The leading effect of $SU(2)$ isospin breaking will be introduced later on.

The Skyrmion is just the embedding of the $SU(2)$ one into $SU(3)$, so the configuration we need is\footnote{In this expression we have dropped a factor Exp$[-\frac{i}{f_{\pi}}\sqrt{\frac23} \delta]$, where $\delta$ is connected to the vacuum value of the singlet, because it gives a subleading correction in $m_{\pi}^2/m_{WV}^2, m_{\pi}^2/m_{K}^2$ to our result.
}
\be
U = U_s = e^{i \vec\lambda \cdot \hat x F(r)}\,,
\label{48}\ee
where with $\vec \lambda$ we denote the vector composed by the first three Gell-Mann matrices.
The function $F$ is the same as for the $N_f=2$ model reported in section \ref{sectionthecouplings}.
One can calculate the quark contribution to the nucleon mass by setting $\lambda_{\theta}=0$, which corresponds to setting to zero the axion field, and putting (\ref{lagrS3}) on-shell on the Skyrmion solution.
The result is
\be\label{deltaM3}
\delta M_N = \frac{ \alpha\pi (m_{\pi}^2+2m_{K}^2)}{6 e^3 f_{\pi}}\,,
\ee
where $\alpha$ is given in (\ref{alpha}). Note that in this formula we have not taken the chiral limit. In the degenerate mass case, eq.~(\ref{deltaM3}) exactly reduces to the $N_f=2$ expression (\ref{37}).

In order to extract the axion couplings we expand (\ref{lagrS3}), on-shell on the Skyrmion, up to second order in $\lambda_{\theta}$.
In the ``chiral limit'', the result turns out to be very simple
\be\label{interme}
{\cal L}_M = \frac{\lambda_{\theta}^2 f_{\pi}^2}{2 m_{\pi}^2}(1-\cos{F})\,.
\ee
In the same limit, we have \cite{dixon}
\be
\lambda_{\theta}= \frac12 \beta m_{\pi}^2 \theta\,,
\label{49}\ee
where
\be\label{beta}
\beta = \frac{4x}{(1+x)^2}
\,, \quad {\rm with}\quad  x\equiv \frac{m_u}{m_d}
\,.
\ee
At leading order the only effect of $SU(2)$ symmetry breaking is encoded in the parameter $\beta$.
Therefore, from (\ref{interme}) we obtain the couplings
\be\label{bargaNNS3}
\bar g_{aNN} = 2 \theta f_a \bar g_{a^2 NN} = \frac{\theta}{f_a} \frac{ \alpha \pi \beta m_{\pi}^2}{8 e^3 f_{\pi}}\,. 
\ee
Apart from the factor $\beta$ accounting for isospin symmetry breaking, the result is the same, in form, of the $N_f=2$ case.
In the present case the relation with the mass correction (\ref{deltaM3}) is
\be
\bar g_{a NN} = \frac38\frac{m_{\pi}^2}{m_K^2} \frac{\theta}{f_a}\beta \delta M_N\,.
\label{50}
\ee
From the results in \cite{dixon} one can as well extract the CP-breaking $\eta$ and $\eta'$ couplings to the nucleons.
In fact, the relevant Lagrangian term in the ``chiral limit" reads\footnote{Note the different convention on the sign of $\theta$ with respect to \cite{dixon}.}
\be\label{lagrmix}
{\cal L} = -\theta \frac{\beta m_{\pi}^2 f_{\pi}}{2 \sqrt{3}}\left( 1- \cos{F} \right)(\pi^8+\sqrt{2}\, \eta')\,.
\ee
At the order we are working, it is sufficient to consider the $\eta'-\pi^8$ mixing in (\ref{12}), where the mixing angle is $\theta_{\eta\eta'} \approx -\pi/9$ (see e.g.~\cite{dixon}).
Using these expressions in (\ref{lagrmix}) we obtain
\bea \label{bargetap3}
&& \bar g_{\eta'NN}=-\theta \frac{\alpha\pi \beta m^2_{\pi}}{4 e^3 f_{\pi}^2}\frac{\sqrt{2}\cos\theta_{\eta\eta'}+\sin\theta_{\eta\eta'}}{\sqrt{3}}\,, \\
&& \bar g_{\eta NN}=-\theta \frac{\alpha\pi \beta m^2_{\pi}}{4 e^3 f_{\pi}^2}\frac{\cos\theta_{\eta\eta'}-\sqrt{2}\sin\theta_{\eta\eta'}}{\sqrt{3}} \,.  \label{bargeta3}
\eea
Therefore, apart from the last result, which takes into account the $\eta-\eta'$ mixing, the $N_f=3$ case in the ``chiral limit" gives the same formal outcome of the $N_f=2$ model for the axion couplings.
This suggests that, not unexpectedly, the corrections due to the inclusion of the strange quark are small.

\section{Instanton moduli quantization}\label{instantonmoduli}


We remind the reader that the quantum description of the instanton solution in the WSS model, is obtained in analogy with what is done for the Skyrmion \cite{anw}. Since $M_{B0}=8\pi^2 \kappa \propto \lambda N_c\gg1$, a non relativistic limit can be taken in which the system reduces to a quantum mechanical model for the time-dependent instanton (pseudo) moduli. The slowly rotating instanton solution around $z\approx0$ is given by
\begin{equation}
U_{\rm{cl}}\rightarrow\mathbf{a}(t)U_{\rm{cl}}\mathbf{a}(t)^{\dagger}\,,
\label{isorotation}
\end{equation}
where, setting $\vec{X}=0$ we have 
\be
U_{\rm{cl}} =-\cos\alpha\unit + i\sin\alpha\frac{x^i}{r}\tau^i\,.
\label{umatrix}
\end{equation}
Here,
\begin{equation}
\alpha=\alpha(r)=\frac{\pi}{\sqrt{1+\rho^2/r^2}}\,,
\label{alphadef}
\end{equation}
 with $r=|\vec x|$.

Time-dependent rotations act non-trivially on the non-Abelian components of the gauge field. Consistency with the equations of motion, in turn, induces a non trivial time dependence on the Abelian components too, except that on $\widehat A_0$ \cite{hss}.

In the $z\ll1$ region, the time dependent non-Abelian field components read
\bea
A_0 &=& -i (1-f(\xi)) \mathbf{a}\,\mathbf{\dot a}^{-1} + i(1-f(\xi)) {\dot X}^M \mathbf{a}(g^{-1}\partial_M g)\mathbf{a}^{-1}\,,\nonumber \\
A_M &=& -i(1-f(\xi)) \mathbf{a}(g^{-1}\partial_M g) \mathbf{a}^{-1}\,.
\label{eqnonab}
\eea
The Abelian field components are given in turn by
\begin{equation}
\widehat{A}_z=-\frac{N_c}{8\pi^2\kappa} \left[ \frac{\xi^2+2\rho^2}{(\xi^2+\rho^2)^2}\dot{Z} + \frac{\rho^2}{(\xi^2+\rho^2)^2} \left(\frac{\chi^j x^j}{2}+\frac{\dot{\rho}z}{\rho}\right) \right]\,,
\label{hataz}
\end{equation}
and
\begin{equation}
\widehat{A}_i=-\frac{N_c}{8\pi^2\kappa} \left[ \frac{\xi^2+2\rho^2}{(\xi^2+\rho^2)^2}\dot{X}^i + \frac{\rho^2}{(\xi^2+\rho^2)^2} \left(\frac{\chi^a}{2}(\epsilon^{iaj} x^j-\delta^{ia}z)+\frac{\dot{\rho}x^i}{\rho}\right) \right]\,.
\label{hatai}
\end{equation}
The quantum Hamiltonian for the baryon states is obtained by substituting the above time-dependent solution into the WSS action.

Notice that the structure of the instanton solution above recalls that of the Skyrme model with vector mesons examined in \cite{meissner}. There,  the time-dependent soliton solution includes non-trivial expressions for the time component of the iso-triplet ($\rho_0$), the space components of the iso-singlet vector ($\omega_i$) and the $\eta$-type meson. The contribution of the latter, which turns out to be proportional to the angular velocity of the spinning soliton, turns out to be crucial for many instances. In \cite{bini} it has been used to compute the mass splitting between neutron and proton in the WSS model with non-degenerate quark masses. It is crucial for  providing non zero values for both the CP even coupling $g_{\eta'NN}$ and the CP-odd $\bar g_{\pi NN}$ one. In turn, these couplings are related, using the mixing terms, to further axion-nucleon couplings.

In the present setup, the angular velocity is given by
\begin{equation}
\chi^j=-i\mathrm{Tr}(\tau^j\mathbf{a}^{-1}\mathbf{\dot{a}})=\frac{J^{j}}{4\pi^2\kappa\rho^2}\,,
\label{chij}
\end{equation}
where $J^j$ is the spin operator
\be
J_k = \frac{i}{2} \left( - y_4 \frac{\partial}{\partial y_k} + y_k \frac{\partial}{\partial y_4} - \epsilon_{klm} y_l \frac{\partial}{\partial y_m}\right)\,,
\label{spin}
\ee
with $y_I\equiv\rho a_I$ and $k,l,m=1,2,3$.

In turn, the isospin operator
\begin{equation}
I_k = \frac{i}{2} \left( + y_4 \frac{\partial}{\partial y_k} - y_k \frac{\partial}{\partial y_4} - \epsilon_{klm} y_l \frac{\partial}{\partial y_m}\right)\,,
\label{isospin}
\end{equation}
can also be rewritten as
\be
- i \Tr (\tau^a \mathbf{a}\,\mathbf{\dot{a}}^{-1}) = \frac{I^a}{4\pi^2\kappa\rho^2}\,,
\label{59}\ee
and it is such that only states with $I=J$ appear in the spectrum.

Instantons are identified with fermions via an anti-periodicity condition $\psi(a^I) = -\psi(-a^I)$ to be implemented on the baryon wave functions. The related states have $I=J=\ell/2$ with $\ell=1,3,5,\cdots$ positive odd integers.

The quantum state for a baryon, $|B, s\rangle$, depends on the (iso)spin and on further quantum numbers $n_\rho$ and $n_Z$ which describe excited states or resonances; the case $\ell=1$, $n_\rho = n_Z = 0$ corresponds to the unexcited proton (with isospin component $I_3=1/2$) and neutron ($I_3=-1/2$) and the corresponding wave functions are
\bea
&& |p\uparrow\rangle \propto R(\rho) \psi_Z(Z) (a_1 + i a_2)\,,\quad  |p\downarrow\rangle \propto R(\rho) \psi_Z(Z) (a_4 - i a_3)\,,\nonumber\\
&& |n\uparrow\rangle\propto R(\rho) \psi_Z(Z) (a_4+ i a_3)\,,\quad  |n\downarrow\rangle \propto R(\rho) \psi_Z(Z)(a_1 - i a_2)\,,
\label{pnwave}
\eea
with
\be
R(\rho)=\rho^{-1+2\sqrt{1+N_c^2/5}}e^{-\frac{M_{B0}}{\sqrt{6}}\rho^2} \, ,\qquad \psi_Z(Z)=e^{-\frac{M_{B0}}{\sqrt{6}}Z^2} \,.
\label{pnwave2}
\end{equation}
Generalizations to larger values of $\ell, n_{\rho}, n_{Z}$ can be found in \cite{SS-baryons}.

\section{CP-odd couplings in the WSS model with $N_f=3$}\label{wssnf3}

For the $N_f=3$ case, we can proceed as in the Skyrme model.\footnote{The quark-mass correction to baryon masses has been computed in \cite{Hashimoto:2009st}.}
Starting from the general Lagrangian (\ref{skyrme2flav}) with $N_f=3$, which as stated before is found also in the holographic case, one can diagonalize for the physical modes as in (\ref{mixingswpions}).
In this case the ``chiral limit" $m_{\pi} \ll m_{WV},m_K$ gives
\be
\chi = \frac{m_u m_d}{m_u+m_d}\frac12  B_0 f_{\pi}^2\,,
\label{68}\ee
for the susceptibility and
\bea
&&\eta'_{p} = \eta' + \frac{f_{\pi}}{\sqrt{6}} \frac{a}{f_a}\,,\nonumber \\
&& \pi^a_{p} = \pi^a + \delta^{a8}\frac{f_{\pi}}{2\sqrt{3}} \frac{a}{f_a}\,,
\label{69}\eea
for the axion content of the $\eta, \eta'$.
Here and in the following, we have made use of the GMOR relations \cite{Hashimoto:2009st}
\be\label{WSSgmor3}
m_{\pi}^2 f_{\pi}^2 = 2 c (m_u+m_d)\,, \quad m_{K^\pm}^2 f_{\pi}^2 = 2 c (m_u+m_s)\,, \quad  m_{K^0}^2 f_{\pi}^2 = 2 c (m_d+m_s)\,,
\ee
and
\be
c=B_0 f_{\pi}^2/4\;.
\label{70}\ee
We now put the mass part of the Lagrangian (\ref{skyrme2flav}) on-shell on the nucleon instanton solution and use eq. (\ref{69}) to obtain
\be\label{WSS3flav}
{\cal L}_M = c {\rm Tr} \left[Me^{i\frac{a}{f_a}\left(\frac13 \unit_3 +\frac{1}{2\sqrt{3}}\lambda^8 \right)}\left(U_{cl}- \unit_3\right)+ h.c.  \right]\,.
\ee
At the order we are working, the $SU(3)$ instanton solution is the same as the (two-degenerate flavors) $SU(2)$ one
\be
U_{cl} = e^{ i \vec\lambda \cdot \hat x \alpha(r)}\,,
\label{71}\ee
where $\vec\lambda$ includes the first three Gell-Mann matrices.
By expanding (\ref{WSS3flav}) to quadratic order in $a/f_a$ and using the first relation in (\ref{WSSgmor3}) we obtain
\be\label{resWSS3flav}
{\cal L}_M = \frac{m_{\pi}^2 f_{\pi}^2}{8} (1-\cos{\alpha(r)}) \frac{a}{f_a} + ...\,,
\ee
where the dots stand for a term in $x^3 \sin{\alpha(r)}$ which is vanishing once we integrate the result in space.
Up to the latter term, the result (\ref{resWSS3flav}) is exactly the same one of the $N_f=2$ computation, so the couplings are still the ones in section \ref{WSSbg}.
This is consistent with what we found in the Skyrme model.
As it happened there, the ``chiral limit" $m_{\pi} \ll m_{WV},m_K$ renders subleading the corrections due to the third quark flavor.

\section{Alternative numerical estimates of the couplings}
\label{appendixnumbers}

In this Appendix we collect some alternative numerical estimates of the couplings calculated in the main body of the paper in the Skyrme and WSS models, using a different parameterization with respect to the one employed in section \ref{sectionnumbers}.
Here we plug in the formulae for the couplings the standard values of the parameters usually employed in the two theories we have considered.

Let us begin from the Skyrme model.
For the $N_f=2$ case the coefficient of the Skyrme term is usually fixed to $e \sim 4.84$ by fitting the masses of baryons and the pion; the fit also sets $f_{\pi} \sim 54$ MeV \cite{an}.
Therefore, with these values of parameters the estimate for the magnitude of the axion couplings (\ref{cngen}) in the mass degenerate case of the $N_f=2$ Skyrme model is
\be\label{estimateS2}
\bar g_{a NN} = 2\theta f_a\, \bar g_{a^2 NN} \approx \theta \left( \frac{21\ {\rm MeV}}{f_a}\right)\,. 
\ee
For the coupling of the $\eta'$ in (\ref{cpoddskyrme}), we obtain that
\be\label{bargetapNNSnum}
\bar g_{\eta' NN} \approx  - 0.8\, \theta\,.
\ee

For what concerns $\bar g_{\pi NN}$, the standard choice of parameters in the extended model of  \cite{meissner,Jain:1989kn}, to which we refer for the notations, is
\bea
&&f_{\pi}=93.34\ {\rm MeV}\,,\quad \bar g_{VV\phi} \equiv \sqrt{2} f_\pi  g_{VV\phi} = 1.9\,, \nonumber\\
&&\tilde h \equiv 2 \sqrt{2} f_{\pi}^3 h = 0.4\,, \quad \delta = -1.5\cdot 10^7 ({\rm MeV})^4\,,
\eea
from which \cite{Jain:1989kn}
\be
\frac{\Delta_{\delta}}{\Theta} = 0.70\ {\rm MeV}\,,
\ee
and finally in (\ref{cpoddskyrme}) 
\be\label{bargpiNNSnum}
\bar g_{\pi NN} = -0.005\ \theta\,.
\ee
It is worth noticing that a first estimate of the same coefficient, in chiral perturbation theory with $N_f=3$, gave $|g_{\pi NN}| \approx 0.027 |\theta|$ \cite{crewther}.

The parameters extracted from hadron mass fittings in the $N_f=2+1$ case are instead \cite{dixon}
\be
e = 3.87\,, \quad f_{\pi} =44.5\ {\rm MeV}\,,
\ee
so, remembering that $\alpha \approx 18.25$ and using $x=0.48$ \cite{villadoro}, the numerical estimate of the axion coupling (\ref{bargaNNS3}) in the $N_f=2+1$ Skyrme model is
\be\label{numberS3}
\bar g_{a NN} \approx  \theta \left( \frac{46\ {\rm MeV}}{f_a}\right)\,. 
\ee
The sizable numerical difference between (\ref{estimateS2}) and (\ref{numberS3}) is due to the difference in the two cases of the values of the parameters $f_{\pi}, e$ usually employed to fit nuclear data.
This is a known drawback of the Skyrme model.
For $\eta$ and  $\eta'$ in (\ref{bargeta3}), (\ref{bargetap3}) we obtain
\be \label{Skyrmebargetapi}
{\bar g}_{\eta NN}\approx -1.6\, \theta\,,  \qquad {\bar g}_{\eta' NN}\approx - 1.1\, \theta\,.
\ee

\vspace{0.5cm}
Coming to the holographic WSS model, fitting the model parameters to the experimental values
\be\label{parametersh0}
f_{\pi} = 92\ {\rm MeV} \,, \quad m_{\pi}=135\ {\rm MeV}\,,\quad m_{\rho}=776\ {\rm MeV}\,,
\ee
implies setting \cite{SS1}
\be\label{parametershapp}
\lambda = 16.63\,,\quad M_{KK} = 949\ \rm{MeV}\,.
\ee

In the classical limit where $Z=0$ and $\rho=\rho_{\rm{cl}}$  (\ref{rhocl}), we obtain for the axion derivative couplings (\ref{dercouplingsWSS})
\be
\hat g_A \approx 0.812\,,\quad g_A\approx 0.697\,.
\ee
A class of important $1/N_c$ corrections to the results for the couplings in the WSS model at large $\lambda$ comes from the quantization of the instanton size.
This implies that the nucleon wave function contains a $\rho$-dependent factor \cite{SS-baryons}
\be
R(\rho) = \rho^{-1+2\sqrt{1+N_c^2/5}}e^{-\frac{M_{B0}}{\sqrt{6}}\rho^2}\,.
\ee
Therefore, in the formulae we replace $\rho_{cl}^n$ with
\be
\langle \rho^n\rangle =  \frac{\int \rho^{3+n} R(\rho)^2 d\rho}{\int \rho^3 R(\rho)^2 d\rho}\,.
\ee
Setting $N_c=3$ and fixing the parameters as above one finds for the quantum corrected couplings
\be
\hat g_A \approx 0.527\,,\quad g_A\approx 0.734\,.
\ee

Concerning the axion non-derivative coupling (\ref{cngenwss}) in the mass-degenerate case,
by keeping $\rho$ to its classical value
we obtain the estimate
\be
\bar g_{a NN} \sim  \theta \left( \frac{9\ {\rm MeV}}{f_a}\right)\,.
\ee
Including the quantum corrections as above we get instead
\be
\bar g_{a NN} \sim  \theta \left( \frac{19\ {\rm MeV}}{f_a}\right)\,.
\label{wssquant}
\ee
Analogously,
we get for the $\eta'$ coupling (\ref{bargetawss})
\be
{\bar g}_{\eta' NN}\approx -0.18\, \theta\,,\ee
in the classical case and
\be \label{wssetanucleon}
{\bar g}_{\eta' NN}\approx -0.4\, \theta\,,
\ee
with the quantum corrections.
We finally find for the pion coupling (\ref{gbarpiNNWSS}) 
\be
\bar g_{\pi NN} \approx -0.035\, \theta\,,
\ee
in the classical limit and
\be
\bar g_{\pi NN}\approx -0.041\, \theta\,,
\ee
taking into account the quantization of the instanton moduli.


\newpage


\end{document}